\documentclass[twocolumn]{aastex631}
\submitjournal{ApJ}
\shorttitle{Concordance of Weak Lensing and Escape Velocity Cluster Masses}
\shortauthors{Rodriguez}
\usepackage{amsmath}
\usepackage{amssymb}

\usepackage[utf8]{inputenc}
\DeclareUnicodeCharacter{2212}{-}

\usepackage{enumitem}
\setlist[enumerate,1]{label=\arabic*.}
\usepackage{hyperref}
\def\arnote#1{{\color{black}{#1}}}

\usepackage{natbib}
\usepackage [autostyle, english = american]{csquotes}
\usepackage{mathtools}
\usepackage{amsmath}
\usepackage{amssymb}

\begin{document}
\title{The Concordance of Weak Lensing and Escape Velocity Mass Estimates for Galaxy Clusters}

\author[0000-0002-8813-9116]{Alexander Rodriguez}
\affiliation{Department of Astronomy, University of Michigan, Ann Arbor, MI, USA}
\email{alexcrod@umich.edu}

\author[0000-0002-1365-0841]{Christopher J. Miller}
\affiliation{Department of Astronomy, University of Michigan, Ann Arbor, MI, USA}
\affiliation{Department of Physics, University of Michigan, Ann Arbor, MI, USA}
\email{christoq@umich.edu}

\begin{abstract}
{In the $\Lambda$CDM paradigm, the masses of the galaxy clusters inferred using background galaxies via weak-lensing shear should agree with the masses measured using the galaxy projected radius-velocity phase-space data via the escape velocity profile. However, prior work indicates that the correlation between caustic-inferred escape masses and weak lensing masses is statistically consistent with zero. Based on recent advancements in the measurement of the escape edge and its physical interpretation, we conduct a revised comparison between these two independent mass inference techniques for 46 galaxy clusters between $0.05 \le z \le 0.3$ and over an order of magnitude in mass, $14.4 \le {\rm log}_{10} M/M_{\odot} \le 15.4$. We find excellent agreement, with a correlation ($0.679^{+0.046}_{-0.049}$), and a mean relative difference between the two mass measurements consistent with zero (0.02 $\pm$ 0.02 dex). The observed scatter between these direct mass estimates is 0.17 dex and is consistent with the reported individual mass errors, suggesting that there is no need for an additional intrinsic component. We discuss the important practical consequences of these results, focusing on the systematic uncertainties inherent to each technique, and their implications for cosmology.}

\end{abstract}

\section{Introduction}
\label{sec:Introduction}


Galaxy clusters are the largest virialized objects in the universe and are excellent laboratories for studies of dynamics on cosmological scales. With modern spectroscopic surveys and multiplexed instruments, we are finally collecting sufficient data to conduct detailed and/or statistical studies of cluster galaxy radius-velocity phase-spaces. From these analyses, we can infer properties like mass-density profiles, dynamical histories, population segregation, \arnote{among other properties}. \citep{Geller13, Rines2013, Rines+2016, stark2016, Rhee2017, sartoris+2020, Coenda2022, Rodriguez+2024}.


Past work on the dynamics within clusters from cosmological simulations has warned us to be wary of systematics. Dark matter (DM) particles are dense enough to experience many-body gravitational interactions which can accelerate particles to speeds faster than the potential allows or slow them down via dynamical friction \citep{Behroozi+2013}. In turn, ensemble statistics like the velocity dispersion ($\sigma_v$) can show biases depending on how tracers are selected in the phase-space. For instance, DM particle dispersions can be smaller than sub-halo dispersions and red ``galaxies'' can have different dispersions than ``blue'' ones \citep{Biviano92,  Gifford_2013, Saro2013, Barsanti2016, Bayliss2017}.  The intracluster medium (ICM) is also a dynamical characterization of a cluster. Hydrodynamic simulations show that the ICM is rarely in perfect hydrostatic equilibrium and is instead affected by non-gravitational physics through cooling, pressure gradients, and shocks. This hydrostatic bias is a subject of intense study\citep{Rasia+2006,Nagai+2007,Lau+2009,Vazza+2009,Nelson+2014}.While the systematics error on cluster mass estimates from these effects is not large (typically $\lesssim 10\%$ on mass inference), their presence limits the accuracy and precision of dynamically inferred cluster masses. 

One solution to the above systematic issues is to use a phase-space velocity field surface rather than the ensemble statistics like velocity dispersion or X-ray temperature. This has been done through so-called  caustic profiles \citep{diaferio+1997, Diaferio1999, Diaferio2005, Serra2011, Geller13, Gifford_2013, Pizzardo2023A&A}. By measuring the maximal velocity boundary in radial-velocity phase-spaces, one has a map from the escape velocity to the potential profile \arnote{(caustic estimates can also be combined with weak lensing data for joint-likelihood mass estimation, such as in \citet{umetsu2025})}.
Just as weak lensing shear enables an instantaneous density field estimate, the escape velocity enables an instantaneous estimate of the potential. Galaxies at escape speeds rarely interact with other galaxies, negating impacts from dynamical friction \citep{Gnedin2003, aguilar+2008, binney+2008}. Due to the sparsity of the galaxies in the cluster volume, acceleration \arnote{from} three-body interactions is negligible. These assumptions have been tested using sub-halos in N-body simulations \citep{miller+2016} where caustic-inferred cluster masses have been shown to be good tracers of the halo mass \citep{Gifford_2013, Gifford2013, Gifford_2017, Pizzardo2023A&A}.

Given that both weak lensing and the escape velocity should be independent of the dynamical state, we expect good agreement between these two independent mass measurement techniques. Simulations suggest that the scatter between cluster caustic or weak lensing mass and the true halo mass is $\sim$ 25\% \citep{Gifford_2013, Pizzardo2023A&A, Becker+2011, Bahe2012}.  These same works suggest those biases to be small at $\sim$ 5-10\%. Over the past 10 years, researchers have used the highest quality and largest quantity of cluster imaging and spectroscopy data to compare caustic and weak lensing masses of clusters \citep{Geller13, Hoekstra2015, Herbonnet2020}. The results have been disappointing, with very poor agreement between the masses and no satisfactory explanation for the general discrepancy. \arnote{\citet{Herbonnet2020} (hereafter H20) suggest that unaccounted for scatter and bias in the dynamical masses could be to blame, although another possibility is that weak lensing mass estimates are also subject to systematic uncertainties, particularly in dynamically disturbed or un-relaxed clusters \citep{Simet_2016,Lee_2023}.} 

In this work, we propose that the primary issue has been with how the escape profile has been inferred and interpreted. Our work builds upon the work of \citet{stark2016} and \citet{miller+2016} who re-visited the theoretical interpretation of escape profile in terms of inferring cluster masses and potentials; \citet{halenka+2020} who reformulated the primary systematic of the edge measurement: its statistical suppression; and that of \citet{Rodriguez+2024} who developed a technique to quantify edge measurement uncertainties in order to measure the escape mass of galaxy cluster Abell S1063.  We expand that sample to nearly 50 clusters in order to conduct a statistical comparison with weak lensing masses.

The paper is structured as follows: in~\S\ref{sec:Escape Velocity Theory}, we present a theoretical overview of the escape velocity methodology. In~\S\ref{sec:Suppression}, we present a detailed analysis on how to infer the 3D cluster mass from the observed phase-spaces.  In~\S\ref{sec:Results} we apply our algorithms on real data. In~\S\ref{sec:Discussion}, we discuss the cosmological consequences of our findings and implications.

\section{Escape Velocity Theory}
\label{sec:Escape Velocity Theory}

As derived for the non-linear field equations in \citet{nandra+2012}, the acceleration of a test particle inside a massive object, which itself is embedded in an accelerating cosmological background, will be determined from the inward pull towards the massive object and an outward pull from the expanding spacetime
\begin{align}
\label{eq:nabla_phi}
    a_{\text{eff}}&=\nabla\Phi_{\text{eff}}(r)=\nabla\Psi(r)+q(z)H^2(z) r\hat{r}
\end{align}
Here, $\Phi_{\text{eff}}(r)$ is the effective potential profile, $\Psi(r)$ is matter-only potential profile, $q$ is the deceleration parameter, and $H$ is the Hubble expansion rate.
Unlike in a non-accelerating spacetime, there is a radius relative to a cluster's gravitational center where the acceleration due to gravity balances the outward acceleration \arnote{from} the expansion of universe given by 

\begin{align}
\label{eq:r_eq}
    r_{\text{eq}}&=\left(\frac{GM}{-q(z) H^2(z)}\right)^{1/3}
\end{align}
where $M$ is the enclosed mass, and for a flat universe, $H(z)=H_0\sqrt{\Omega_\Lambda+\Omega_M(1+z)^3}$, and $q(z)=\frac{1}{2}\Omega_M(z)-\Omega_\Lambda(z)$.

The corresponding Poisson equation allows us to determine the gravitational potential governing the inward pull as
\begin{equation}
\Psi(r) =-{\rm 4\pi G}\Big{[} \frac{1}{r}\int_{0}^{r}\rho(r')r'^2dr' + \int_{r}^{r_{eq}}\rho(r')r'dr'\Big{]},
\label{eq:phi_poisson}
\end{equation}
where $\rho(r)$ is the matter density profile integrated to $r_{eq}$. This ensures the physical requirement of balanced forces so that the escape speed,  $v_{\text{esc}}^2(r)=-2\Phi_{\text{eff}}$, for any massive tracer is zero at this radius. 

We then have,
\begin{align}
\label{eq:vesc_final}
    v_{\text{esc}}^2(r)&=-2(\Psi(r)-\Psi(r_{\text{eq}})) -q(z)H^2(z) (r^2-r^2_{\text{eq}}).
\end{align}
While the phase-space density may show local variations \arnote{as a result of} sub-structure, anisotropies in the tracer velocity vectors, \arnote{and other reasons (see~\S\ref{subsec:testing})}, the extrema of the phase-space tracer velocities will always be bounded by the escape surface. Equation \ref{eq:vesc_final} results in a lower escape profile than in a non-accelerating spacetime and is therefore valid at times after dark energy begins to dominate the energy density of the universe. We can use the extrema of the phase-space galaxy velocities as a direct constraint on the effective potential profile, which includes both cluster mass and cosmology.


Modification of the 3D radial escape profile away from Newtonian expectations has been characterized in $\Lambda$CDM simulations \citep{Behroozi+2013, miller+2016}. However, 
\citet{halenka+2020} found that the observed escape profile will be suppressed to the one defined by the effective potential \arnote{from} the under-sampling of the phase-space. 
\begin{figure*}[ht]
    \centering
    \begin{minipage}{\textwidth}
        \includegraphics[width=.33\textwidth]{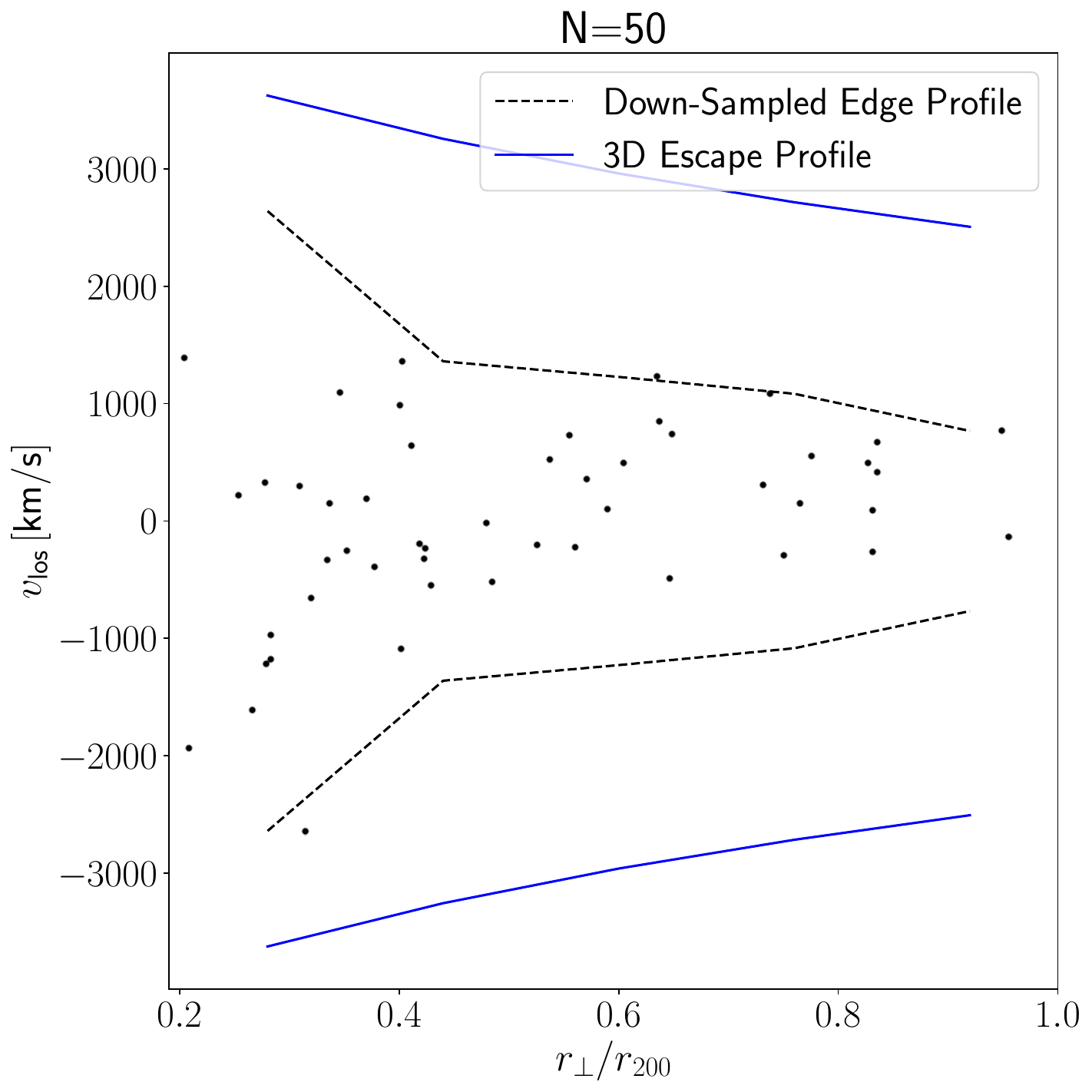}%
        \includegraphics[width=.33\textwidth]{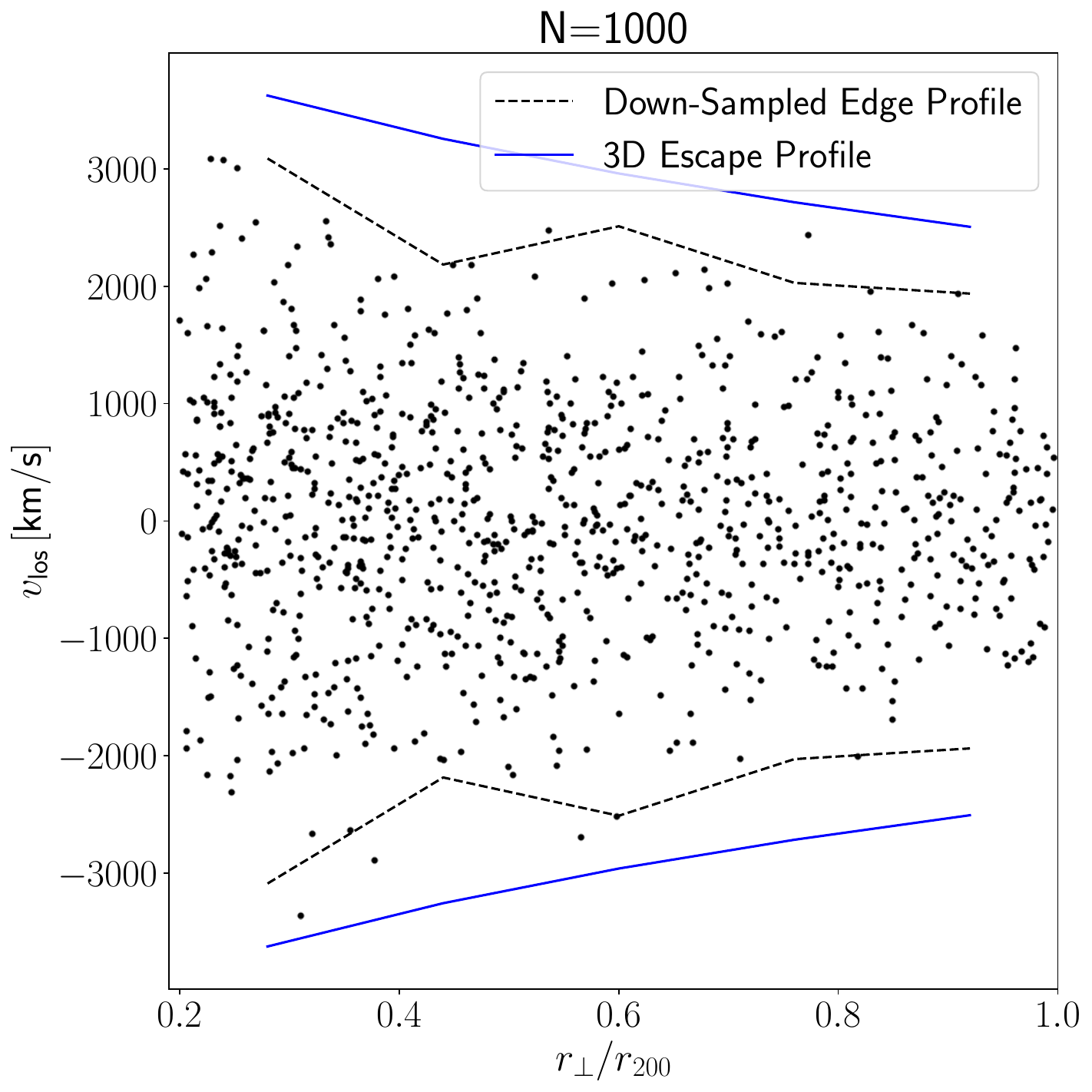}
        \includegraphics[width=.33\textwidth]{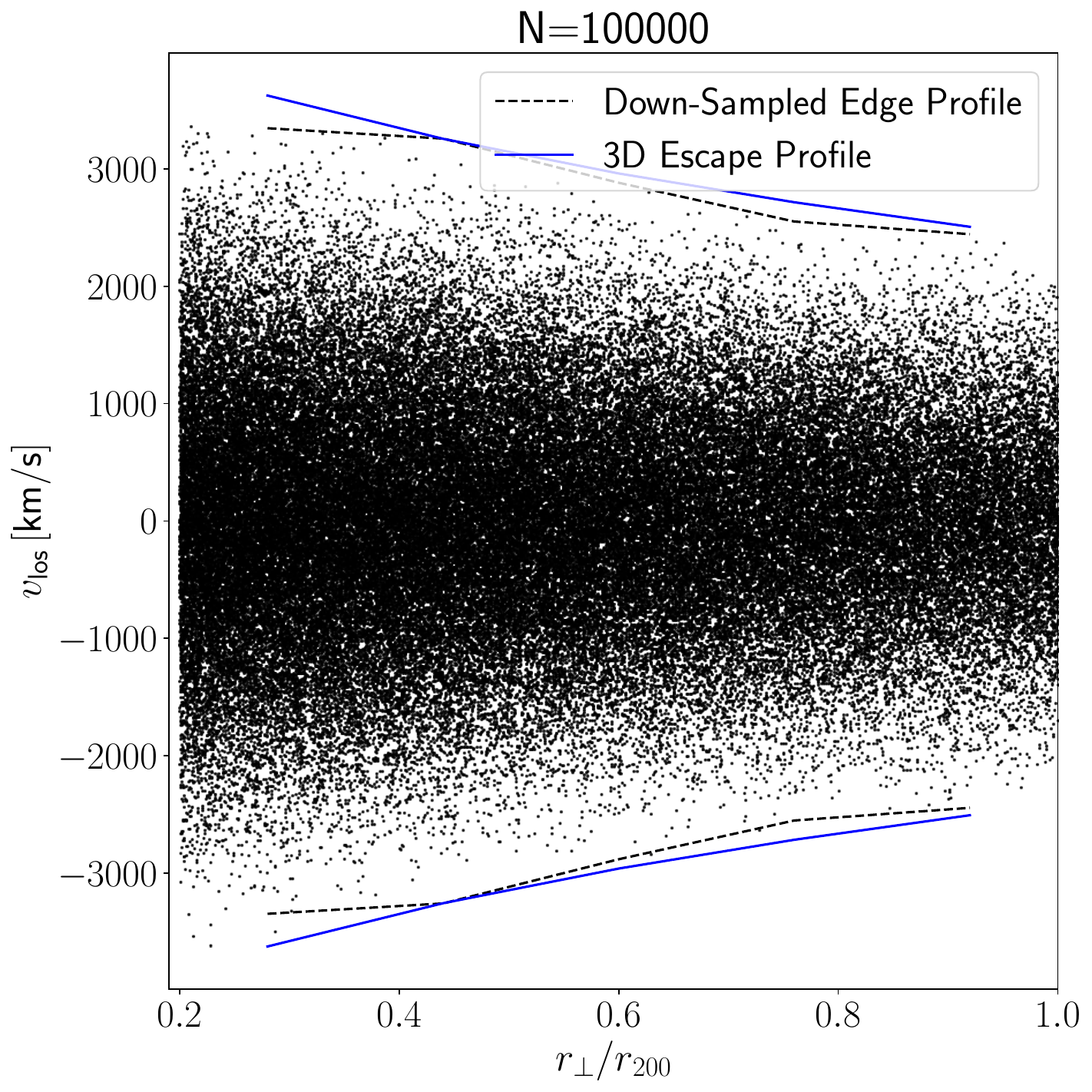}
    \end{minipage}

\caption{\label{fig:Phase_space_AGAMA} A cluster phase-space sampled with an increasing number of tracers. The data were generated using the AGAMA framework (see text) for a $M_{200} = 10^{15}\,M_{\odot}$ cluster at redshift $z=0.01$ with $\beta = 0.25$. The black points are the inferred line-of-sight velocities from a single line-of-sight draw. The down-sampled edge profile (black dotted lines) corresponds to the maximum absolute velocity within each radial bin. The blue (solid) lines correspond to the 3D escape profiles (equation \ref{eq:vesc_final}). Note the increased suppression of the edge as sampling decreases.
} 
\end{figure*}
The suppression of the 3D edge profile (denoted $Z_v$) has been shown to be parametrized by a function corresponding to the number of tracers, $N$, between between $0.2\,r_{200} \le r \le r_{200}$, given by \citep{halenka+2020,Rodriguez+2024}
\begin{align}
\label{eq:vesc_down}
\langle v_{\text{esc, \text{down-sampled}}}\rangle(r_{\perp})&= \frac{\langle v_{\text{esc}}\rangle(r_{\perp})}{\langle Z_v(N) \rangle}
\end{align}
where $\langle \cdot \rangle$ is over many lines-of-sight and radii between $0.2r_{200}$ and $r_{200}$ inferred from the critical density. We show this effect explicitly in Figure \ref{fig:Phase_space_AGAMA}. The suppression is the primary systematic uncertainty when using the observed phase-space edge (at low sampling) to infer the true escape velocity profile.

For the rest of this paper, we refer to the observed phase-space edge profile as the down-sampled edge profile, unlike earlier works \citep{halenka+2020,Rodriguez+2024} which refer to it as a projected or line-of-sight edge profile\footnote{While the phase-spaces are generated from data along the line-of-sight, the suppression of the edge profile is a result of sparse sampling and has nothing to do with radius or velocity vectors along the line-of-sight.}
In order to physically interpret the down-sampled escape edge in equation \ref{eq:vesc_down}, we need this suppression function $Z_v$, the cluster density/potential model parameters $\Psi$, and the cosmology, $q$, $H$, and $r_{\text{eq}}$. 

In this work, we generally use a flat $\Lambda$CDM universe with $\Omega_m = 0.3$ and $H_0 = 70$km s$^{-1}$Mpc$^{-1}$, except when we compare to the Millennium simulation or otherwise stated.
We use the \citet{dehen+1993} potential parametrization to model the 3D escape velocity in equation \ref{eq:vesc_final}:
\begin{equation}
\label{eq:Dehnen}
\Psi(r) =
\begin{cases}
\frac{GM_{\text{tot}}}{r_s} \frac{-1}{2-\gamma} \left[ 1 - \left(\frac{r}{r+r_s}\right)^{2-\gamma} \right], & \text{if } n \neq 2, \\
\frac{GM_{\text{tot}}}{r_s} \log\left(\frac{r}{r+r_s}\right), & \text{if } n = 2.
\end{cases}
\end{equation}
where the total mass $M_{\text{tot}}$ is a normalization factor, $r_s$ is the scale radius, and $\gamma$ is the Dehnen index. When only the mass at $r_{200, \text{critical}}$ is available, we use the mass-concentration relation from \citealt{duffy+2008} and numerically map the Dehnen density profile to the NFW \citep{navarro+1997}. 

We estimate $r_\text{{eq}}$ assuming an \citet{duffy+2008} mass-concentration relation and then minimizing the $\chi^2$ difference in the Dehnen and NFW forms over the range $0.2 \le r/r_{200} \le 1$. We integrate the corresponding Dehnen density profile out to radius r, interpolating the radius at which the inward gravitational acceleration and the outward cosmological acceleration are equal as $r_{\text{eq}}$.  



\section{The Down-sampled Projected Escape Profile}
\label{sec:Suppression}
The suppression of the true escape edge in a cluster, $Z_v$ in equation \ref{eq:vesc_down}, was introduced by \citet{halenka+2020}. It is a power-law function with a simple $N$ dependence, where $N$ is the phase-space galaxy count over some prespecified projected radial range. \citet{Rodriguez+2024} noted the radial dependence of suppression as well as the effects of binning. In this section, we explore $Z_v$ in more detail and pay specific attention to its distribution function (as opposed to its mean). 

Our procedure starts with isolated spherical cluster projected phase-spaces to build our model for $Z_v$. We then test our model against projected semi-analytic galaxy data from an N-body simulation which has more realistic phase-spaces. Finally, we incorporate observation effects onto the simulation galaxies in order to conduct an end-to-end test of the $Z_v$ model. 

\subsection{Modeling the Suppression}
\label{subsec:Analytic Construction}


We use the Action-based Galaxy Modeling Architecture (AGAMA) \citep{Vasiliev2019} to create galaxy positions and velocities given a density profile \citep{dehen+1993} and velocity anisotropy.  We use a constant velocity anisotropy parameter $\beta=0.25$, comparable to the measured values from simulations \citep{Lemze+2012, Stark+2019}. We test this assumption in the next subsection. We remove any galaxies above the allowable escape speed defined in equation \ref{eq:vesc_final} for our cosmology and redshift. We keep all tracers to within $10\times r_{200}$ and project the positions and velocities along a line-of-sight \citep{Gifford2013,halenka+2020,Rodriguez+2024}. The line-of-sight velocities are with respect to the mean velocity of all tracers over the range $0.2 \le r_{\perp}/r_{200} \le 1$. The radial coordinates are with respect to the center of the density profile. As a consequence of the projection, the phase-spaces will have cluster members with 3D radii that are larger than or equal to their projected radii. So there is contamination from cluster members, but there is no contamination from non-member galaxies (e.g., from nearby structure). 
 
For the sampling of the phase-spaces, we use a range of $N$ between $50<N<1200$, where $N$ is defined as the number of tracers between $0.2\,r_{200} \le r_{\perp} \le r_{200}$, following the definition from \citealt{Rodriguez+2024}. This range is chosen to cover the range of possible sampling in typical observational samples.  We then divide the phase-space into 5 bins between $0.2\,r_{200}$ and $r_{200}$ and estimate the down-sampled edge profile from the maximum absolute velocity in each radial bin. Figure \ref{fig:Phase_space_AGAMA} was produced with this framework for a single line-of-sight.


\begin{figure}
\includegraphics[width=.4\textwidth]{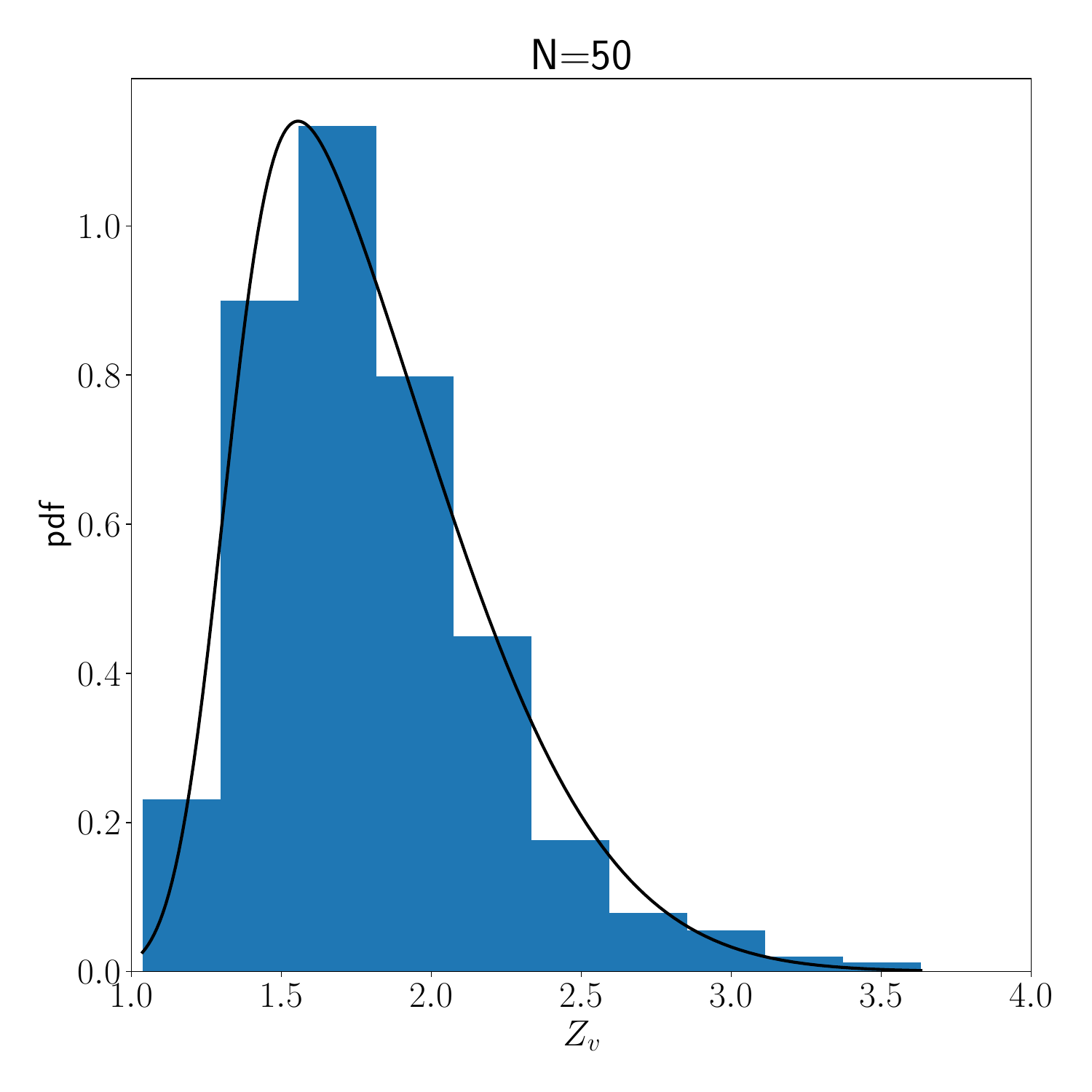}\\
\includegraphics[width=.4\textwidth]{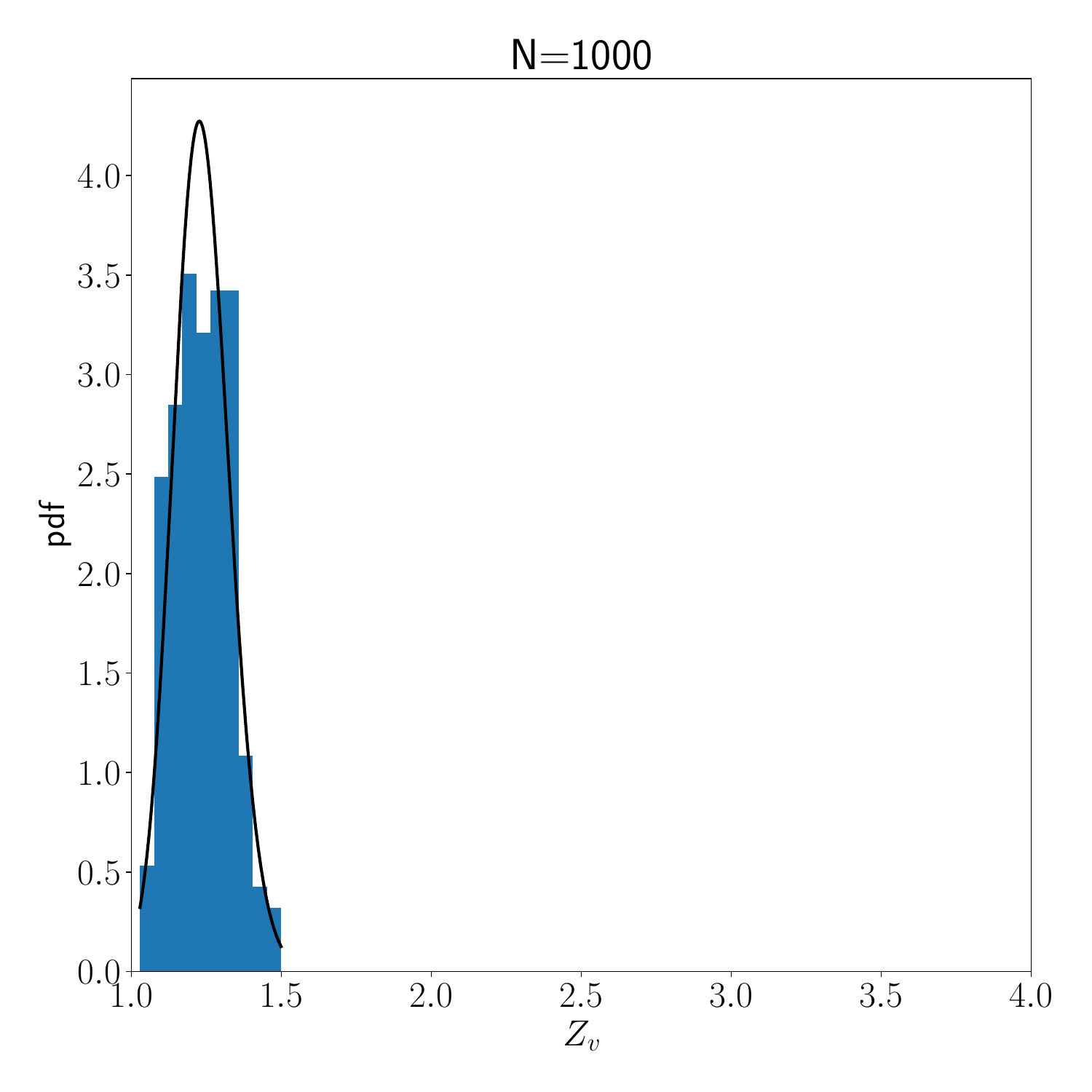}

\caption{\label{fig:Zv_50_1000} The normalized distributions of the suppression ($Z_v$) in AGAMA for a $N=50$ (top panel) and $N=1000$ (bottom panel) for a $10^{15}\,M_{\odot}$ cluster at redshift $z=0.01$, inferred for 1000 different viewing angles in the innermost bin. The black lines correspond to fits to the distributions using the skewness, location, and scale \citep{Azzalini+2009}. The poorly sampled system exhibits high suppression values and a long tail. 
} 
\end{figure}

In Figure \ref{fig:Zv_50_1000}, we show the distribution of the suppression ($Z_v$) in the inner-most bin for $N=50$ (top panel) and $N=1000$ (bottom panel) for a $10^{15}\, M_{\odot}$ cluster at redshift $z=0.01$ measured from 1000 lines-of-sight. The well-sampled system is nearly Gaussian while the more poorly-sampled system has a distribution which is not only much wider, but also skewed to higher values of $Z_v$. The shape of the $Z_v$ distribution characterizes the scatter in edge measurements inherent in the lines-of-sight. As we increase the number of tracers in the phase-space the suppression decreases and large values become rare. The large suppression values occur when member interlopers are used to identify the edge. These interlopers are actually much further out (in their 3D radius) where the potential is much smaller. As we increase the phase-space sampling, less common tracers near the true escape velocity become more prevalent at any radius. The $Z_v$ distribution tends towards a delta function at $Z_v = 1$ as sampling increases.

We model the $Z_v$ distribution as a skewed-normal according to 
\begin{align}
\label{eq:Zv_pdf}
    Z_v(N; \xi, \omega, \alpha) = \frac{2}{\omega} \phi\left(\frac{N - \xi}{\omega}\right) \Phi\left(\alpha \frac{N - \xi}{\omega}\right)
\end{align}
where $\xi$ is the location parameter, $\omega$ is the scale parameter, and $\alpha$ is the skewness parameter, and $\phi(z)$ is the standard normal PDF and $\Phi(z)$ is the standard normal CDF. In equation \ref{eq:Zv_pdf}, $Z_v$ depends on the phase-space sampling $N$. 

\citet{halenka+2020} found a small dependency on the cluster mass attributed to keeping $N$ fixed against varying $r_{200}$. We also find small dependencies on mass and redshift, again at the percent level and attribute these to variations in the density and phase-space window used in the count. \arnote{As a result of these dependencies}, equation \ref{eq:Zv_pdf} is measured on a grid which includes $N$, mass, and redshift -- the skewed Gaussian parameters at fixed mass and redshift are linearly related to $N$ \arnote{(see Figure \ref{fig:a1})}. We store a table which contains the best-fit slope and intercept values for $\xi$, $\omega$, and $\alpha$ as a function of $N$, $M_{200}$, and redshift $z$, in each radial bin. We will investigate the percent-level dependencies on mass and redshift in a future work.


\subsection{Testing the Suppression Model}
\label{subsec:testing}
Our model for $Z_v$ stems from phase-spaces generated directly from analytic (spherical) gravitational theory for isolated potentials. We can test the accuracy and precision of our parametrization of $Z_v$ using more realistic clusters from an N-body simulation. In doing so, we assess potential systematic biases which could be caused by locally varying cosmological backgrounds, internal cluster substructure, cluster mergers, asphericities, hyper-escape-speed galaxies, variable velocity anisotropies, non-cluster interlopers, etc. None of these are present in the AGAMA phase-spaces.

We use the same halo and simulated galaxy sample as originally defined in \citet{Gifford_2013}. The data are from the Millennium simulation with galaxies populated using semi-analytic techniques \citep{Springel2005, Guo2013}.  The halo sample consists of 100 clusters, with redshifts all below $z=0.15$, and masses between $\sim 10^{14}\,M_{\odot}$ and $\sim 10^{15}\,M_{\odot}$. Each cluster is inside a box extending to 60$h^{-1}$Mpc where we use all of the DM particles to constrain the halo density profiles with the Dehnen parametrization. For the phase-spaces, we use the semi-analytic galaxy population as described in \citet{halenka+2020}.

In generating the projected phase-space data, we follow nearly the same approach as AGAMA, using all galaxy data out to $10\times r_{200}$. We center on each cluster's density peak and use a mean velocity corresponding to the redshift of the cluster. A critical difference is that we do not cull galaxies given that the Millennium simulation already contains the cosmological acceleration \citep{miller+2016}. Another important difference to the AGAMA phase-spaces is that the large box ensures that localized large-scale structure will be present in the projected phase-spaces (i.e., non-cluster interlopers).  

We then apply an interloper rejection algorithm based on the shifting-gapper technique \citep{fadda+1996,girardi+1996,adami+1997, Wing2013}, which requires an initial velocity gap choice and also a binning scheme (where we fix the number of galaxies to same number per bin). Due to the presence of non-cluster interlopers at random locations in the Millennium projected phase-spaces, the choice of binning and velocity gap size could have an effect on the interloper removal and thus the edge identification. \arnote{Hence}, these could be hyper-parameters which require fine tuning.
\begin{figure}
\includegraphics[width=.5\textwidth]{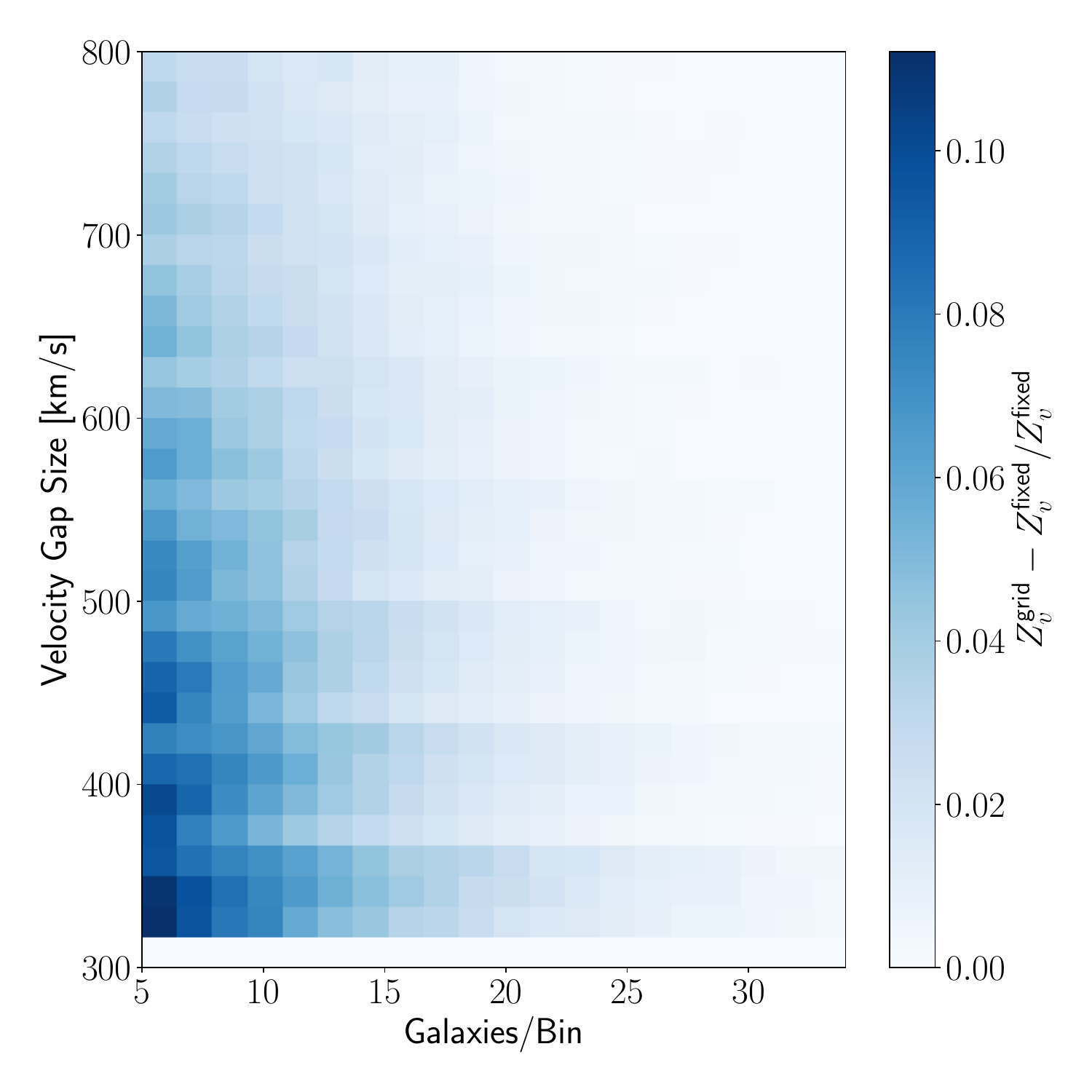} 
\caption{\label{fig:Interloper_grid} Fractional differences between the suppression values of a grid of velocity gap and binning choices for the shifting-gapper \citep{Gifford2013}, and a fixed fiducial measurement, analogous to \citet{Rodriguez+2024}. There is an obvious trend towards increasing stability of the parameter choice, which is used for robust identification of interlopers.}
\end{figure}

To choose the shifting-gapper velocity gap and binning scheme, we follow the procedure in \citet{Rodriguez+2024}. We choose fiducial values for the number of galaxies per bin and the velocity gap and then apply the shifting-gapper to remove interlopers. We next measure the edge as the absolute maximum velocity in five radial bins between $0.2 \le r/r_{200} \le 1$. We also enforce a rule that the down-sampled edge profile be monotonically decreasing (outside of the first 2 bins), given the monotonic nature of gravitational potentials. Note that there is no need for this step in the AGAMA phase-spaces since they lack non-cluster interlopers. Finally, we measure the ratio of the true escape edge to the observed (and projected) down-sampled edge while varying the initial gap size and galaxy per bin count.

In Figure \ref{fig:Interloper_grid} we show this ratio as a function of the initial gap size and binning. Following \citet{Rodriguez+2024}, we identify a region inside the range of explored gap and binning where the ratio is robust such that no fine\ tuning is required. This analysis was also done for Abell S1063 in the above work, except they focused on the five most massive and well sampled Millennium phase-spaces with $N > 500$ and similar phase-space sampling to the data. Here, we use the entire set of 100 Millennium clusters over the range in mass from $10^{14} \le M/M_{\odot} \le 10^{15}$. This mass range and phase-space sampling ($\langle N \rangle = 180$) more closely resembles the data we will analyze. Not surprisingly, the gap and binning values where the edge measurement is robust are smaller than what was used in \citet{Rodriguez+2024}.

We choose 600 km/s for the velocity gap and 20 galaxies per bin when running the shifting-gapper throughout the rest of this paper. The scatter induced onto the edge measurement from variations of $\pm{5}$ galaxies per bin and $\pm{100}$km/s in the initial gap is small ($< 1\%$). \arnote{We find} our interloper identification algorithm is not a source of systematic uncertainty in the edge measurement and no fine-tuning is required.

\subsection{Edge Measurement and Error}
\label{subsec:edge measurement}
In observed data, galaxy redshifts will have uncertainties. Typical redshift errors for modern spectroscopic surveys range from $30\,\text{km/s}\le \frac{c\sigma_z}{1+z}\le 140$ km/s \citep{Laureijs+2011,Bolton+2012,Guzzo+2014}. We use the Millennium data to understand how redshift measurement error carries through into an error on the edge measurement. For a given cluster, we build a fiducial phase-space from a random line-of-sight and remove interlopers using the shifting-gapper parameters defined above. For each galaxy in the phase-space, we then add velocity noise by sampling from a Gaussian with a pre-specified variance. We then re-measure the edge as before. We repeat this exercise 10000 times and measure the variance in each radial bin. 

Nominally, this variance should be same as the input choice of variance we randomly added to each galaxy in the fiducial dataset. However, the enforcement of monotonic edges acts to smooth the edge. We create monotonicity in the edge by ensuring that each edge be equal to or less than the edge measurement of its next nearest radially inward bin. \arnote{Hence}, moving outwards radially in the phase-space, when we encounter a bin where the maximum absolute galaxy velocity is higher than that measured just inward, we infer the edge as the velocity of the edge of the next inner bin. 
\begin{figure}[h]
\includegraphics[width=.5\textwidth]{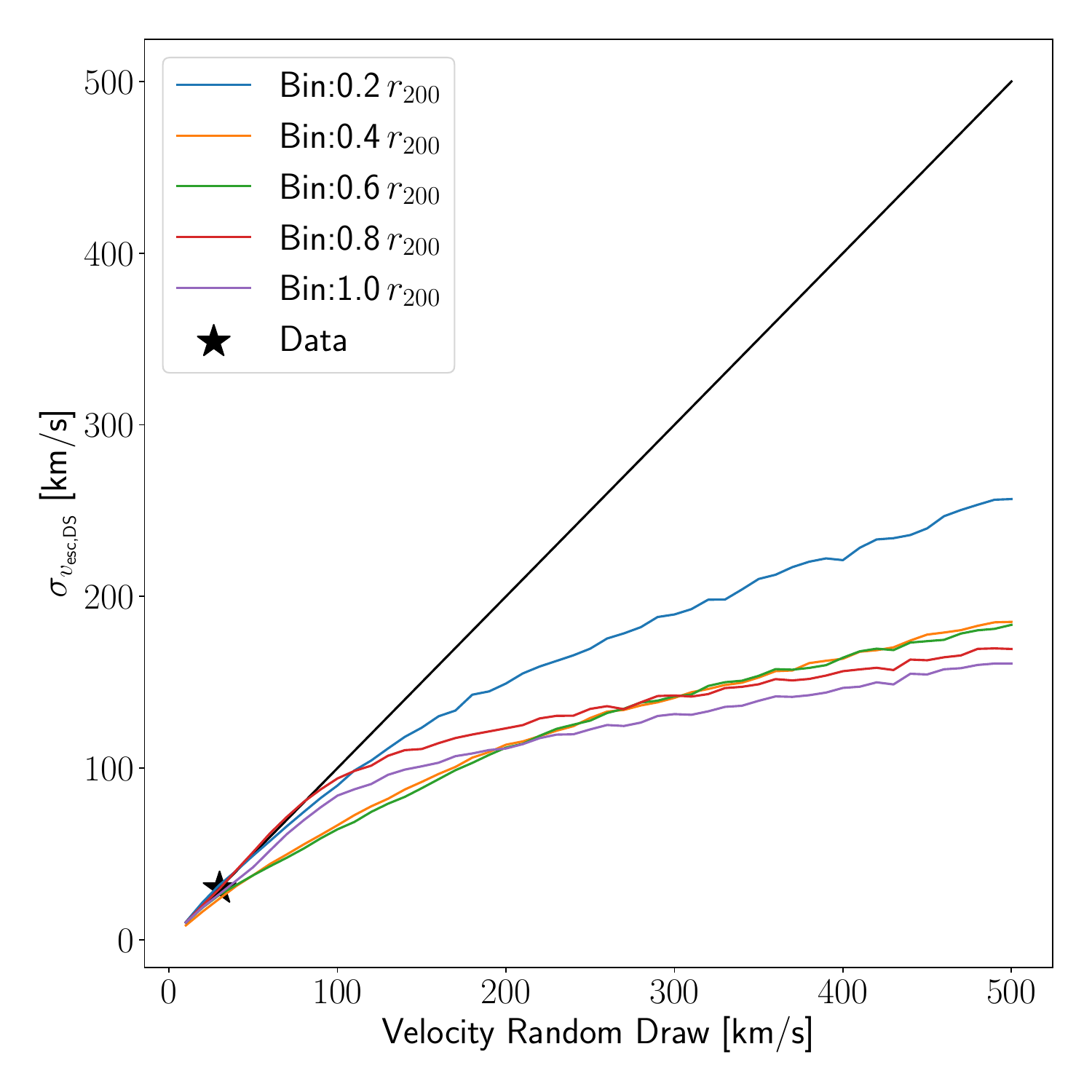} 

\caption{\label{fig:sigma_err_draws} Random velocity draws (assumed to be from a normal distribution) that are added to the galaxy velocities in an example halo in Millennium, then used to measure the standard deviation in the resulting down-sampled edge profile ($\sigma_{v_{\text{esc,DS}}}$), over 10000 velocity draws. These are shown for different bin numbers, as shown by the different colored lines. The uncertainty in the down-sampled edge profile is always less than or equal to the error on the velocity itself. The star shows the value of $\sigma=30$ km/s assumed in the data. The black solid line shows unity slope.
} 
\end{figure}

We note that by adding a random error into the galaxy velocities we can expect the identification of edge galaxies to sometimes change. For instance, if the Gaussian draw adds a positive shift to a positive line-of-sight velocity such that it becomes larger than its nearest inward bin, then the monotonic rule is activated and that galaxy (with its large shift) is no longer used to define the edge. If the shift is negative, the original edge galaxy shifts down and it is possible that another galaxy in that bin is defined as the new maximum. In a sense, the monotonic rule acts as a secondary interloper rejection tool. The net result of this is that the variance placed on the phase-space galaxies need not translate to the same variance in the edge.

In Figure \ref{fig:sigma_err_draws}, we plot the square root of the bin variances (y-axis) for the edge measurements after adding random Gaussian errors to the galaxy redshifts (x-axis). The five colors are for each radial bin, which the highest induced edge uncertainty being from the inner most bin and with a decreased induced edge uncertainty with increasing radius. We note that the standard deviation of the induced edge uncertainty is always less than the induced velocity error on the galaxies from the standard deviation of the random additions.

When small errors are added to the galaxies ($<100$km/s), the induced edge uncertainty has a standard deviation that matches the standard deviation of the galaxy velocity errors. \arnote{As a result of} the small galaxy redshift error, the same galaxies are used to the define the edge and the error translates directly from the galaxy to the edge.  However, as the error on the galaxy velocities increases, it becomes more likely that different galaxies are used to define the edge (compared to before the errors are added). For large errors, it becomes more likely that the monotonicity requirement needs to be enforced on the edge inference and interloper rejection is happening more often. In addition, as we move outward in radius, the chance increases that the monotonicity rule is activated at some point in the inner bins. Each time the rule is activated, the overall inferred edge is smoothed resulting in a lower edge variance. In the absence of monotonicity a similar effect is still present where the variance is always less than the input variance, although bin dependence of this effect vanishes.

Note that even the first radial bin has a smaller induced edge uncertainty for large velocity errors $ \gtrsim 200$km/s. This zeroth bin is not smoothed by the monotonic rule. For this bin, smaller induced edge uncertainty stems from the re-identification of the edge galaxy after the errors are added.

We also examine the error distribution on the edges after including galaxy redshift errors. We find that the edge uncertainties in each radial bin are consistent with Gaussian distributions \citep{Normal1973}. \arnote{Or put differently}, after inducing Gaussian velocity errors, the edge measurement errors remain Gaussian.

\subsection{Edge Measurement Summary}
\label{subsec:edge_error_summary}
In summary, in~\S\ref{subsec:Analytic Construction}, ~\S\ref{subsec:testing} and~\S\ref{subsec:edge measurement}, we revisit the suppression function and find it to be represented by a skewed Gaussian with its location, scale, and skewness primarily dependent on the projected number of galaxies in the phase-space, $N$. We find a very weak dependence on the cluster mass and redshift which we include in our suppression model. The skewed shape of the distribution of suppression values at fixed $N$ is \arnote{from} line-of-sight variations. At small sampling, the function is more skewed to higher values than when the sampling is high, where it is nearly Gaussian. 

We then tested whether this function holds when there are non-cluster interlopers identified and removed using the shifting-gapper technique. Since this technique is not perfect, we also adjusted the edge measurement algorithm to enforce radial monotonicity in the edge. We find that variations in the shifting-gapper parameters induces almost no additional scatter into the edge measurement. \arnote{In summary}, the edge measurement is robust to how we account for interlopers. Finally, we add in measurement uncertainties from a normal distribution to the galaxy redshifts and find that our edge algorithm smooths out the induced edge uncertainty. However, the edge uncertainties remain Gaussian.

We express the total edge measurement uncertainty as:
\begin{equation}
\label{eq:edge_error}
\sigma^2_{edge} = \sigma^2_{los} + \sigma^2_{inter} + \sigma^2_{cz}
\end{equation}
where the terms refer to the line-of-sight scatter in the edge, the scatter induced by the interloper rejection algorithm, and the scatter induced by galaxy redshift errors. The line-of-sight uncertainty is encoded in the scale and skewness of the suppression function. The interloper scatter is small enough that we will ignore it for the rest of this paper. Redshift errors will be included in our analysis as needed.

\subsection{Recovering the Mass}
\label{subsec:likelihood}
We now conduct an end-to-end test of our algorithm to infer cluster masses from projected phase-space galaxy data using galaxies in halos from the Millennium N-body simulation. We use the same data as in~\S\ref{subsec:testing} and~\S\ref{subsec:edge measurement}. As we see shortly, we will find a lack of bias when applying the AGAMA model in N-body simulations--a critical component of the analysis, since being rooted in analytic theory without calibration to simulations is a large advantage of our methodology.

We use equations \ref{eq:vesc_final}, \ref{eq:vesc_down} and \ref{eq:Dehnen} as our escape model to compare to the observed edge. 
We use Bayes' theorem and Goodman \& Weare’s affine invariant Markov chain Monte Carlo (MCMC) Ensemble sampler to model the $M_{200}$ posteriors \citep{emcee}. We use a Gaussian likelihood of the form
\begin{align}
\label{eq:likelihood}
\mathcal{L}(\mu, \sigma \mid v_{\text{esc,DS}}) &= \prod_{i=1}^{5} \frac{1}{\sqrt{2\pi} \sigma} \exp\left(-\frac{(v_{\text{esc,DS,i}} - \mu_i)^2}{2\sigma_{cz}^2}\right)
\end{align}
For each cluster, we sum the log likelihood over 5 radial bins, where $v_{\text{esc,DS}}$ is the down-sampled edge profile, $\mu$ is the suppressed theoretical escape profile, and $\sigma_{cz}$ is the error on $v_{\text{esc,DS}}$ (see~\S\ref{subsec:edge measurement}). When galaxy redshift errors are present, we use a Gaussian prior on each $v_{\text{esc,DS}}$ with the mean from the edge measurement and a dispersion from Figure \ref{fig:sigma_err_draws}. 
\begin{figure*}
\includegraphics[width=.5\textwidth]{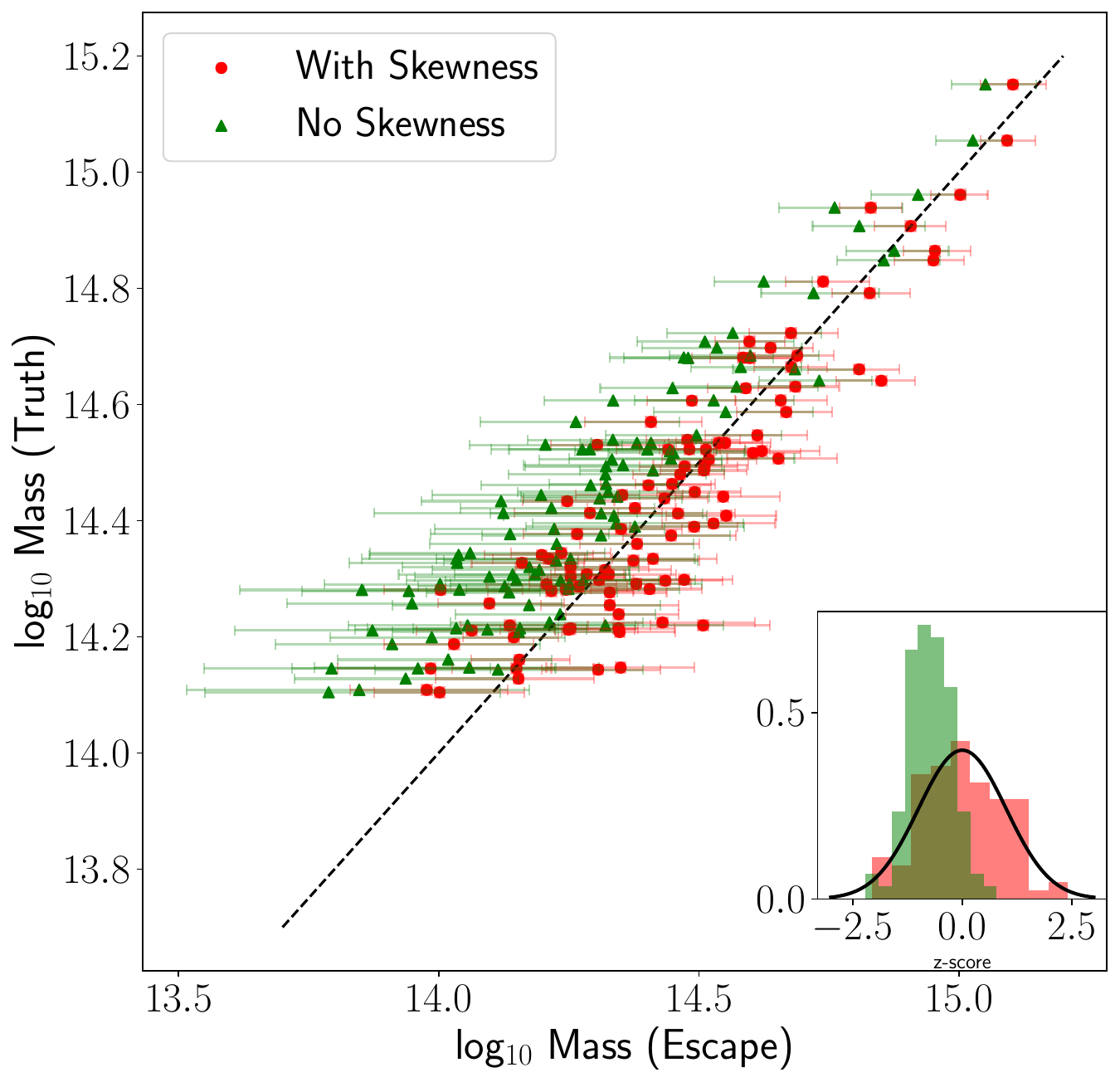}
\includegraphics[width=.5\textwidth]{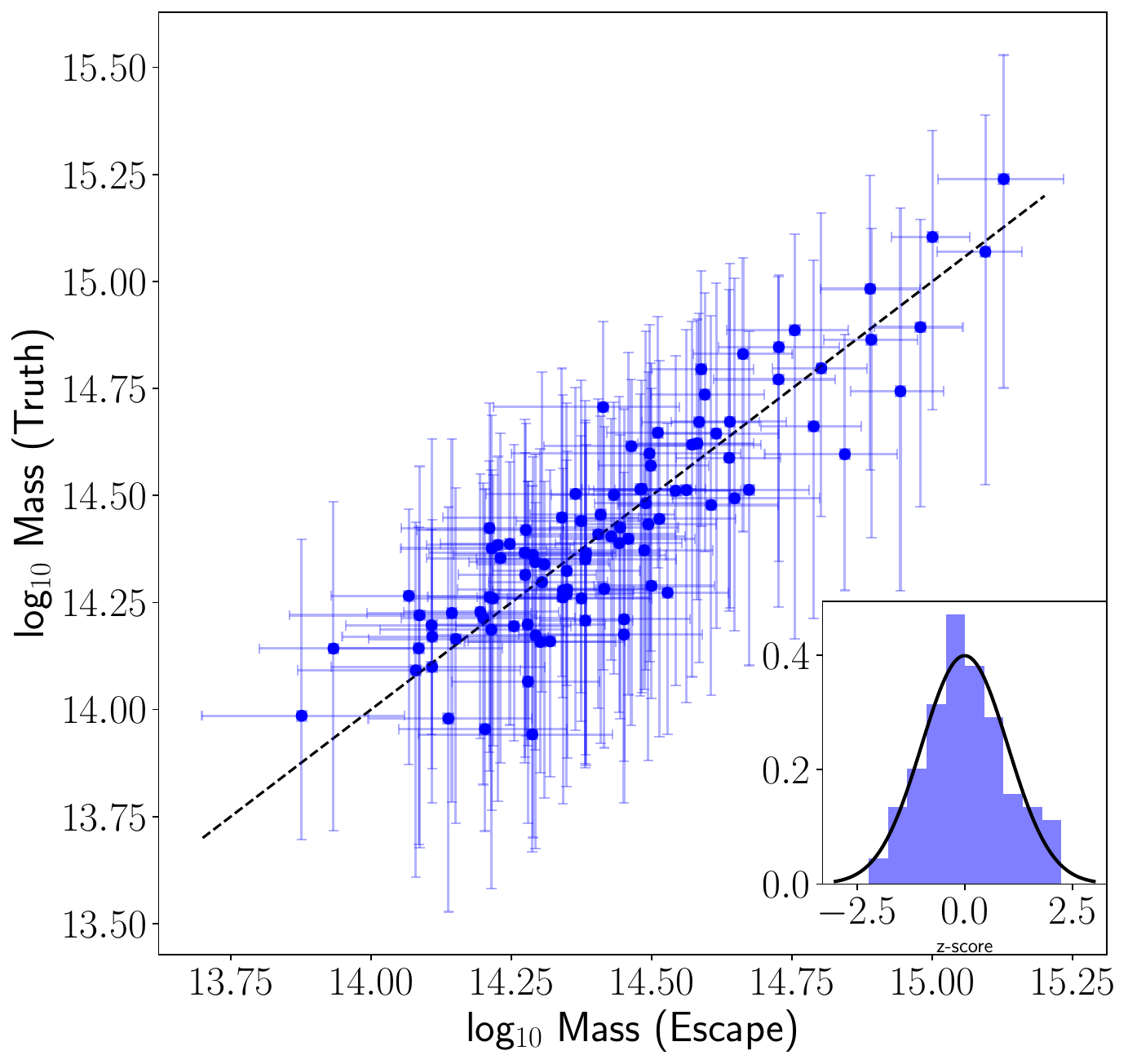}

\caption{\label{fig:Millennium_bias} \textit{Left panel:}
We infer unbiased escape masses for the halos in the Millennium simulation using the AGAMA analytic model for the suppression. When the skewness in the suppression distribution is ignored, the bias increases by 0.1 (\arnote{green} points). The embedded histograms show the distributions of z-scores, compared to the expectation of a Gaussian centered at $\mu=0$ with $\sigma=1$. The dashed line shows unity slope.
\textit{Right panel}: Same as the left panel except when the true mass has a 0.6 dex uncertainty. The bias is statistically consistent.}
 
\end{figure*} 

At each step in the MCMC chain we draw an $M_{200}$ from a uniform distribution and convert to a Dehnen escape profile (equation \ref{eq:Dehnen}) using a mass-concentration relation \citep{duffy+2008}\footnote{We minimize the $\chi^2$ difference in the two forms over the range $0.2 \le r/r_{200} \le 1$.}. We then suppress the chain's theoretical escape edge in order to compare to the projected phase-space escape edge for a halo in the simulation ($v_{esc,DS}$).
We apply our radial projected suppression function to the theoretical escape edge:

\begin{equation}
    \begin{split}
        {Z}^j_v(r_{bin}) \sim &\ \alpha_{\epsilon,\zeta}(r_{bin}| \hat{N}), \\
                    &\ \xi_{\epsilon,\zeta}(r_{bin}| \hat{N}), \\
                    &\ \omega_{\epsilon,\zeta}(r_{bin}| \hat{N}).
    \end{split}
\label{eq:Zv_hat}
\end{equation}
For each $j^{th}$ simulation cluster, the skewness parameters ($\alpha$, $\xi$, $\omega$ from~\S\ref{subsec:Analytic Construction}) are approximated with linear fits as function of $N$. These parameters are compressed to be interpolated via a slope ($\epsilon$) and intercept ($\zeta$). An example of this linearity for a specific radial bin is shown in Figure \ref{fig:Skewness_Fits}.

In order to apply equation \ref{eq:Zv_hat} to the model, we need to know $\hat{N},M_{200}$ and $z$. As discussed in~\S\ref{subsec:Analytic Construction}, $Z_v$ is only weakly dependent on the mass and redshift. We include that dependency by calculating equation \ref{eq:Zv_hat} over a grid of mass and redshift. To measure $\hat{N}$, we first need a binning scheme.

\citet{Rodriguez+2024} found that there is no analytical mass bias in the escape masses so long as the phase-space data and the AGAMA-based suppression calculations follow a similar binning scheme (see their Figure 3).
The binning scheme is normalized between the model and the data by using $r_{200}$ as a scaling parameter for the radial component of the phase-space data.  We \arnote{will} discuss this in the next subsection.

\subsubsection{Test 1: Suppression}
\label{sec:subsub_supp}
For our first end-to-end test, we ignore redshift errors on the galaxies and assume that the cluster's $r_{200}$ are known (from the particle data). This means that the inferred cluster $M_{200}$ uncertainties should come directly from the line-of-sight scatter encoded in $Z_v$ as determined from AGAMA. Any additional scatter must then come from something we do not account for in AGAMA 
(e.g., asphericity, radial-dependent velocity anisotropy, \arnote{among the other factors listed in~\S\ref{subsec:testing}}). The results are shown in the left panel of Figure \ref{fig:Millennium_bias}. 

Denoting the bias $B$ to be the average log$_{10} M_{200,\text{truth}}-$log$_{10}M_{200,\text{inferred}}$ for the 100 cluster sample from Millennium, we find $\text{B}=0.0\pm 0.01$ (stat) $\pm 0.01$ (sys). The statistical error is the error on the mean and the standard deviation between the masses is 0.11 dex. The systematic error is calculated by repeating the analysis over 20 different lines-of-sight. Compared to AGAMA where the statistical error is 0.05 dex (for the same range of $N$ in the sample), this likely leaves interlopers and asphericity as being the main drivers in the increase in scatter. More importantly, no bias is introduced when incorporating all of the additional complexities of the N-body data.

We calculate the mass errors by using 67\% of the area under the posterior around the median (i.e., they can be asymmetric). In the sub-panel we plot the distribution of the z-scores (red) which should be represented by a Gaussian with a standard deviation of one (black). No additional scatter other than what is quantified in the posterior is required. 
We repeated the analysis for velocity anisotropies with a constant $\beta=-0.5$ or $\beta=0.5$ in AGAMA and found no change to the bias or scatter. This is consistent with \citet{halenka+2020} who found that velocity anisotropy has a sub-percent level effect on $Z_v$.

From the lack of bias or additional scatter in this analysis, we conclude that our edge measurement algorithm allows halos in the N-body simulation to be accurately and precisely modeled with a suppression function which is created from spherical and isolated potentials.

\begin{figure}
\includegraphics[width=.45\textwidth]{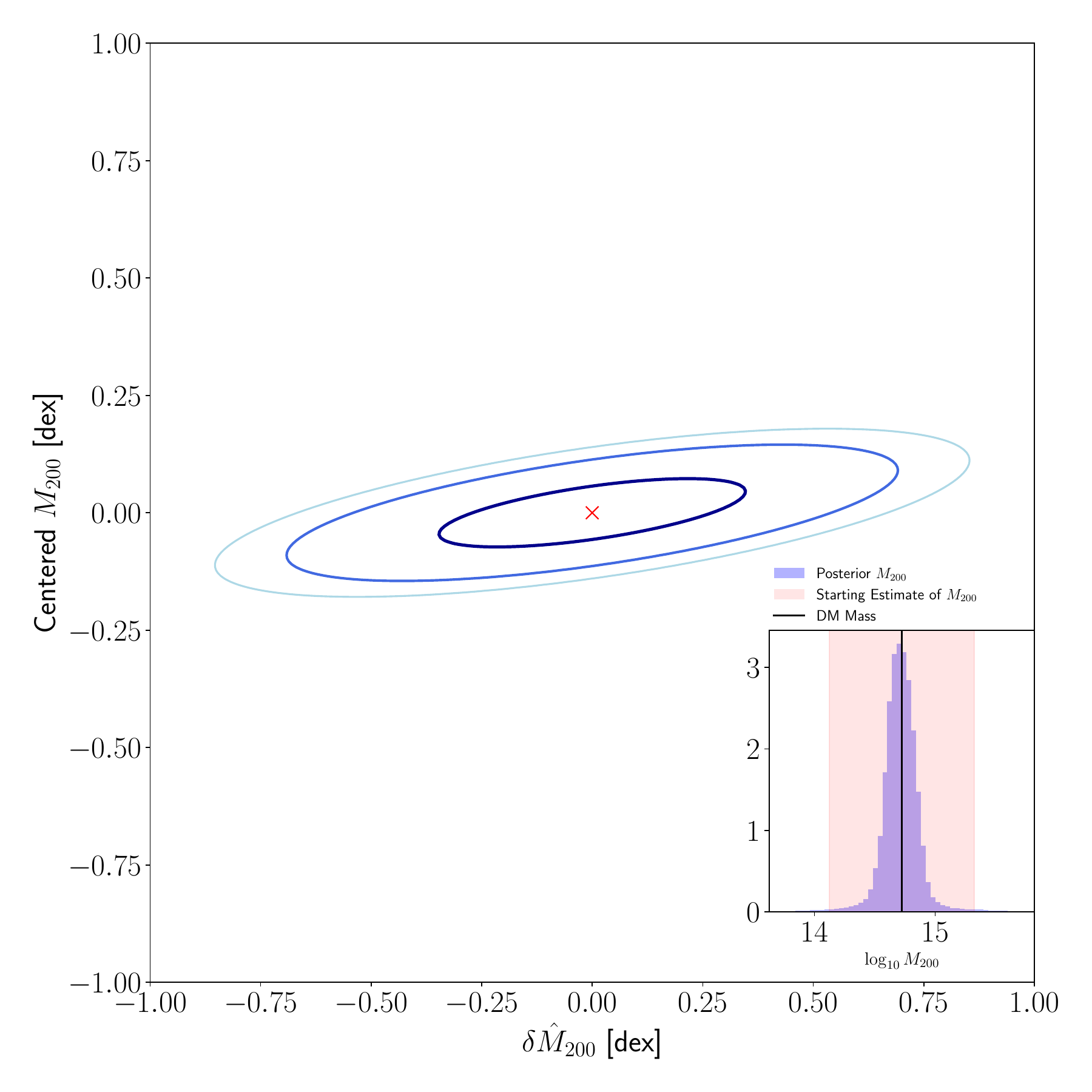}
\caption{\label{fig:M200_r200_covariance}
We show the affect of the choice of the initial estimate of $r_{200}$ on the inferred escape mass (y-axis). The initial $r_{200}$s are determined after scatting the true mass by $\delta M$. The posterior $M_{200}$ is centered about its mean for visual purposes. 
We use $N=100$ different phase-space realizations of $\hat{M}_{200}$ for a fixed line-of-sight projection. The inner blue circle denotes the $1\sigma$ confidence level, with the outer ones denoting $2\sigma$ and $3\sigma$. The ellipse is nearly horizontal such that the induced correlation while present, is very weak. The embedded histogram shows the marginalized posterior. The shaded region defines the range of initial $r_{200}$s used, which has little influence on the mass.
}
 
\end{figure}

We note that incorporating the skewness of $Z_v$ is an essential component of the analysis (as shown by the \arnote{green, triangular} points), which increases the bias  to $\text{B}=0.10\pm 0.01$ (stat) $\pm 0.01$ (sys) when it is not folded into the analysis. This is again \arnote{a result of} the nature of $Z_v$, which becomes highly skewed (especially in outer radial bins) at low sampling. The orange histogram shows the distribution of z-scores, which is skewed towards underestimated masses. Thus, when we ignore the long tail of $Z_v$, the posterior mass distribution under-represents the true mass.

\subsubsection{Test 2: Galaxy redshift uncertainties}
\label{sec:subsub_redshifts}
For our second test, we add errors to the galaxy redshifts in the simulation clusters. We add them stochastically from a normal with a standard deviation of 150 km/s. Per Figure \ref{fig:sigma_err_draws}, we incorporate this into the likelihood using an edge uncertainty of $\sim100$ km/s and add a prior to the edge measurements of $v_{esc,DS} = \mathcal{N}(\hat{v}_{esc,DS},100)$, where $\hat{v}_{esc,DS}$ is the measurement from the phase-space data as described earlier. We then infer the escape masses. We find a bias of $0.02\pm{0.01}$ and a standard deviation of $0.11$ dex, which is statistically unchanged from the perfect galaxy redshifts used in Figure \ref{fig:Millennium_bias} (left). We note that such an error is $\sim 5\times$ the spectroscopic error in observational data in~\S\ref{sec:Results}, where we obtain a bias that is statistically zero when using an error of 30 km/s.

\subsubsection{Test 3: $\hat{r}_{200}$ prior}
\label{sec:subsub_r200}
Finally, we assess the impact of requiring an initial estimate of $r_{200}$ (or ${M}_{200}$) to define the binning in equation \ref{eq:Zv_hat}. For the prior two tests, we used the known $r_{200}$s from the simulation data. Instead of changing $r_{200}$, we could assign a mass offset such that $r^{\prime}_{200} \propto (M_{200} + \delta M) ^{1/3}$. For example, a 50\% positive error at $M_{200} = 1\times10^{15}$M$_{\odot}$ would lead to a $\sim 14\%$ increase in $r_{200}$ used to define the lower and upper edges of the phase-space window for $\hat{N}$. The area of that revised scaled phase-space would increase slightly less, around 10\%. For a uniformly sampled phase-space with a perfectly flat escape edge, the count $\hat{N}$ would then increase by this same fraction. However, because of monotonicity, it increases by a smaller amount. We used AGAMA to estimate this and found that it is closer to a 5\% increase to $\hat{N}$. This small change in $\hat{N}$ causes sub-percent changes in $Z_v$. In summary, a 50\% error in the assumed prior mass for the $r_{200}$ binning has a sub-percent effect on our suppression value $Z_v$. 

An increase in $r_{200}$ will also systematically move each of the bin locations outward by a small amount in physical coordinates. If the escape edge was horizontal (or sampling were infinite), this would not be a problem. However, they are not and so the (incorrectly) re-binned escape edges will by systematically inflated with an artificial increase in the $r_{200}$ used to estimate $\hat{N}$. This then alters the measured $v_{esc,DS}$ and results in a covariance between the mass (or radius) error and the inferred escape mass. 

The results are shown in Figure \ref{fig:M200_r200_covariance} where we sample from a distribution of initial $r_{200}$s corresponding to a large mean mass uncertainty in $\delta M$ of 200\%. A horizontal ellipse would be expected in the case that the initial $r_{200}$ did not induce a correlation with the escape mass estimate. We find a correlation of $R=0.46$, but the effect on the mass inference is extremely weak. In the 
inset histogram we show the range of the initial masses used for $\hat{N}$ when inferring the escape mass (pink) compared to the posterior probability from the MCMC. We find that our initial choice for $r_{200}$ does not influence our final mass estimate.

For our last end-to-end test, we add errors to the true masses in the simulation. While the clusters in our data (presented in the next section) have weak lensing mass errors of around 0.15 dex, we use quadruple this value to scatter the true masses as shown in the right panel of Figure \ref{fig:Millennium_bias}. We then measure the phase-space count $\hat{N}$ based on the $r_{200}$ inferred from the scattered ``true'' mass to constrain the individual escape masses. For the sample of 100 simulation clusters. we find a bias compared to the scattered masses $\text{B}=0.01\pm 0.01$ (stat) $\pm 0.01$ (sys). Again, we find that the escape mass errors are consistent with the observed scatter. Combined, these analyses indicate that our measured escape masses are nearly independent of any initial estimate of $r_{200}$ used to bin the phase-space data and measure the count $\hat{N}$.

\subsubsection{Summary of Simulation Tests}
\label{sec:subsubmasstests}
In summary, we have shown that our mass modeling algorithm recovers the $M_{200}$s of halos in the Millennium N-body simulation with good precision and excellent accuracy. We observe an increase in the scatter in the inferred masses when applying the AGAMA-based suppression model to dynamically complex simulation halos. We see the impact of a carefully modeled non-Gaussian suppression function, without which the masses would be underestimated. Finally, we identify a small but important systematic from binning at the level of 0.01 dex (2\%).



\section{Escape versus Weak Lensing Masses}
\label{sec:Results}

There has been steady improvement over the past decade in the quality and quantity of weak lensing mass estimates for galaxy clusters \citep{Umetsu+2014,Applegate+2014,Schrabback+2018}. There has also been an increase in the number of spectroscopic instruments with significant multiplexing capabilities such as Hectospec on the MMT , M2FS on Magellan, and VIMOS on the VLT \citep{LeFeve+2003,Fabricant+2005,Mateo+2012}. Such instruments enable efficient observing programs to collect galaxy redshifts along the lines-of-sight of galaxy clusters for phase-space analyses. 

\begin{figure}[b]
\includegraphics[width=.5\textwidth]{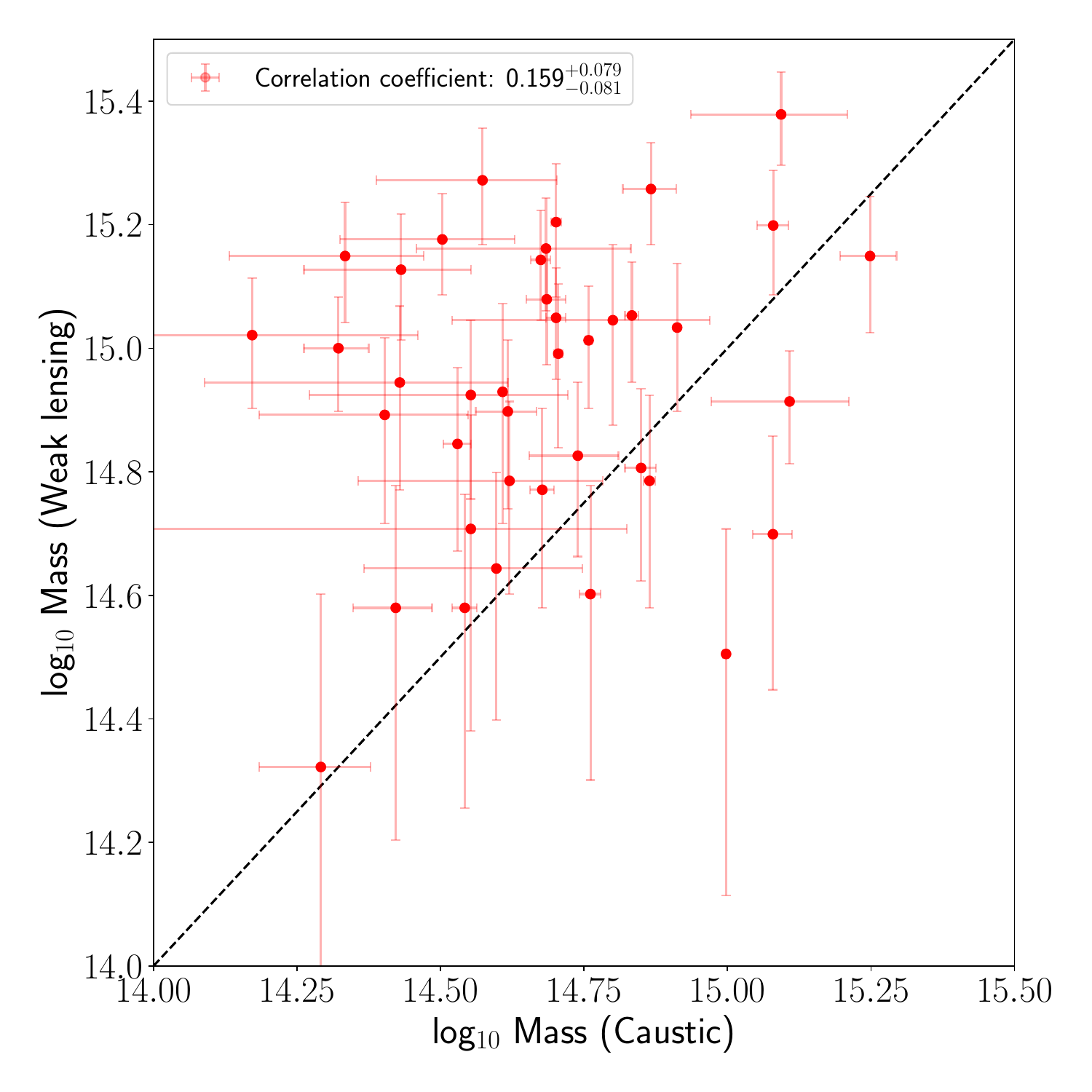}

\caption{\label{fig:Caustic_M_M_plot} Agreement between weak lensing and the caustic masses for the H20 sample.  There is a very large observed bias and poor correlation (see Table \ref{table:cluster_correlations}), despite the fact that we should see better agreement between caustics and weak lensing, as discussed in~\S\ref{sec:Introduction}.
} 
\end{figure}

Recently, comparisons have been made between weak lensing masses and dynamical masses inferred from the caustic technique \citep{diaferio+1997}. In principle, the caustic technique infers mass from the escape velocity of the phase-space data.  In Figure \ref{fig:Caustic_M_M_plot}, we plot the comparison between weak lensing masses and caustic masses for 40 clusters. This figure was first presented in H20, where the caustic masses are taken from \citet{Rines2013} and \citet{Rines+2016} while the weak lensing masses are from H20.

Both \citet{Hoekstra2015} and H20 noted the \arnote{negligible} correlation between caustic and weak lensing cluster masses. They also both noted that the lensing masses were generally higher. Neither conducted a statistical comparison, but as noted in the Introduction, prior research based on simulations had suggested we should expect both small biases and relatively low scatter. 

In order to quantify the correlation between weak lensing and caustic masses, we employ a Monte Carlo error propagation method that accounts for the asymmetric uncertainties in both measurements. For each cluster, we sample from truncated normal distributions defined by the asymmetric uncertainties in both the lensing mass and caustic measurements, generating multiple realizations of the dataset. The correlation coefficient for each realization is computed, yielding a distribution that incorporates the full measurement uncertainties. The final correlation coefficient is taken as the median of this distribution, with uncertainties derived from the 16th and 84th percentiles (corresponding to 1$\sigma$ confidence intervals).

We infer a correlation coefficient $0.159^{+0.079}_{-0.081}$. The percent chance of observing this correlation due to random chance is $32.09\%$ -- or no statistically significant correlation. As noticed by H20, there is also a significant bias B, (defined to be $M_{200,\text{WL}}-M_{200,\text{Caustic}}$) of $\text{B}=0.25\pm 0.05$ (stat). 
The discrepancy between the two mass techniques is much larger than the several percent level systematics on the lensing masses, where H20 suggest dynamical masses can suffer from large biases and scatter. However, no correlation between the dynamical state of the clusters and mass differences was found. What else could be causing the lack of correlation and large bias? 

As discussed in detail in \citet{halenka+2020} and as shown in \citet{Rodriguez+2024}, there are numerous flaws in the interpretation and implementation of the standard caustic techniques for mass inference. Except for the use of galaxy radius and velocity data, the techniques we employ to measure escape edges and infer cluster masses have nothing in common with the caustic technique. Therefore, we will re-assess this situation using our technique to infer dynamical cluster masses from down-sampled escape profiles. The results of~\S\ref{sec:Suppression} suggest that if the weak lensing masses are accurate and reasonably precise \arnote{(in~\S\ref{subsec:systematics} we discuss how to interpret agreement of the two methods from their systematics)}, we should find excellent agreement between these two independent mass measurement techniques. 
 
\subsection{The Sample}
\label{subsec:sample}

We start with the sample of clusters with weak lensing masses from H20 and we add in additional data from the weak lensing study of \citet{Okabe+2016} (hereafter O16). The former uses galaxy shapes and photometric redshifts from the CFH12K and Megacam cameras on the Canada-France-Hawaii Telescope (CFHT) and the latter uses SuprimeCam on the Subaru telescope.  We cross-match these to cluster fields with galaxy spectroscopic redshifts from \citet{Rines2013} and \citet{Rines+2016}. The majority of the redshifts come from the Hectospec Cluster Survey (HeCS),  which is an extensive spectroscopic survey of galaxy clusters on the MMT \citep{Fabricant+2005}. We use the $M_{200}$ weak lensing masses from Table A2 in H20 and from Appendix B in O16. We also include Abell S1063 and its available spectroscopy as described in \citet{Rodriguez+2024}. Here, we convert the mass to the fiducial cosmology of this work by using the \citet{duffy+2008} mass-concentration relation to generate the corresponding density profile, and interpolate the cumulative mass density to identify the radius which is 200$\times$ the critical density and report the corresponding mass within that radius as $M_{200}$.


We do not keep all of the clusters in H20 or O16.  Twelve clusters from O16 are also in the H20 list. We of course exclude clusters in the weak lensing catalogs not observed by HeCS. We apply a minimum phase-space sampling of constraint of $N=50$ galaxies. We also find some clusters with a published caustic mass \citep{Rines2013, Rines+2016} that lack any phase-space data when centered on the weak lensing position. These are A119, A1650, A2142, A2670, A85, ZWCL1215, and we exclude them from our analysis. Two clusters lack galaxy redshift data in their outskirts at $r_{200}$: A2111 from H20 and A2537 from O16. We note that this can occur in systems where even though there is adequate data beyond $r_{200}$, but our algorithm requires galaxies in every radial bin to make an edge measurement. Finally, there is the double system of A750 and MS0906 which was noted in \citet{Geller13}. This is a rare line-of-sight double system which we also exclude.


In the tables we provide two centers and two redshifts for each cluster. We start with the centers and cluster redshifts provided by the lensing catalogs. The escape edge, which represents the potential, must be symmetric about the $v_{esc} = 0$ axis. So we calculate a mean redshift and sky position which is centered on galaxies in the range of interest $0.2\,r_{200}$ and $r_{200}$. We then re-build the phase-space and make a second estimate of the cluster mean redshifts and positional centers. We iterate this process ten times to reach convergence, which usually occurs on the fourth or fifth iteration. The only cases where convergence was not achieved is in the aforementioned double system of A750 and MS0906, which we have excluded from the analysis. 

The final revised centers and mean redshifts are provided in Table \ref{table:cluster_info} and Table \ref{table:cluster_info2}. In the tables, we include the offsets from the weak lensing centers to the dynamic centers via $|\delta v|$ (km/s) and $\delta s$ (Mpc).  The median and standard deviation of the positional offsets are $182\ \&\ 174$kpc. In terms of the mean weak lensing $r_{200}$ for our sample (1.86Mpc), this offset corresponds to $0.085\times r_{200}$. Positional offsets at this level can shift the weak lensing masses by less than $\sim 0.01$ dex \citep{Zhang2019}.  Abell 2029 has the largest revised center which is offset $\sim 800$kpc from the lensing center.  

In terms of velocity offsets, we note that the H20 sample used the brightest cluster galaxy (BCG) as the cluster redshift, while the O16 redshifts come from a variety of sources, including BCGs, galaxy means, and unknown explanations via private communications (e.g., Abell 773). Therefore, we only compare redshifts to the former sub-sample. We find the (absolute) mean and standard deviation of the cluster velocity offsets in our final sample to be $ 514\ \&\ 469$km/s. Others have studied the peculiar velocities of central galaxies in data and in simulations \citep{Malmuth92, Oegerle2001, Martel2014}. In the large study by \citet{Coziol2009}, they find a mean absolute peculiar velocity of the BCG to be 44\% of the cluster dispersion. Their result is consistent with our velocity offset for an estimated 1D velocity dispersion at our median weak lensing mass ($\sim 1100$km/s). 

We note that our final sample extends to $z \sim 0.3$ and that the clusters have lower sampling compared to Abell S1063 \citet{Rodriguez+2024} which had $N>600$. The median $N$ of our sample is just $N=103$ and is also lower than the median for the Millennium clusters in~\S\ref{subsec:testing} ($N=180$). 

We assign line-of-sight velocities to each galaxy following
\begin{align}
\label{eq:los}
    v_{\text{los}}&=c\,\frac{z_g-z_c}{1+z_c},
\end{align}
where $z_g$ is the galaxy redshift, $z_c$ is the mean cluster redshift, and $c$ is the speed of light. We cull all galaxies with velocities $>|4500|$ km/s as these are readily identified as non-cluster members. We use the shifting-gapper \arnote{(see~\S\ref{subsec:Analytic Construction})} to identify interlopers, using 20 galaxies/bin and a velocity gap of 600 km/s. These parameters are chosen following~\S\ref{subsec:testing}. We remind the reader that these parameters propagate to a down-sampled edge uncertainty, as shown in equation \ref{eq:edge_error}. However, this component of the edge error was determined to be negligible given our analysis in Figure \ref{fig:Interloper_grid}.

The projected radius for each galaxy is calculated for our chosen cosmology and the galaxy redshifts:
\begin{align}
\label{eq:r_perp}
    r_{\perp}&=r_\theta\left(\frac{1}{1+z_c}\frac{c}{H_0}\int_{0}^{z_g}\frac{dz'}{E(z')}\right),
\end{align}
where $r_\theta$ and $r_\perp$ are the angular and projected physical separation between the galaxy and the center of the cluster, and $E(z)=\left[\Omega_\Lambda +\Omega_M (1+z)^3\right]^{1/2}$ for a flat $\Lambda \text{CDM}$ universe.

\begin{table*}
\centering
\scriptsize
\setlength{\tabcolsep}{3pt}
\caption{Basic information on the clusters used in this work, along with the lensing and escape velocity mass estimates. Column 1 denotes the cluster name, Column 2 denotes the phase-space sampling $N$, Column 3 denotes the RA prior to re-centering, Column 4 denotes the RA after re-centering, Column 5 denotes the DEC prior to re-centering, Column 6 denotes the DEC after re-centering, Column 7 denotes the separation on the sky $\delta s$ between our new and old centers in kpc, Column 8 denotes the cluster redshift prior to re-centering, Column 9 denotes the cluster redshift after re-centering, Column 10 denotes the absolute value of the velocity shift, $|\delta v|$, in km/s, Column 11 denotes the weak lensing mass $M_{200,\text{WL}}$ we use in $\log_{10}M_{\odot}$(taken from H20), and Column 12 denotes the escape velocity mass we obtain, $M_{200,\text{Esc.}}$, also in $\log_{10}M_{\odot}$. We note that clusters A267 through A2261 appear both in this sample and in O16, separated by the horizontal line.}
\label{table:cluster_info}
\begin{tabular}{lcccccccccccc}
\hline
Cluster & N & RA bef. $[^\circ]$ & RA aft.  $[^\circ]$ & DEC bef.  $[^\circ]$ & DEC aft. $[^\circ]$  & $\delta s$ (kpc) & z bef. & z aft. & $|\delta v|$ (km/s) & $M_{200,\text{WL}}$ & $M_{200,\text{Esc.}}$ & Ref. \\
\hline\hline
A7 & 120 & 2.939 & 2.934 & 32.416 & 32.388 & 195.221 & 0.106 & 0.103 & 863.69 & $14.64^{+0.16}_{-0.25}$ & $14.73^{+0.14}_{-0.15}$ & H20 \\
A21 & 124 & 5.154 & 5.164 & 28.659 & 28.676 & 122.642 & 0.095 & 0.095 & 122.43 & $14.79^{+0.13}_{-0.18}$ & $14.94^{+0.13}_{-0.13}$ & H20 \\
A646 & 64 & 125.540 & 125.560 & 47.098 & 47.096 & 172.749 & 0.129 & 0.127 & 564.62 & $14.58^{+0.18}_{-0.32}$ & $14.62^{+0.20}_{-0.20}$ & H20 \\
A655 & 105 & 126.371 & 126.354 & 47.134 & 47.157 & 237.183 & 0.127 & 0.127 & 147.19 & $14.77^{+0.13}_{-0.19}$ & $14.70^{+0.16}_{-0.15}$ & H20 \\
A795 & 103 & 141.022 & 141.032 & 14.173 & 14.176 & 90.568 & 0.136 & 0.138 & 708.92 & $15.20^{+0.09}_{-0.12}$ & $14.76^{+0.16}_{-0.12}$ & H20 \\
A961 & 75 & 154.095 & 154.159 & 33.638 & 33.636 & 523.267 & 0.124 & 0.127 & 831.57 & $14.85^{+0.12}_{-0.17}$ & $14.66^{+0.18}_{-0.15}$ & H20 \\
A990 & 78 & 155.916 & 155.937 & 49.144 & 49.138 & 196.394 & 0.144 & 0.142 & 622.24 & $15.15^{+0.09}_{-0.11}$ & $15.04^{+0.18}_{-0.17}$ & H20 \\
A1033 & 94 & 157.935 & 157.924 & 35.041 & 35.051 & 115.479 & 0.126 & 0.123 & 1014.26 & $14.93^{+0.14}_{-0.21}$ & $14.69^{+0.15}_{-0.13}$ & H20 \\
A1132 & 72 & 164.599 & 164.547 & 56.795 & 56.784 & 457.993 & 0.136 & 0.135 & 202.68 & $15.05^{+0.08}_{-0.10}$ & $14.95^{+0.17}_{-0.16}$ & H20 \\
A1361 & 56 & 175.915 & 175.924 & 46.356 & 46.344 & 116.109 & 0.117 & 0.116 & 195.19 & $14.68^{+0.15}_{-0.23}$ & $14.40^{+0.20}_{-0.18}$ & H20 \\
A1413 & 59 & 178.825 & 178.820 & 23.405 & 23.430 & 228.543 & 0.143 & 0.141 & 511.75 & $15.03^{+0.10}_{-0.14}$ & $14.98^{+0.20}_{-0.20}$ & H20 \\
A1795 & 102 & 207.219 & 207.244 & 26.593 & 26.694 & 455.973 & 0.062 & 0.063 & 420.51 & $15.14^{+0.08}_{-0.10}$ & $14.97^{+0.24}_{-0.18}$ & H20 \\
A2029 & 147 & 227.734 & 227.715 & 5.745 & 5.898 & 815.283 & 0.077 & 0.078 & 182.78 & $15.26^{+0.07}_{-0.09}$ & $14.98^{+0.12}_{-0.11}$ & H20 \\
A2050 & 74 & 229.075 & 229.087 & 0.089 & 0.108 & 186.376 & 0.118 & 0.120 & 681.48 & $14.66^{+0.16}_{-0.25}$ & $14.81^{+0.17}_{-0.18}$ & H20 \\
A2055 & 56 & 229.690 & 229.674 & 6.232 & 6.248 & 157.804 & 0.102 & 0.103 & 255.30 & $14.46^{+0.21}_{-0.42}$ & $14.64^{+0.21}_{-0.28}$ & H20 \\
A2065 & 129 & 230.622 & 230.648 & 27.708 & 27.719 & 143.148 & 0.073 & 0.073 & 144.78 & $15.08^{+0.09}_{-0.11}$ & $15.01^{+0.13}_{-0.12}$ & H20 \\
A2069 & 92 & 231.031 & 231.031 & 29.889 & 29.886 & 36.376 & 0.116 & 0.114 & 574.35 & $14.51^{+0.20}_{-0.39}$ & $14.78^{+0.18}_{-0.27}$ & H20 \\
A2440 & 113 & 335.987 & 335.997 & -1.583 & -1.587 & 62.898 & 0.091 & 0.091 & 107.77 & $14.99^{+0.11}_{-0.15}$ & $14.64^{+0.15}_{-0.14}$ & H20 \\
A2443 & 57 & 336.533 & 336.512 & 17.357 & 17.384 & 252.125 & 0.108 & 0.110 & 671.28 & $15.13^{+0.09}_{-0.11}$ & $14.73^{+0.19}_{-0.15}$ & H20 \\
A2495 & 68 & 342.582 & 342.593 & 10.904 & 10.904 & 72.849 & 0.078 & 0.079 & 306.59 & $14.32^{+0.28}_{-1.02}$ & $14.39^{+0.19}_{-0.20}$ & H20 \\
RXJ2344 & 68 & 356.076 & 356.075 & -4.380 & -4.390 & 62.571 & 0.079 & 0.079 & 96.29 & $14.58^{+0.20}_{-0.38}$ & $14.51^{+0.18}_{-0.20}$ & H20 \\
A1246 & 84 & 170.995 & 170.992 & 21.479 & 21.483 & 61.326 & 0.190 & 0.191 & 353.14 & $14.79^{+0.14}_{-0.21}$ & $14.97^{+0.16}_{-0.16}$ & H20 \\
A2259 & 72 & 260.040 & 260.063 & 27.669 & 27.676 & 237.081 & 0.164 & 0.160 & 1076.25 & $14.83^{+0.12}_{-0.16}$ & $14.70^{+0.24}_{-0.19}$ & H20 \\
\hline
A267 & 117 & 28.175 & 28.171 & 1.007 & 0.995 & 174.060 & 0.230 & 0.229 & 273.76 & $14.81^{+0.13}_{-0.18}$ & $14.85^{+0.16}_{-0.15}$ & H20 \\
A963 & 92 & 154.266 & 154.265 & 39.047 & 39.038 & 123.829 & 0.206 & 0.204 & 485.67 & $15.01^{+0.09}_{-0.11}$ & $14.96^{+0.15}_{-0.13}$ & H20 \\
A1689 & 133 & 197.875 & 197.871 & -1.342 & -1.339 & 63.489 & 0.183 & 0.184 & 378.29 & $15.38^{+0.07}_{-0.08}$ & $15.21^{+0.18}_{-0.15}$ & H20 \\
A1763 & 126 & 203.834 & 203.834 & 41.001 & 41.014 & 184.854 & 0.223 & 0.232 & 2646.35 & $15.15^{+0.10}_{-0.12}$ & $15.33^{+0.14}_{-0.16}$ & H20 \\
A2219 & 183 & 250.083 & 250.093 & 46.711 & 46.711 & 134.624 & 0.226 & 0.225 & 162.09 & $14.91^{+0.08}_{-0.10}$ & $15.05^{+0.12}_{-0.13}$ & H20 \\
A586 & 106 & 113.084 & 113.096 & 31.634 & 31.609 & 287.502 & 0.171 & 0.170 & 248.01 & $14.60^{+0.18}_{-0.30}$ & $14.69^{+0.14}_{-0.14}$ & H20 \\
A697 & 85 & 130.740 & 130.737 & 36.366 & 36.363 & 68.472 & 0.282 & 0.281 & 203.06 & $15.05^{+0.12}_{-0.17}$ & $15.22^{+0.15}_{-0.16}$ & H20 \\
A1758N & 74 & 203.189 & 203.176 & 50.543 & 50.535 & 222.636 & 0.279 & 0.277 & 486.62 & $15.18^{+0.07}_{-0.09}$ & $15.42^{+0.14}_{-0.15}$ & H20 \\
A1835 & 121 & 210.258 & 210.265 & 2.879 & 2.884 & 122.962 & 0.253 & 0.253 & 221.94 & $15.20^{+0.09}_{-0.11}$ & $15.36^{+0.13}_{-0.14}$ & H20 \\
A1914 & 148 & 216.486 & 216.498 & 37.816 & 37.850 & 370.064 & 0.171 & 0.167 & 1163.32 & $15.05^{+0.09}_{-0.11}$ & $14.95^{+0.13}_{-0.12}$ & H20 \\
A2111 & 82 & 234.919 & 234.920 & 34.424 & 34.428 & 56.712 & 0.229 & 0.228 & 378.69 & $14.90^{+0.12}_{-0.16}$ & $14.80^{+0.20}_{-0.22}$ & H20 \\
A2261 & 129 & 260.613 & 260.628 & 32.133 & 32.107 & 382.226 & 0.224 & 0.226 & 478.49 & $15.27^{+0.08}_{-0.10}$ & $15.21^{+0.17}_{-0.19}$ & H20 \\
\hline

\hline
\end{tabular}
\end{table*}

\begin{table*}
\centering
\scriptsize
\setlength{\tabcolsep}{3pt}
\caption{Same as Table \ref{table:cluster_info}, except Column 11 denotes the weak lensing mass $M_{200,\text{WL}}$ we use in $\log_{10}M_{\odot}$is taken from O16. We note that clusters A267 through A2261 appear both in this sample and in H20, separated by the horizontal line.}
\label{table:cluster_info2}
\begin{tabular}{lcccccccccccc}
\hline
Cluster & N & RA bef. $[^\circ]$ & RA aft.  $[^\circ]$ & DEC bef.  $[^\circ]$ & DEC aft. $[^\circ]$  & $\delta s$ (kpc) & z bef. & z aft. & $|\delta v|$ (km/s) & $M_{200,\text{WL}}$ & $M_{200,\text{Esc.}}$ & Ref. \\
\hline\hline
A0773 & 96 & 139.498 & 139.477 & 51.706 & 51.729 & 392.629 & 0.217 & 0.218 & 440.69 & $15.15^{+0.05}_{-0.05}$ & $15.29^{+0.15}_{-0.14}$ & O16 \\
ZwCl0949.6+5207 & 57 & 148.205 & 148.198 & 51.885 & 51.879 & 120.708 & 0.214 & 0.216 & 450.06 & $14.82^{+0.11}_{-0.12}$ & $14.89^{+0.18}_{-0.19}$ & O16 \\
ZwCl1021.0+0426 & 57 & 155.915 & 155.897 & 4.186 & 4.179 & 309.641 & 0.291 & 0.289 & 599.05 & $14.89^{+0.08}_{-0.09}$ & $14.91^{+0.17}_{-0.16}$ & O16 \\
A1423 & 92 & 179.344 & 179.364 & 33.655 & 33.621 & 493.839 & 0.213 & 0.214 & 268.38 & $14.81^{+0.10}_{-0.11}$ & $14.68^{+0.15}_{-0.15}$ & O16 \\
A1682 & 87 & 196.728 & 196.737 & 46.556 & 46.533 & 319.574 & 0.226 & 0.227 & 257.79 & $15.11^{+0.06}_{-0.06}$ & $14.99^{+0.17}_{-0.17}$ & O16 \\
A2009 & 113 & 225.081 & 225.079 & 21.369 & 21.378 & 98.966 & 0.153 & 0.152 & 327.19 & $15.06^{+0.15}_{-0.13}$ & $14.89^{+0.15}_{-0.14}$ & O16 \\
RXJ1720.1+2638 & 223 & 260.037 & 260.050 & 26.635 & 26.620 & 195.664 & 0.164 & 0.160 & 1134.59 & $14.89^{+0.14}_{-0.14}$ & $14.80^{+0.11}_{-0.11}$ & O16 \\
RXJ2129.6+0005 & 86 & 322.419 & 322.421 & 0.097 & 0.102 & 94.265 & 0.235 & 0.234 & 322.42 & $14.85^{+0.13}_{-0.14}$ & $14.72^{+0.17}_{-0.17}$ & O16 \\
A2631 & 71 & 354.421 & 354.406 & 0.276 & 0.273 & 230.057 & 0.278 & 0.276 & 426.70 & $15.03^{+0.11}_{-0.11}$ & $15.06^{+0.22}_{-0.18}$ & O16 \\
A2645 & 58 & 355.320 & 355.320 & -9.027 & -9.036 & 114.056 & 0.251 & 0.250 & 214.68 & $14.79^{+0.10}_{-0.12}$ & $14.74^{+0.22}_{-0.33}$ & O16 \\
\hline
A0267 & 127 & 28.217 & 28.170 & 1.046 & 0.996 & 909.418 & 0.230 & 0.229 & 332.62 & $14.95^{+0.08}_{-0.08}$ & $15.01^{+0.13}_{-0.14}$ & O16 \\
A0586 & 139 & 113.085 & 113.104 & 31.634 & 31.585 & 545.239 & 0.171 & 0.170 & 270.23 & $14.99^{+0.12}_{-0.12}$ & $14.78^{+0.15}_{-0.13}$ & O16 \\
A0697 & 92 & 130.736 & 130.738 & 36.362 & 36.365 & 46.769 & 0.282 & 0.281 & 202.39 & $15.16^{+0.11}_{-0.10}$ & $15.19^{+0.15}_{-0.15}$ & O16 \\
A0963 & 97 & 154.308 & 154.264 & 39.025 & 39.030 & 537.673 & 0.205 & 0.204 & 210.37 & $15.03^{+0.08}_{-0.08}$ & $14.96^{+0.14}_{-0.13}$ & O16 \\
A1689 & 122 & 197.873 & 197.870 & -1.341 & -1.335 & 77.880 & 0.183 & 0.184 & 319.98 & $15.21^{+0.06}_{-0.06}$ & $15.17^{+0.14}_{-0.13}$ & O16 \\
A1758N & 65 & 203.188 & 203.187 & 50.542 & 50.542 & 13.926 & 0.280 & 0.277 & 809.84 & $14.94^{+0.10}_{-0.11}$ & $14.90^{+0.30}_{-0.24}$ & O16 \\
A1763 & 140 & 203.826 & 203.852 & 40.997 & 41.019 & 459.660 & 0.228 & 0.232 & 1240.28 & $15.40^{+0.08}_{-0.07}$ & $15.35^{+0.14}_{-0.14}$ & O16 \\
A1835 & 121 & 210.260 & 210.265 & 2.880 & 2.884 & 96.447 & 0.253 & 0.253 & 70.02 & $15.18^{+0.07}_{-0.07}$ & $15.35^{+0.13}_{-0.13}$ & O16 \\
A1914 & 159 & 216.507 & 216.493 & 37.827 & 37.853 & 302.825 & 0.171 & 0.167 & 1282.02 & $15.11^{+0.08}_{-0.09}$ & $15.20^{+0.17}_{-0.19}$ & O16 \\
A2111 & 82 & 234.934 & 234.920 & 34.416 & 34.428 & 237.625 & 0.229 & 0.228 & 380.09 & $14.86^{+0.18}_{-0.15}$ & $14.79^{+0.20}_{-0.23}$ & O16 \\
A2219 & 225 & 250.089 & 250.107 & 46.706 & 46.711 & 241.660 & 0.228 & 0.226 & 713.76 & $15.19^{+0.08}_{-0.08}$ & $15.29^{+0.11}_{-0.14}$ & O16 \\
A2261 & 125 & 260.613 & 260.627 & 32.134 & 32.108 & 390.082 & 0.224 & 0.226 & 491.73 & $15.25^{+0.07}_{-0.07}$ & $15.18^{+0.15}_{-0.15}$ & O16 \\
\hline

\hline
\end{tabular}
\end{table*}

\begin{table*}
\centering
\scriptsize
\setlength{\tabcolsep}{3pt}
\caption{Same as Tables \ref{table:cluster_info} and \ref{table:cluster_info2}, except Column 11 denotes the weak lensing mass $M_{200,\text{WL}}$ we use in $\log_{10}M_{\odot}$ is taken from \citet{Rodriguez+2024} (the relevant weak lensing mass is from \citet{gruen2013} (G13)).}
\label{table:cluster_info3}
\begin{tabular}{lcccccccccccc}
\hline
Cluster & N & RA bef. $[^\circ]$ & RA aft.  $[^\circ]$ & DEC bef.  $[^\circ]$ & DEC aft. $[^\circ]$  & $\delta s$ (kpc) & z bef. & z aft. & $|\delta v|$ (km/s) & $M_{200,\text{WL}}$ & $M_{200,\text{Esc.}}$ & Ref. \\
\hline\hline
AS1063\footnote{This cluster was already studied in-depth using an earlier version of the technique in this work in \citet{Rodriguez+2024}. We simply repeat the analysis using the updates to the technique described in earlier sections. This weak lensing mass was converted to the cosmology assumed in this work. All masses are measured in $\log_{10}M_{\odot}$.} & 618 & 342.183 & 342.197 & -44.531 & -44.531 & 117.0 & 0.345 & 0.345 & 66.48 & $15.37^{+0.09}_{-0.11}$ & $15.40^{+0.09}_{-0.06}$ & G13 \\\hline

\hline
\end{tabular}
\end{table*}

\begin{table*}
\centering
\scriptsize
\setlength{\tabcolsep}{3pt}
\caption{Summaries of the bias, scatter, and correlation for the Caustic masses in Figure \ref{fig:Caustic_M_M_plot}, the H20 sample, the O16 sample, the full sample (H20+O16+\citet{Rodriguez+2024}) assuming a fiducial cosmology (a flat universe with $\Omega_M=0.3$ and  $h=0.7$), the full sample assuming a cosmology associated with the CMB \citep{Planck+2018}, the full sample assuming a cosmology associated with Type 1a Supernovae/Cepheids \citep{Brout+2022}, and the Millennium sample (left panel of Figure \ref{fig:Millennium_bias}).
}
\label{table:cluster_correlations}
\begin{tabular}{lcccccccccccc}
\hline
 & Bias [dex] & Scatter [dex] & Correlation \\
 \hline\hline
AGAMA & $0.00\pm0.01$ & 0.05 & $0.986^{+0.002}_{-0.002}$ \\
\hline
Millennium (perfect) & $0.00\pm0.01$  & 0.11 &  $0.883^{+0.010}_{-0.010}$ \\
\hline
Millennium (scattered) & $0.01\pm0.01$ & $0.13$ & $0.659^{+0.042}_{-0.046}$ \\
\hline
Full Sample (CMB Cosmology) & $0.02\pm0.02$ & 0.17 & $0.679^{+0.046}_{-0.049}$ \\
\hline
Full Sample (Fiducial Cosmology ($\Omega_M=0.3, h=0.7$)) & $0.04\pm0.03$ & 0.17 & $0.693^{+0.043}_{-0.048}$ \\
\hline
Full Sample (Type 1a Supernovae/Cepheids Cosmology) & $0.06\pm0.03$ & 0.17 & $0.683^{+0.046}_{-0.049}$ \\
\hline
H20 (Fiducial Cosmology) & $0.04\pm0.03$ & 0.18 & $0.677^{+0.049}_{-0.056}$ \\
\hline
O16 (Fiducial Cosmology)  & $0.02\pm0.02$ & 0.11 &  $0.787^{+0.052}_{-0.060}$ \\
\hline
Caustic Sample & $0.25\pm0.05$ & 0.30 & $0.159^{+0.079}_{-0.081}$ \\
\hline

\hline
\end{tabular}
\end{table*}


\subsection{Cluster Escape Mass Estimates}
We infer the cluster mass estimates using the same techniques applied to the Millennium phase-space data (~\S\ref{subsec:edge measurement},~\S\ref{subsec:edge_error_summary}, and~\S\ref{subsec:likelihood}). We do so for the H20 and O16 samples independently, followed by an overlap of the two samples. In our sample of $\sim5000$ spectroscopic members we use in the analysis, the median spectroscopic error is 30 km/s. This is the only contribution to the assumed $\sigma$ in the likelihood, as from equation \ref{eq:edge_error}, the line-of-sight component of the edge error is incorporated from our model for $Z_v$.

We also need an initial $r_{200}$ to estimate each cluster's phase-space count and define the binning scheme. To accomplish this, we draw 50 uniform random radii from a cluster's weak lensing mass and uncertainty and create separate escape-mass posteriors. For the final cluster masses, we combine the posteriors for each different initial $r_{200}$ and report the median. The asymmetric mass uncertainties cover 67\% of the posterior. We note that the posteriors of the individual clusters are often symmetric, which is reflected in the error bars.

As a consequence of our algorithm, while the galaxy projected positions and line-of-sight velocities are fixed, there is not a single instantiation of each cluster's escape edge or phase-space count, but instead many realizations based on the 50 initial $r_{200}$ values. Hence in the tables, all corresponding columns (N, RA, DEC, $\delta s$, \arnote{...}) are simple averages over this range of possible $\hat{r}_{200}$. The phase-space diagrams for all 46 clusters with the corresponding down-sampled edge profiles, lensing profiles, and dynamical fits are presented in Appendix \ref{sec:Phase_spaces} (note that AS1063 is not shown, see \citet{Rodriguez+2024} \arnote{for its phase-space}). These figures are provided to visualize the phase-space data and the suppressed mass models. However it is important to keep in mind that \arnote{as a result of} the statistical nature of the algorithm, the mass models use a suppression function which is determined from the average phase-space count.

In terms of the width of the uniform distribution for initial $r_{200}$ used to create each cluster's posterior samples, we have explored ranges from 1-4 $\sigma_\text{WL}$.
We saw no effect on final the bias when varying this range. In terms of the scatter we found that 2.5 and 3.5 $\sigma_{\text{WL}}$ (for H20 and O16 respectively) produced a relation which is statistically consistent, i.e., where the z-score is consistent with a Gaussian of width 1. This is a very wide range, corresponding to an order of magnitude in $\hat{M}_{200}$ for the average cluster (and similar to Figure \ref{fig:M200_r200_covariance}). So while our algorithm requires some prior knowledge of the cluster mass in order to function, the statistical range on that prior knowledge is large enough to 
avoid inducing any significant covariance into the final mass estimates (e.g. see~\S\ref{sec:subsub_r200} and Figure \ref{fig:M200_r200_covariance}).

We emphasize that while we do require $r_{200}$s to create the binning schemes and the phase-space counts, we do not use those $r_{200}$ as traditional priors in the Bayesian sense. Specifically, we do not directly use them in the likelihood calculations. Their only purpose is to define an initial placement of the centers of the radial bins to enable phase-space counts for the suppression function. The final $r_{200}$ can be anything the sampling chain prefers. As a counter-example, scaling parameter observables are often measured within projected radii pre-determined from a mass (e.g, in the mass-temperature relation). This choice is known to induce a fairly significant positive correlation \citep{Mahdavi2013}, which reduces the observed scatter. Our analysis in~\S\ref{sec:subsub_r200} shows that this is not the case for the escape masses.

\begin{figure}
\includegraphics[width=.45\textwidth]{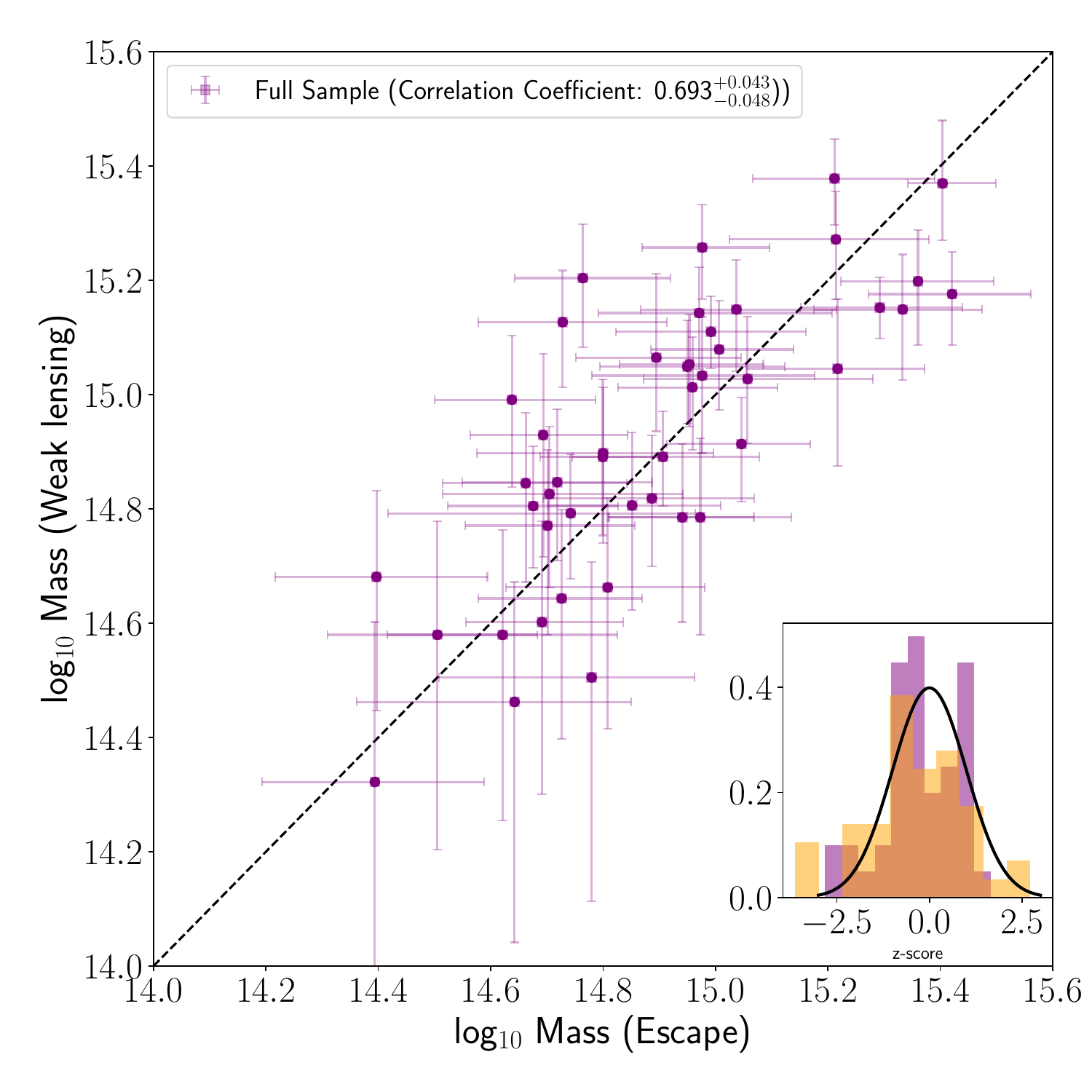}
\caption{\label{fig:Data_M_M_all} Weak lensing and escape masses for the 46 clusters in our sample, assuming our fiducial cosmology.  The bias (lensing vs. escape) and correlation are significantly improved from using caustics in Figure \ref{fig:Caustic_M_M_plot}. The embedded histograms (purple being escape, yellow being lensing) show the distribution of z-scores compared to the expectation of a Gaussian centered at 0 with $\sigma=1$, for which we find very good agreement for the escape velocity (lensing errors are marginally underestimated). This indicates our errors are consistent with the intrinsic scatter.
} 
\end{figure}

Before we conduct a statistical comparison between the weak lensing and escape masses, we will conduct a search for outliers.  We adopt a Bayesian approach to outlier rejection that incorporates a linear relationship combined with a Bernoulli-distributed indicator variable to each cluster which is a probability that it belongs to nominal population \citep{hogg+2010}. We also incorporate measurement uncertainties from both the escape velocity and weak lensing. Unlike simpler outlier rejection schemes, this method provides a full posterior probability distribution for both the fit parameters and the classification of each point as an ``inlier'' or ``outlier''. We classify points as outliers when their posterior probability of belonging to the inlier population falls below 0.7. This threshold provides a conservative classification criterion based on the marginalized posterior probabilities from our hierarchical model.  Using this outlier identification method, we identify no credible outliers, which is visibly consistent from Figure \ref{fig:Data_M_M_all}.

The weak lensing and escape velocity masses are plotted in Figure \ref{fig:Data_M_M_all}, using H20 for the overlapping clusters. 
Compared to Figure \ref{fig:Caustic_M_M_plot}, we find much better agreement. The correlation coefficient has risen to $0.693^{+0.043}_{-0.048}$.
The chance of observing this correlation due to random chance is only $1.25\%$, as opposed to $32.09\%$ of the time using the caustic technique. As in Millennium, we denote the bias as B, (defined to be $\log_{10}M_{200,\text{WL}}-\log_{10}M_{200,\text{Escape}}$) for which we obtain $\text{B}=0.04\pm 0.03$ (stat). 
The observed scatter in the mass estimates is 0.17 dex. 

This improvement in statistical strength of the correlation in conjunction with the decrease in bias from $0.25$ to $0.04$ serve as evidence of significant improvements in the inference and interpretation of escape profiles of galaxy clusters. However, we do still note a small bias where the weak lensing masses are higher on average than the escape masses. This bias decreases to just 0.02 when a \citet{Planck+2018} cosmology is assumed for the escape masses.

We also plot the histogram of z-scores in the sub-panel of Figure \ref{fig:Data_M_M_all}. For the escape masses (purple), we find that individual mass errors are consistent with each other within their observed scatter about the one-to-one line. Our measured errors are representative of the underlying true mass errors. The lensing errors could be slightly underestimated ($\sigma=1.4$ instead of the predicted $\sigma=1$).

\begin{figure*}
\includegraphics[width=.45\textwidth]{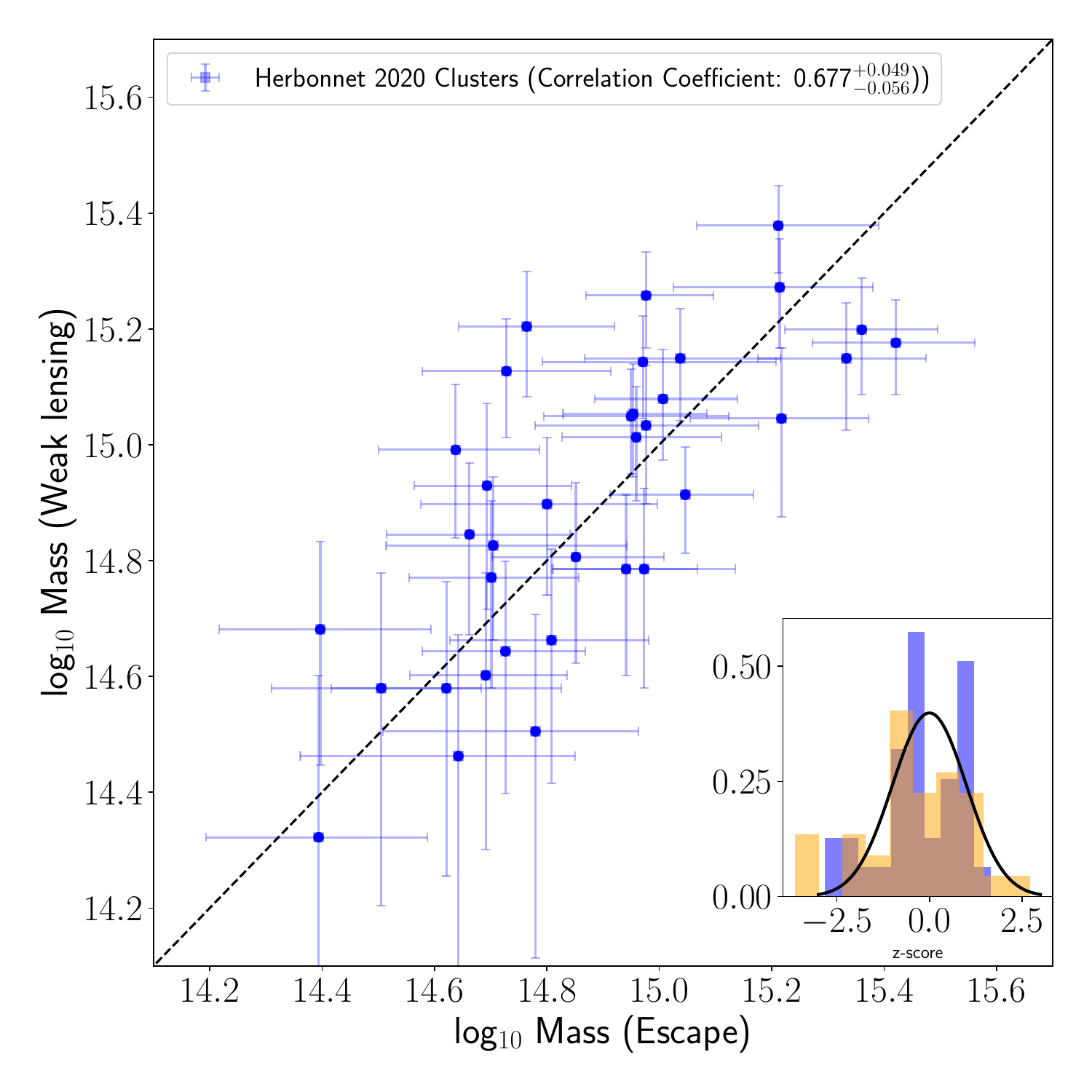}
\includegraphics[width=.45\textwidth]{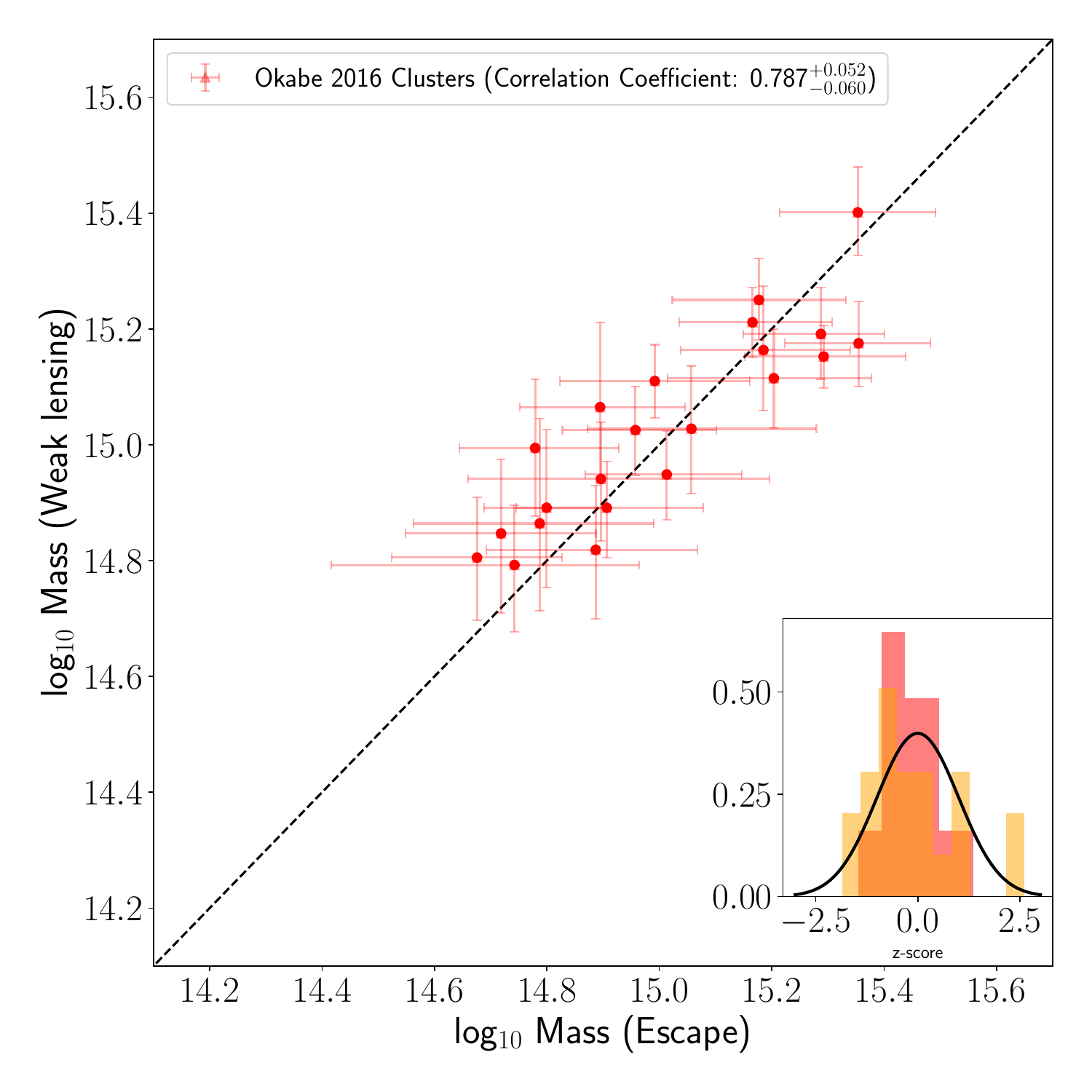}
\caption{\label{fig:Data_M_individuals} Same as Figure \ref{fig:Data_M_M_all}, except for the two full observational samples. Both individual samples have improved biases and correlations over Figure \ref{fig:Caustic_M_M_plot}. The agreement between the samples indicates sample selection is not a significant systematic (see~\S\ref{subsec:systematics}).
} 
\end{figure*}

\subsection{Systematics}
\label{subsec:systematics}

Both the lensing masses and the escape masses have systematics. These are measurement or modeling effects which \arnote{will} bias the inferred mass. Systematics can be related to calibration or tuning variables such that model inferred masses could be systematically low or high. For both weak lensing and escape techniques, prior estimates on the mass systematics have been estimated from simulations and also by using the data itself. Our goal in this subsection is to give the reader an estimate of the level of systematic biases, which if properly accounted for, would change our current conclusion that the escape masses and lensing masses are unbiased with respect to each other. We start with the systematics that both techniques have in common. 
\subsubsection{Centering}  
\label{subsub:sys_center}
In terms of centering, we note that the velocity center is the most important component for the escape mass while the sky center is most important for weak lensing mass. Both H20 and O16 discuss how centering could affect their weak lensing masses and conclude that any unaccounted for biases should be negligible. The argument is that the BCG is known to trace the peak of the density profile well enough while their binning algorithms exclude the core and this avoid offcentering issues. 

The escape masses use centers from galaxies between $0.2 \le r_{\perp}/r_{200} \le 1$, and by doing so ignore the BCG and the region where the galaxy density (or shear profile) is highest. Our iterative approach to define the dynamical potential center is robust, since it requires convergence to the mean velocity. However, as noted in~\S\ref{subsec:sample} and in Table \ref{table:cluster_info}, the final sky positions do differ such that the mean R.A. and decl. of galaxies in our phase-space window is not an accurate representation of the BCG position. Regardless, the median positional offset of the clusters is only $0.085\times\ r_{200}$, which is half the size of the phase-space bin widths. \arnote{Thus}, while galaxies could shift radially in the phase-space by this amount, the maximal velocities in each bin would remain unaltered. We conclude that centering is a negligible component of our systematic error budget for both lensing and escape masses.

\subsubsection{Binning}  
\label{subsub:sys_binning}
There are many ways to radially bin the shear and escape profiles \citep{Mandelbaum+2005, Umetsu+2014}. A concern for weak lensing is that covariance is introduced into neighboring bins, especially in the cluster cores. However, O16 argue that any biases from their binning scheme should be negligible ($\pm{1}$\%). 

For the escape masses, \citet{Rodriguez+2024} showed that the inferred mass is unaffected by the choice of bin size so long as the same scheme is used to quantity $Z_v$.
However, this is only for a highly sampled $(N>600)$  system. In more poorly sampled systems such as our Millennium sample ($\langle N \rangle = 180$) or our observational sample ($\langle N \rangle = 100$), changing $r_{200}$ also systematically moves the bin locations, which is more noticeable with sparse sampling. \arnote{As a result of} the escape edge not being horizontal and/or not infinitely sampled, the (incorrectly) re-binned edges will be systematically inflated, with an artificial increase in the $r_{200}$ used to estimate $\hat{N}$, where the opposite is true for an artificial decrease in $r_{200}$. Using our chosen binning scheme and typical phase-space sampling, with $r_{200}$ uncertainties derived from lensing errors (Tables \ref{table:cluster_info}, \ref{table:cluster_info2}, and \ref{table:cluster_info3}), we found a binning systematic of 0.01 dex $(\sim2\%)$.

\subsubsection{Concentration}  
\label{subsub:sys_conc}
Both our technique and lensing utilize the NFW profile which can be quantified with a mass and concentration. H20 uses the mass--concentration relation from \citet{Dutton_2014} and they suggest that mass biases could be introduced at the 2\% level \citep{Hoekstra2015}. O16 allow the concentration to be a free parameter, but note that it matches many relations in the literature. The \citet{Giocoli24} studied orientation bias using the most massive clusters in the Three Hundred Simulation \citep{Cui22}. They found that clusters observed in an orientation along their major ellipsoidal axis have a boosted shear and a overestimated weak-lensing inferred mass.
This bias is attributed to a concentration inferred from the projected data and can be mitigated to below a few percent by choosing a fixed concentration of $\sim 3$. We note that the concentration for the mean mass of our sample is $c=3.05$ \citep{duffy+2008}.

We use the \citet{duffy+2008} relation for the escape masses. However, \citet{Rodriguez+2024} showed that the marginalized mass versus concentration posterior probability distribution is actually independent of the concentration. We tested this on the Millennium simulation, using a uniform random draw in $c$ between $c=1$ and $c=10$ instead of a mass--concentration relation. We found that this has a sub-\arnote{0.01 dex} impact on the bias and scatter in mass.  Given that the potential profile comes from the integrated density profile it is inherently much flatter. So this is not an unexpected result. We conclude that \arnote{using these well-known} mass-concentration relations contributes a 2\% systematic uncertainty to weak lensing masses, but has negligible impact on escape velocity measurements.
\subsubsection{Sphericity}  
\label{subsub:sys_asphere}
The masses in this work use spherical symmetry in the mass profile modeling. For lensing, it is well known that the non-spherical density profiles bias the inferred (de-projected) 3D mass. H20 quantified the effect from simulations and applied a 3\% correction to their masses. O16 did not account for this correction.

For the escape masses, we modeled the line-of-sight suppression on spherical clusters, but we tested against simulations which have non-spherical halos and we still find unbiased masses. Therefore asphericity for the escape technique is negligible, but is at the 3\% level for the weak lensing.
\subsubsection{Dynamical state}  
\label{subsub:sys_dynamics}

Galaxy clusters are some of the youngest objects still undergoing gravitational collapse. There are many known examples of obvious cluster mergers, e.g. the famous Bullet Cluster \citep{Springel+2007,Wik+2014,Robertson+2017}. For a comparison like the one we conduct here, one could easily avoid obvious merging systems given the extensive data available per cluster (\arnote{ideally a multi-wavelength analysis, comparing the ICM with optical tracers}). For instance, we exclude Abell 750 and MS0906, \arnote{which again as noted by \citet{Geller13} is a rare line-of-sight double system.}

\arnote{We conduct a literature search for clusters with reported evidence of non-equilibrium conditions. To create the disturbed sample, we use \citet{Yuan+2020,Yuan_2022} as a reference for the X-ray morphology indices ($\delta$), where $\delta$ is a parameter that combines information about a galaxy cluster's overall shape and asymmetry to classify its dynamical state. Typically $\delta>0$ implies evidence of non-equilibrium conditions \citep{Yuan+2020}, although for more robust selection we impose $\delta>0.5$. Clusters without morphology indices may still be flagged as mergers with adequate multi-wavelength literature evidence.}

\arnote{There are 10 clusters matching these criterion, and are: A655 \citep{Markevitch_2001}, A2065\citep{Chatzikos+2006},  A2069 \citep{Drabent+2015},  A2440 \citep{Maurogordato+2011}, A2111\citep{Wang_1997}, A1682 \citep{Clarke_2019}, A2631 \citep{Monteiro_Oliveira_2020}, A1758N \citep{Monteiro_Oliveira_2016}, A773 \citep{Barrena_2007}, as well as S1063 \citep{gomez+2012,mercurio+2021}. This constitutes $\sim20\%$ of our sample.} Although 'dynamical state' remains poorly defined in the literature \citep{Haggar2024},
the evidence ranges from sub-structure in the ICM \citep{Rasia+2006,Nelson+2014,Lau+2009}, substructure in the galaxy spatial distribution \citep{Barrena_2007}, non-Gaussian velocity distributions \citep{girardi+1996,fadda+1996}, and highly offset BCGs from the X-ray or weak lensing centers \citep{Martel2014, Coziol2009}. 

For this subset, we find a bias of \arnote{0.00} dex with scatter \arnote{0.18} dex. We see no difference in the bias and scatter with respect to the weak lensing masses when sub-selecting only clusters with evidence for dynamical non-equilibrium. While we did not explicitly conduct this test in the simulations, the sample of 100 halos has a wide range of dynamical states, likely contributing to the increase in scatter compared to the AGAMA realizations, but with no induced bias.


\subsubsection{Sample Dependencies}  
\label{subsub:sys_sample}
In Figure \ref{fig:Data_M_individuals}, we show the comparison between the H20 (left panel) and O16 (right panel) samples. We remind the reader that the selection of the clusters and their respective lensing analysis and modeling pipelines are distinct, yet we still obtain a bias of $\text{B}=0.04\pm 0.03$ ($0.18$ dex scatter) and $\text{B}=0.02\pm 0.02$ ($0.11$ dex scatter) for the respective samples. 12 clusters are overlapping in the two samples, where in our final sample we choose to use H20 masses in favor of O16. If we instead had used O16 lensing masses for these clusters, we note that our bias increases from $0.04 \pm 0.03$ to $0.05 \pm 0.03$, which corresponds to a small sample dependency systematic of 0.01 dex ($\sim 2\%$).



\subsubsection{Cosmology}  
\label{subsub:sys_cosmo}
The derivation of weak lensing masses from observational data inherently depends on the assumed cosmological framework. Critical to this dependence is the calculation of the critical surface density ($\Sigma_{\text{crit}}$), which scales the observed shear signal to physical mass measurements. This quantity incorporates angular diameter distances between observer, lens, and source, all of which are direct functions of the cosmological parameters $H_0$, $\Omega_M$, and $\Omega_\Lambda$. Furthermore, the adopted mass-concentration relation also carries cosmological assumptions from the simulations used for calibration. This corresponds to a sub-percent level systematic when assuming a flat universe. In a non-flat universe, the equation of state parameter, $w$, leads to a percent-level lensing mass systematic \citep{Applegate+2016, stark2017}. 

Angular diameter distances are used to calculate the projected radii for the phase-spaces. We incorporate a small systematic with redshift in the suppression function. Cosmology also plays a role through the evolution of the mass-concentration relationship. However, the low redshift range of our data implies a negligible effect from these issues \citep{Merten2015}. 
Cosmology significantly influences our measurements through equation \ref{eq:vesc_final}, where the escape edge scales strongly with the cosmological parameters via the terms containing $q H^2$. 
A range of $\sim \pm{3}$km$^{-1}$s around our fiducial $H_0 = 70$ km s$^{-1}$ leads to a 0.04 dex ($\sim$ 10\%) difference in the escape masses. Thus, while cosmological parameter uncertainties introduce negligible systematics in weak lensing, they contribute significantly to escape mass uncertainties.

\subsubsection{WL specific systematics}  
\label{subsub:sys_wl_only}
Besides the above systematics which are in common to both mass measurements, weak lensing has its specific issues. For instance, galaxy shape measurement systematics could be present. H20 suggest these are small ($\sim$1 \%) for their sample. O16 suggest the multiplicative shape bias is $\sim$3 \% in their sample. They correct their masses for this bias. H20 also correct for shear bias and claim a systematic uncertainty of 2\% in cluster masses. Magnification and lensing source galaxy background contamination can boost the inferred shear. H20 finds magnification bias to be negligible, while they calculate a boost correction using random sampling techniques. The H20 boost corrections are accurate to 1.8 \% at radii larger than 0.5 Mpc. \citet{Okabe+2008} suggest that their source identification algorithm mitigates boost effects altogether and they do not apply a correction.
For the H20 clusters, the accuracy of the source galaxy photometric redshift distribution is around 2\% leading to a mass systematic of 4.5\%. O16 found their masses decreased by 4\% when they used the photo-z distribution as opposed to individual source photo-zs. 

\subsubsection{Escape specific systematics}  
\label{subsub:sys_esc_only}

The suppression model we use relies on analytic theory, yet we find it agrees well to within sub-percent level bias with the Millennium N-body simulation, which contains locally varying cosmological backgrounds, internal cluster substructure, cluster mergers, asphericities, hyper-escape-speed galaxies, variable velocity anisotropies, non-cluster interlopers, etc., none of which is present in the analytic phase-spaces. Hence, we conclude that this systematic is not significant, at least when considering gravity alone.

An alternative is to test the $Z_v$ model in a simulation such as Illustris TNG \citep{nelson2021}, \arnote{which contains realistic baryonic physics including AGN and Supernova feedback, MHD processes, ICM dynamics, star formation, etc. In the context of suppression, such processes could yield biased tracers of the underlying potential, for instance large scale redistribution of galaxies could occur from AGN feedback.} However, order-of-magnitude estimates suggest displacing a single typical galaxy would take $\sim 10^{60}\,\text{erg}$ of energy, comparable to total AGN output over a Hubble time, hence we conclude this is likely not a \arnote{significant source} for influencing the galaxy velocities. 

Additionally, we note that the $Z_v$ model is calibrated to high precision in AGAMA. The skewed-normal model contains 1000 line-of-sight draws, and from Appendix \ref{sec:Fits}, each free parameter (skewness, location, and scale) follows a tight scaling relation with the sampling, $N$. In terms of observable systematics, we found that of the $\sim 5000$ galaxy redshifts we used in the analysis, there was a mean observed redshift uncertainty of 30 km/s, which was the dominant contribution to our edge errors
(interloper selection errors were negligible and line-of-sight scatter in the edge was accounted for via $Z_v$). Astrometry from SDSS and CFHT is also highly positionally accurate to sub-arcsecond levels, which results in negligible positional offsets at cluster scales. 

Another potential systematic is the completeness of the phase-space sample. \citet{Rodriguez+2024} found that non-uniform sampling variations up to 30\% have no effect on the measured velocity dispersion. In our case, non-uniform sampling directly translates into variations in $Z_v(r)$, which depend on the estimated phase-space count $N$.  Fortunately, the redshifts were targeted from the SDSS photometric catalog which is nearly complete for bright galaxies. The HeCS and HeCS-SZ targeting procedure creates a largely complete magnitude-limited sample of brighter galaxies and their final observed phase-spaces have reasonably uniform sampling as a function of cluster radius \citep{Rines2013,Rines+2016}.  We plot the data in Appendix \ref{sec:Phase_spaces} and there are no obvious examples of non-uniform phase-space sampling. Given all of the above, we have no evidence to suggest that spectroscopic completeness should have any effect on the cluster masses.

\subsubsection{Summary of lensing and escape systematics}  
We quantified a set of systematic uncertainties that could affect weak lensing and escape velocity masses. Certain issues are negligible for our analysis, including centering, binning, sample differences, and dynamical equilibrium.  These would not be issues for other samples or other techniques (e.g., caustic masses or stacked measurements). The systematics which could contribute to the difference between the mean log masses can be summarized as:
\begin{enumerate}[noitemsep,topsep=5pt,parsep=0pt,partopsep=5pt]
    \item WL: shape measurement bias (2--3\%)
    \item WL: density asphericity (0-3\%)
    \item WL: boost factor corrections (0-2\%)
    \item WL: NFW concentration (0-2\%) 
    \item WL: photometric redshift distribution uncertainties (4\%).
    \item ESC: binning (2\%)
    \item ESC: cosmology (10\%)
\end{enumerate}

While weak lensing has more sources of systematic errors, its total error budget is about half that of the escape technique. The dominant systematic for weak lensing stems from the photo-zs, while for the escape masses it stems from our current uncertainty in cosmology through the $qH^2$ terms in equations \ref{eq:r_eq} and \ref{eq:vesc_final}.

\section{Discussion and Final Remarks}
\label{sec:Discussion}

The goal of this paper is to assess the concordance between escape velocity-based masses and shear-based lensing masses of galaxy clusters. For a $\Lambda$CDM universe, these two independent mass inferences should agree if and only if both techniques achieve high levels of precision and accuracy. We demonstrated for the first time that this concordance is seen in the data (Table \ref{table:cluster_correlations} and Figure \ref{fig:Data_M_M_all}), and we discussed in detail the numerous systematics on each technique, including those from observational measurements like galaxy redshifts and shapes, as well as from modeling like concentration and cosmology (see~\S\ref{subsec:systematics}). The levels of these systematics are constrained by comparing with simulations (e.g, galaxy shapes, cluster asphericities, dynamical states) or by using the data to identify differences from expectations (e.g., photo-zs, boost factors, sample differences). 

For the escape technique, our simulation-based tests are described in~\S\ref{sec:Suppression}. We built a suppression function for the escape profile using isolated, spherical, and analytically generated phase-space realizations. The tracers in these fake data have realistic instantaneous orbital positions and velocities, but lack the complexities of non-linear gravitational collapse like substructure, aspherical density distributions, dynamical friction, complex orbital anisotropies, and non-cluster interlopers. So we then validate our model against N-body simulations, which inherently contain all of those components, and find an increase in the scatter of our inferred masses, but no evidence for systematic bias. 
On the data side of systematics control, we subselect clusters with observational evidence for non-equilibrium dynamics and we find no difference in the mean mass bias when compared to lensing. 

For weak lensing, simulations are used to determine corrections for galaxy shape bias and/or asphericity. In terms of data, the boost factor is an example where a systematic correction is made based on an observed background expectation. The combined WL systematics level should be below $\sim 4\%$ (or $\sim 0.02$ dex) in mass (H20 and O16). We conclude that for a fixed cosmology, the data and techniques we use in this work meet the quality threshold for a direct comparison.

We can ask whether ignoring the WL systematic corrections for the boost factor, shapes, and asphericity would affect our comparison. For each case, the lensing masses would be smaller. We estimate that ignoring the first two would lower the H20 masses by 0.1 dex and the O16 masses by 0.02 dex.\footnote{Both the Okabe and Herbonett masses incorporate a correction \arnote{from} the shape measurement bias and the contamination of background galaxies. Okabe's correction raises their masses by 4.2\% and is entirely attributed to the shear correction.  Herbonett does not explicitly state the effect on mass from their corrections, but instead note an uncertainty of $~\sim 2\%$ for each. However \citet{Hoekstra2015}, which uses the same shear pipeline, notes that the effect on mass from ignoring the shear and boost corrections is $\sim 15\%$ and $\sim 10\%$. }  Thus,  while the Okabe clusters would be better aligned with the escape masses, the Herbonett masses would be 2$\sigma$ low. Our comparison supports the need for the systematic correction terms in the technique employed by \citet{Herbonnet+2022}, whereas the technique used by O16 is naturally in good agreement with the escape masses.

We can also ask about the effect on the escape mass from cosmology. 
As shown in Table \ref{table:cluster_correlations}, we find that decreasing $H_0$ increases the inferred escape masses. In fact, the large positive mass bias between the weak lensing and caustic-inferred masses could be explained by an unrealistic $H_0 \approx 50$~km~s$^{-1}$~Mpc$^{-1}$. Mass changes due to variations in $\Omega_M$ are much less significant for clusters at this redshift \citep{stark2017}.  
Hence, the observed tension in the CMB $H_0 = 67.66 \pm 0.42 \, \text{km} \, \text{s}^{-1} \, \text{Mpc}^{-1}$ \citep{Planck+2018} and the Type 1a $H_0 = 73.6 \pm 1.1 \, \text{km} \, \text{s}^{-1} \, \text{Mpc}^{-1}$ \citep{Brout+2022} is a much stronger driver on our masses compared to the disagreement on the respective $\Omega_M = 0.311 \pm 0.006$ and $ \Omega_M = 0.334 \pm 0.018$. The escape technique was not included in the numerous $H_0$ probes compared in \citet{H0tension}. However, as the only known probe (predominantly dynamical) which constrains $qH^2 = -\frac{\ddot{a}}{a}$, a future effort could provide a very interesting direct constraint on the expansion rate and acceleration.

This concordance between these independent mass measurement technique is unlikely to be coincidental. Our analysis combines two fundamental equations that govern the dynamics of tracers in any cosmological potential through equations \ref{eq:nabla_phi} and \ref{eq:phi_poisson}. Formally, the equality in equation \ref{eq:nabla_phi} requires the presence of a non-relativistic stress energy in General Relativity's (GR) field equations. Non-standard cosmologies like Hu-Sawicky $f(R)$ gravity do not have this requirement and in turn require a modification to the potential in the Poisson equation \citep{Daniel2009}. The idea of using the escape profiles of galaxy clusters to test non-GR gravity theory was first presented in \citet{Stark_fR2016}, where they showed that the escape profiles can be enhanced in $f(R)$ compared to GR at fixed cluster mass. The enhancement is a function of the cluster mass \arnote{from} Chameleon screening, where the dynamics in high density (mass) regions match GR. It is very unlikely that the clusters in our sample would show effects from $f(R)$, gravity. If they did and  since photons are not affected in $f(R)$ gravity, we would expect the escape inferred masses to be higher than the weak lensing masses. We see no evidence for this in the data.

We conclude that the dominant systematic in our newly refined escape velocity mass estimated technique is from current uncertainties on cosmology.  With only 46 clusters compared, we have reached a level of accuracy and precision that is cosmologically interesting. Given the \citet{Planck+2018} cosmology, our work places stringent limits on the possible systematics that exist in weak lensing mass estimation techniques. Our newly refined escape mass estimation technique provides a clear path forward to measure precise and accurate cluster masses, to dynamically probe the late-universe spacetime expansion, and to test general relativity on megaparsec scales.

\clearpage

\footnotesize{
\bibliographystyle{natbib}
\bibliography{master}
}

\appendix
\renewcommand{\thefigure}{A\arabic{figure}}
\label{sec:appendix}

\setcounter{figure}{0}
\section{Fits to the Skewed Gaussian}
\label{sec:Fits}
\begin{figure*}[ht]
\label{fig:a1}
    \centering
    \begin{minipage}{\textwidth}
        \includegraphics[width=.33\textwidth]{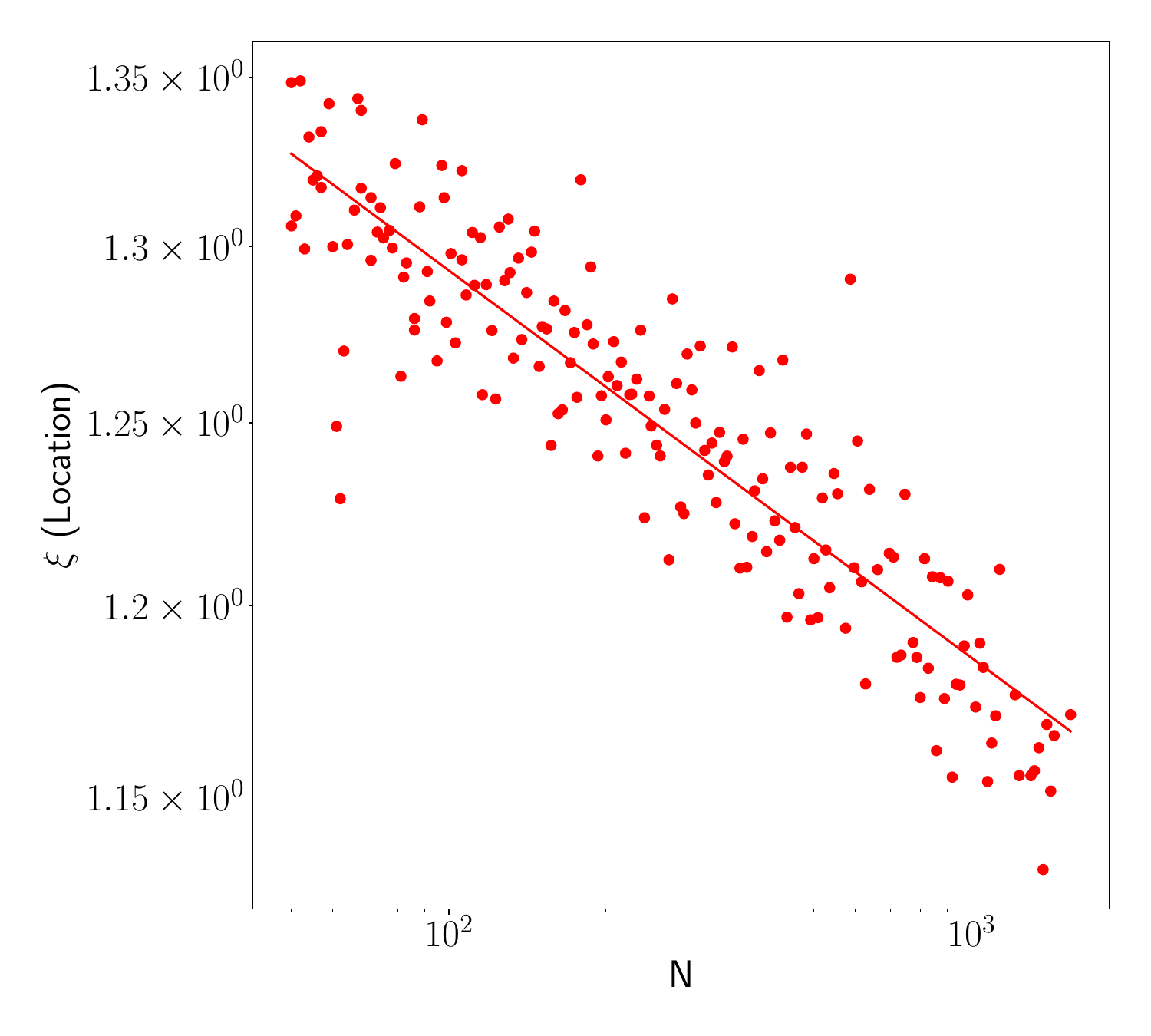}%
        \includegraphics[width=.33\textwidth]{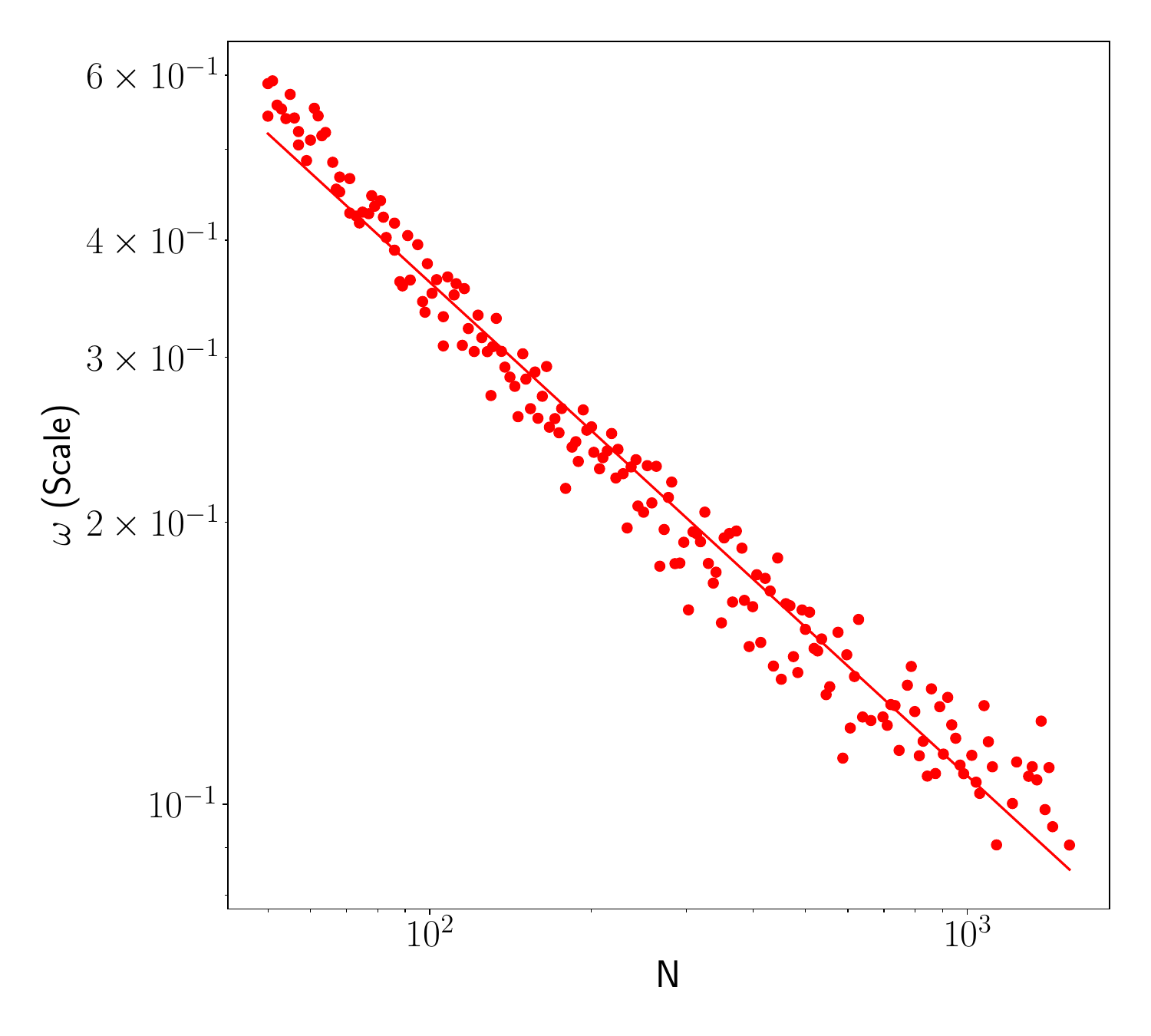}
        \includegraphics[width=.33\textwidth]{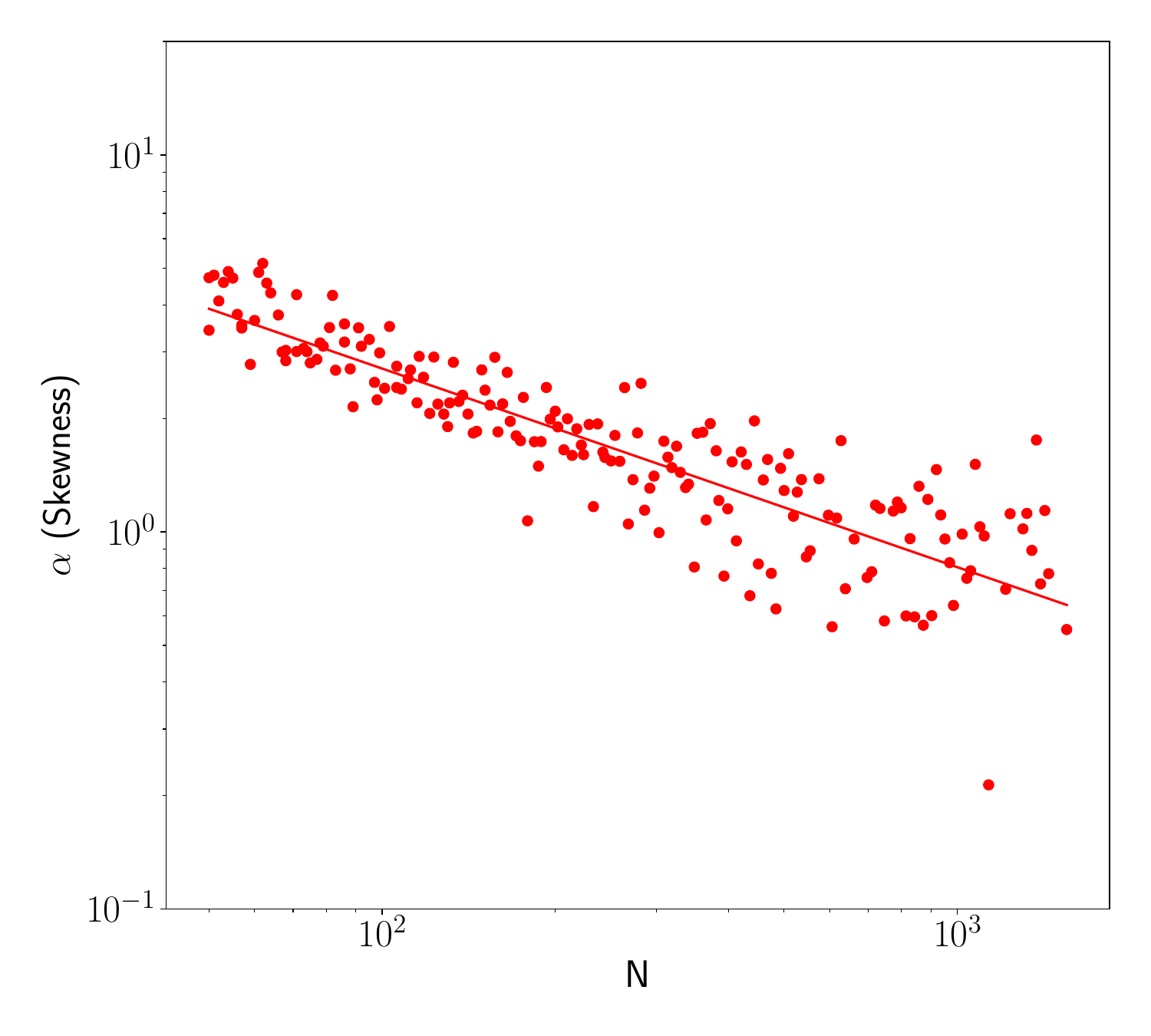}
    \end{minipage}

\caption{\label{fig:Skewness_Fits} $\xi$ (location), $\omega$ (scale), and $\alpha$ (skewness) parameters in AGAMA for $Z_v$, following its representation as a skewed normal distribution (equation \ref{eq:Zv_pdf}), for a $10^{15}\,M_{\odot}$ cluster at redshift $z=0.01$, inferred for 1000 different viewing angles in the innermost bin. Other bin, redshift, and mass choices follow similar trends, with each parameter following a linear scaling in $\log(N)$ space.
} 
\end{figure*}

\section{Phase-Spaces for the Sample}
\label{sec:Phase_spaces}
Below, all phase-spaces we use in the analysis of the final sample from \citet{Herbonnet2020} and \citet{Okabe+2016} are presented. For presentation purposes, the suppression in the diagrams is taken to be the median of 1000 draws of $Z_v$ in each bin, although the actual MCMC chains sample from $Z_v$ stochastically without taking any means or medians. The same suppression is applied to both the lensing estimates (green lines) and the dynamical fits (blue lines). For visual purposes, all phase-spaces are also shown at the centers of the $\hat{r}_{200}$ starting estimate ranges, i.e. the corresponding lensing $r_{200}$. Hence, the relative agreement between the dynamical fits and lensing profiles may not exactly match Tables \ref{table:cluster_info} and \ref{table:cluster_info2}. 

For specific phase-spaces, we note that we impose that no interlopers are identified in the first bin, given the difficulty of projecting galaxies into the core. Visual inspection of the phase-spaces forces us to drop this constraint on the following clusters: A2050, A2055, RXJ2129.6+0005, A1914, and A2390. We also do not enforce the monotonicity constraint on A2443 and ZwCl0949.6+5207. In rare cases such as A2645 and A1689, we find the shifting-gapper does not remove obvious interlopers, so we manually adjust the velocity cut constraint to 2500 and 3000 km/s for these two clusters respectively to ensure proper interloper removal. In the diagrams below, red points represent interloper galaxies.

\setlength{\intextsep}{0pt}
\setlength{\textfloatsep}{0pt}

\setlength{\intextsep}{0pt}
\setlength{\textfloatsep}{0pt}

\begin{figure}
\centering
\includegraphics[width=.19\textwidth,height=.12\textheight,keepaspectratio]{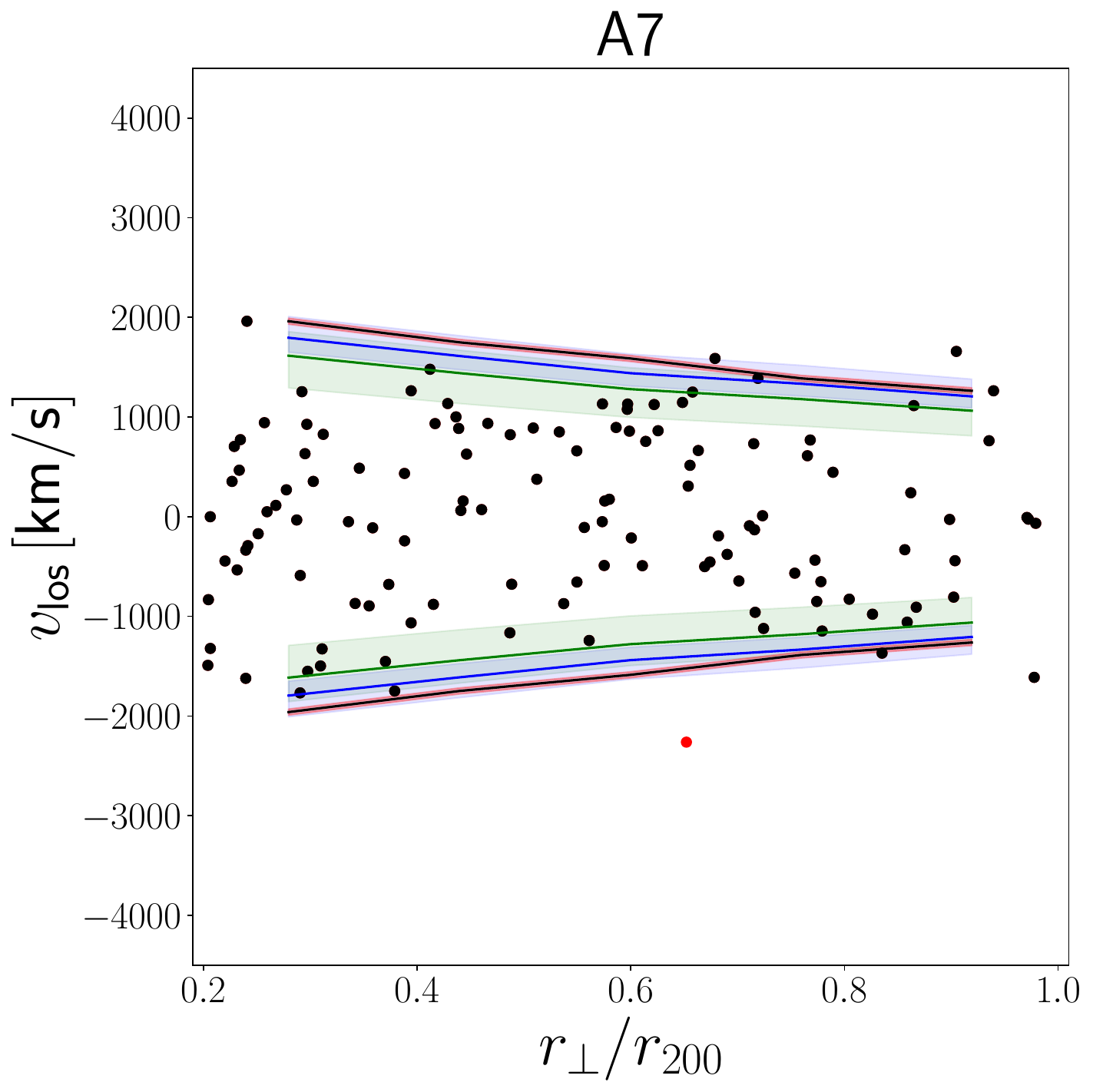}
\includegraphics[width=.19\textwidth,height=.12\textheight,keepaspectratio]{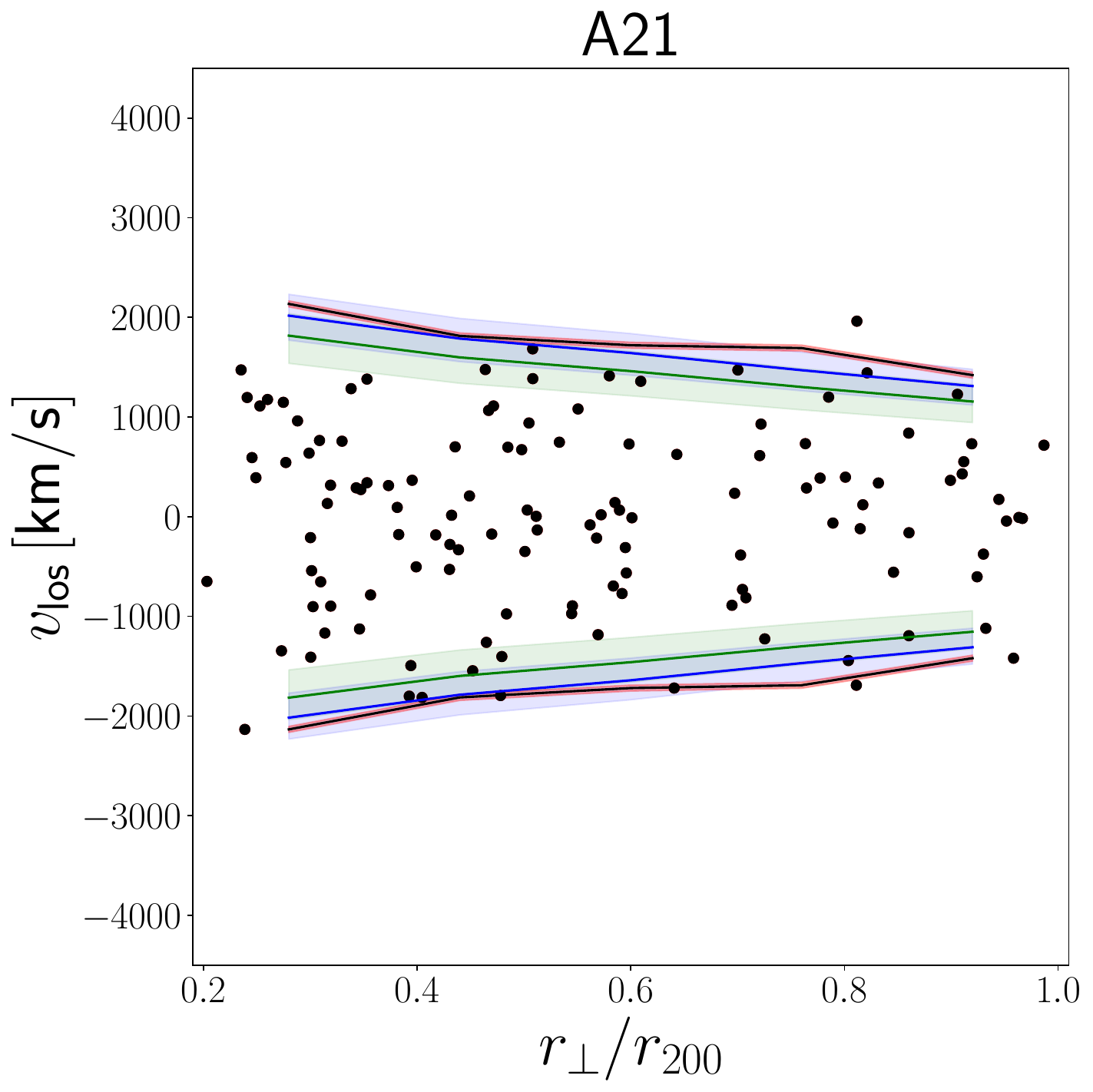}
\includegraphics[width=.19\textwidth,height=.12\textheight,keepaspectratio]{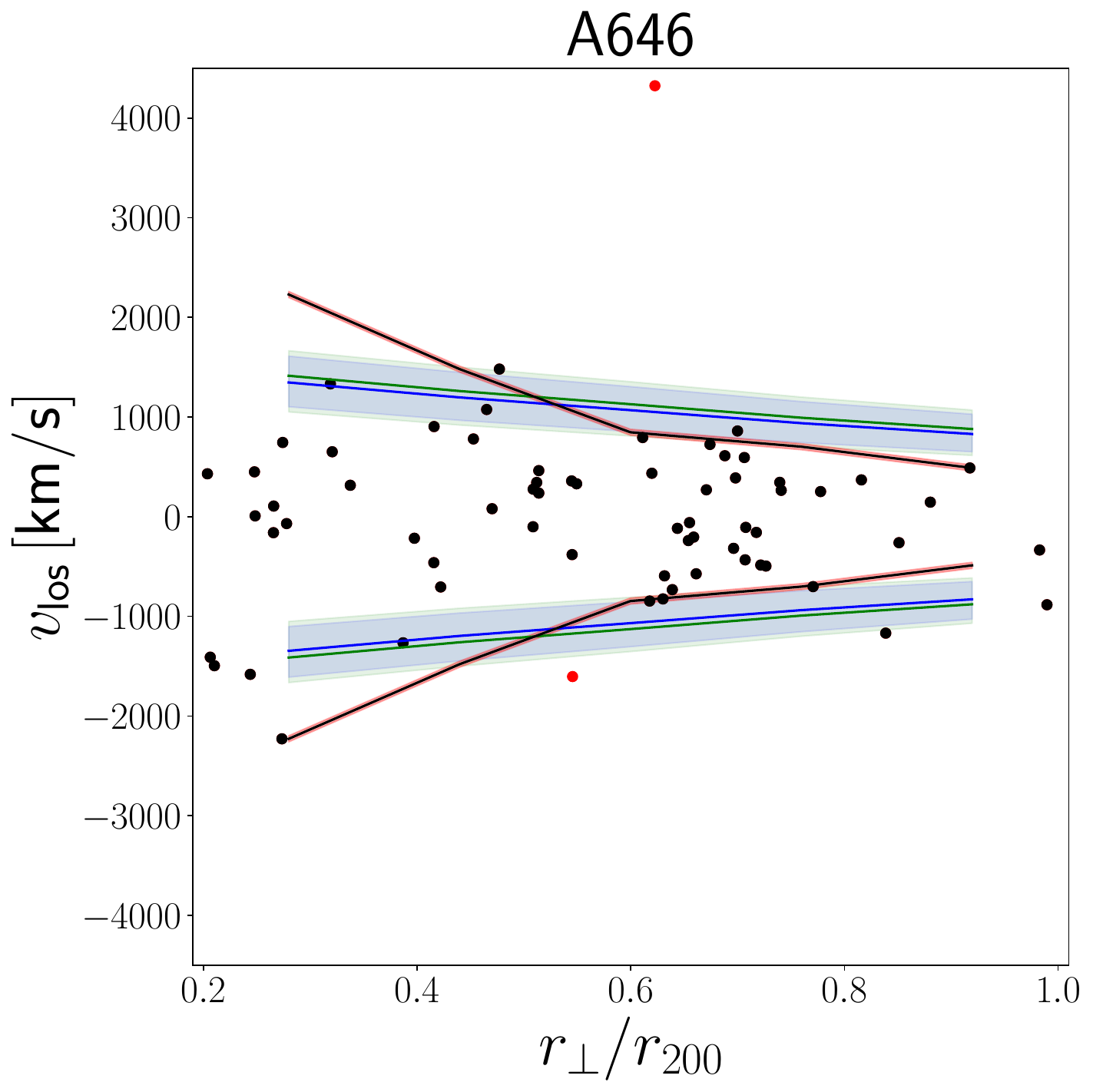}
\includegraphics[width=.19\textwidth,height=.12\textheight,keepaspectratio]{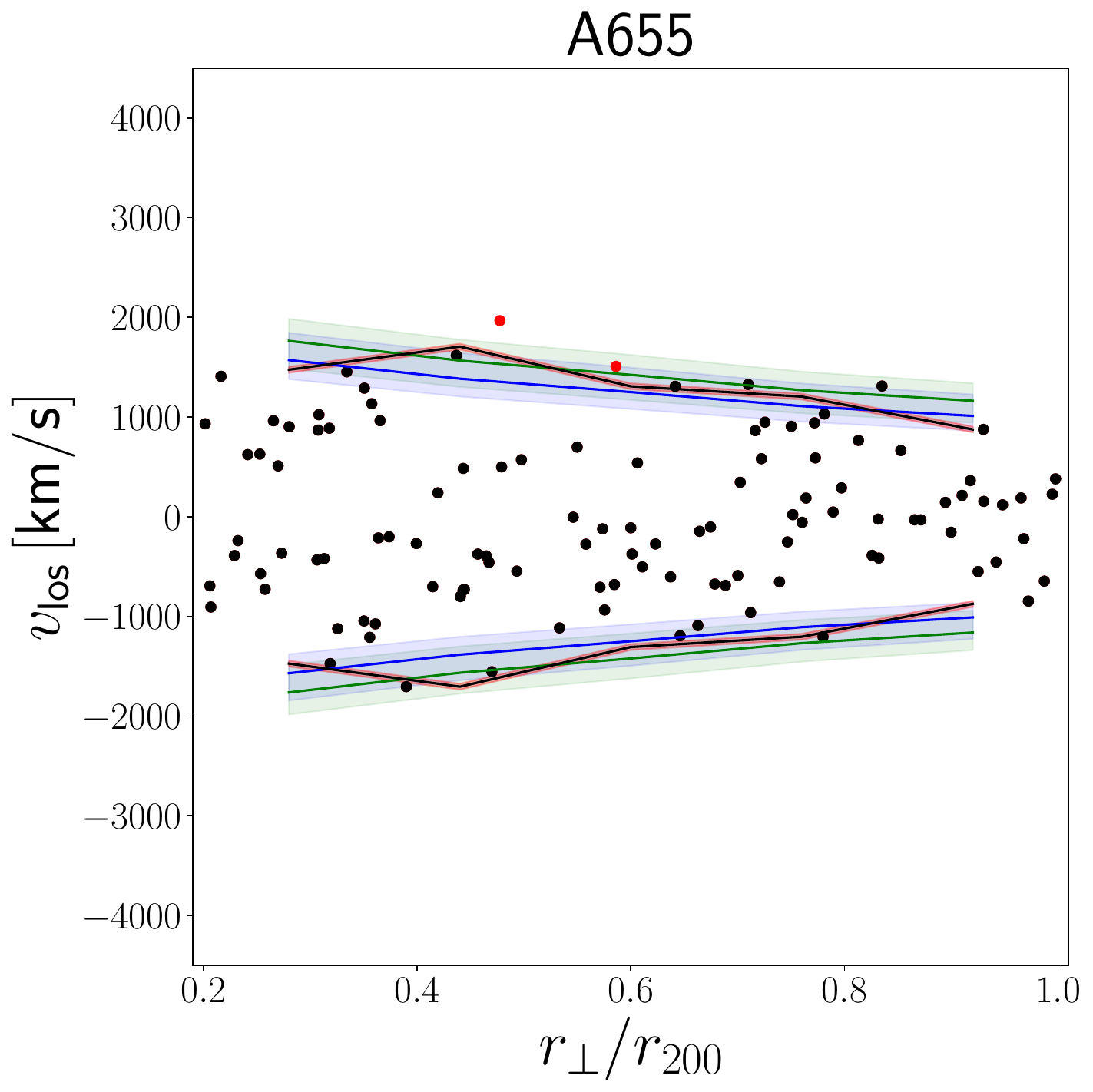}
\includegraphics[width=.19\textwidth,height=.12\textheight,keepaspectratio]{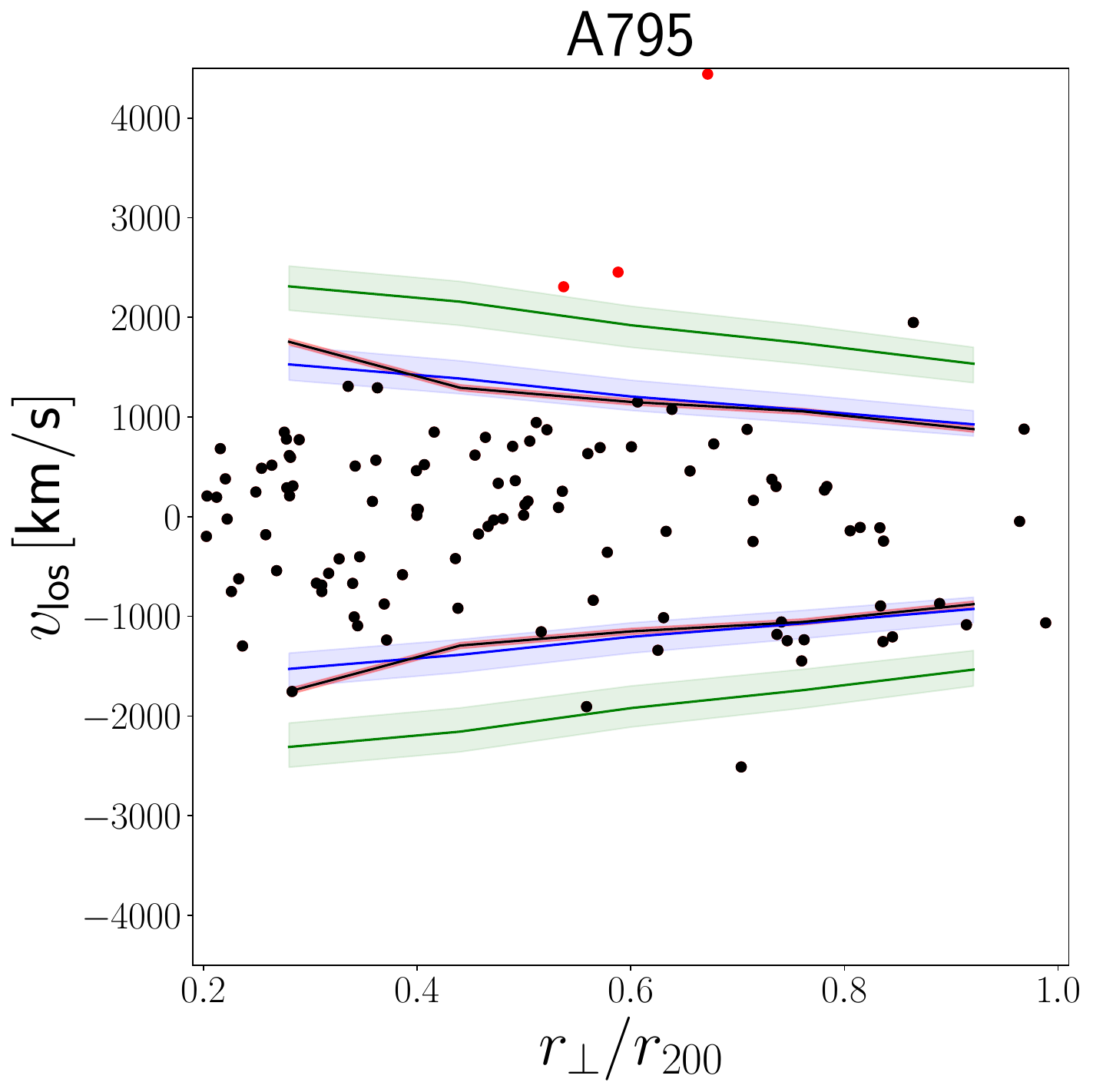}\\[-.7em]
\includegraphics[width=.19\textwidth,height=.12\textheight,keepaspectratio]{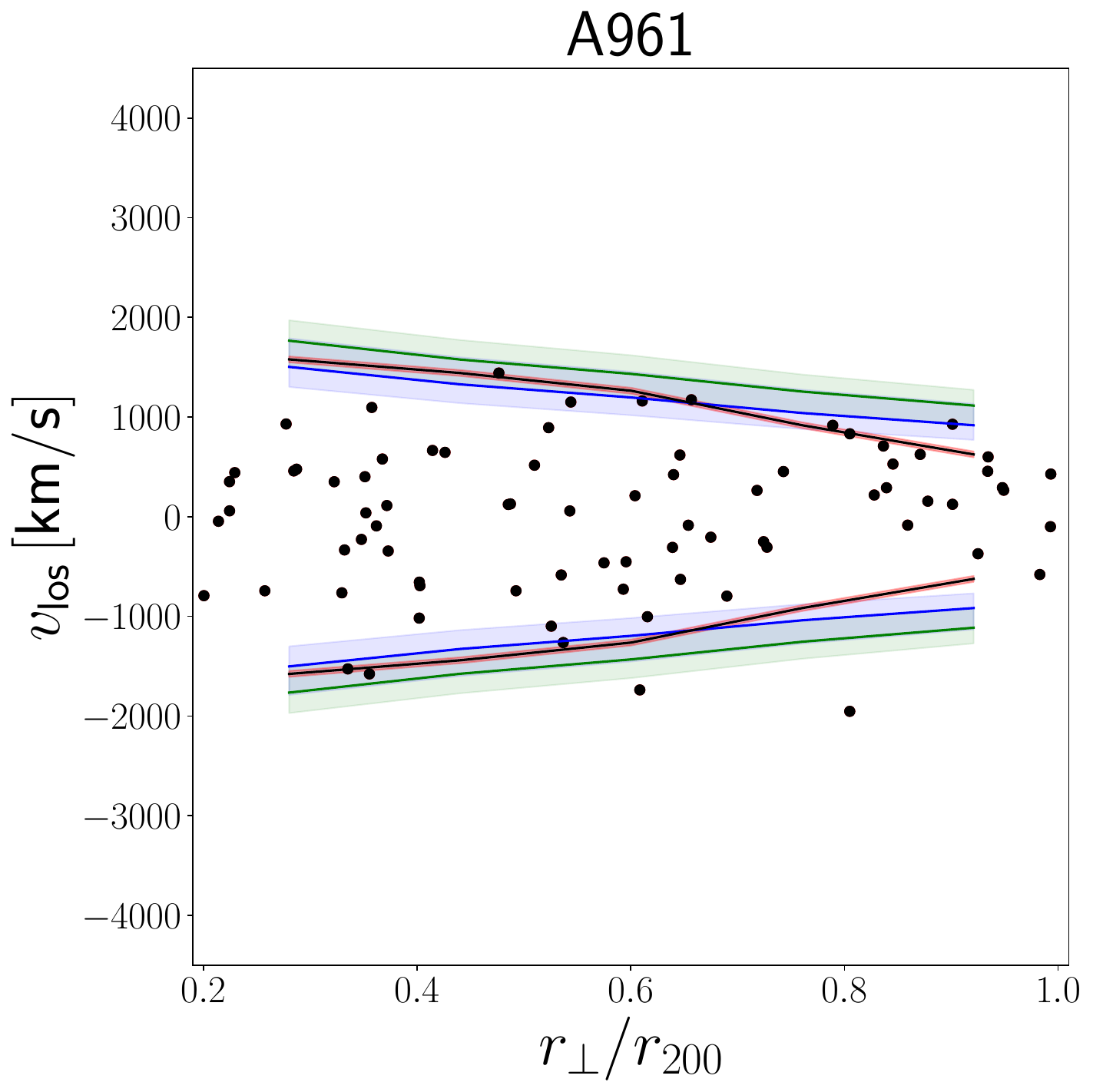}
\includegraphics[width=.19\textwidth,height=.12\textheight,keepaspectratio]{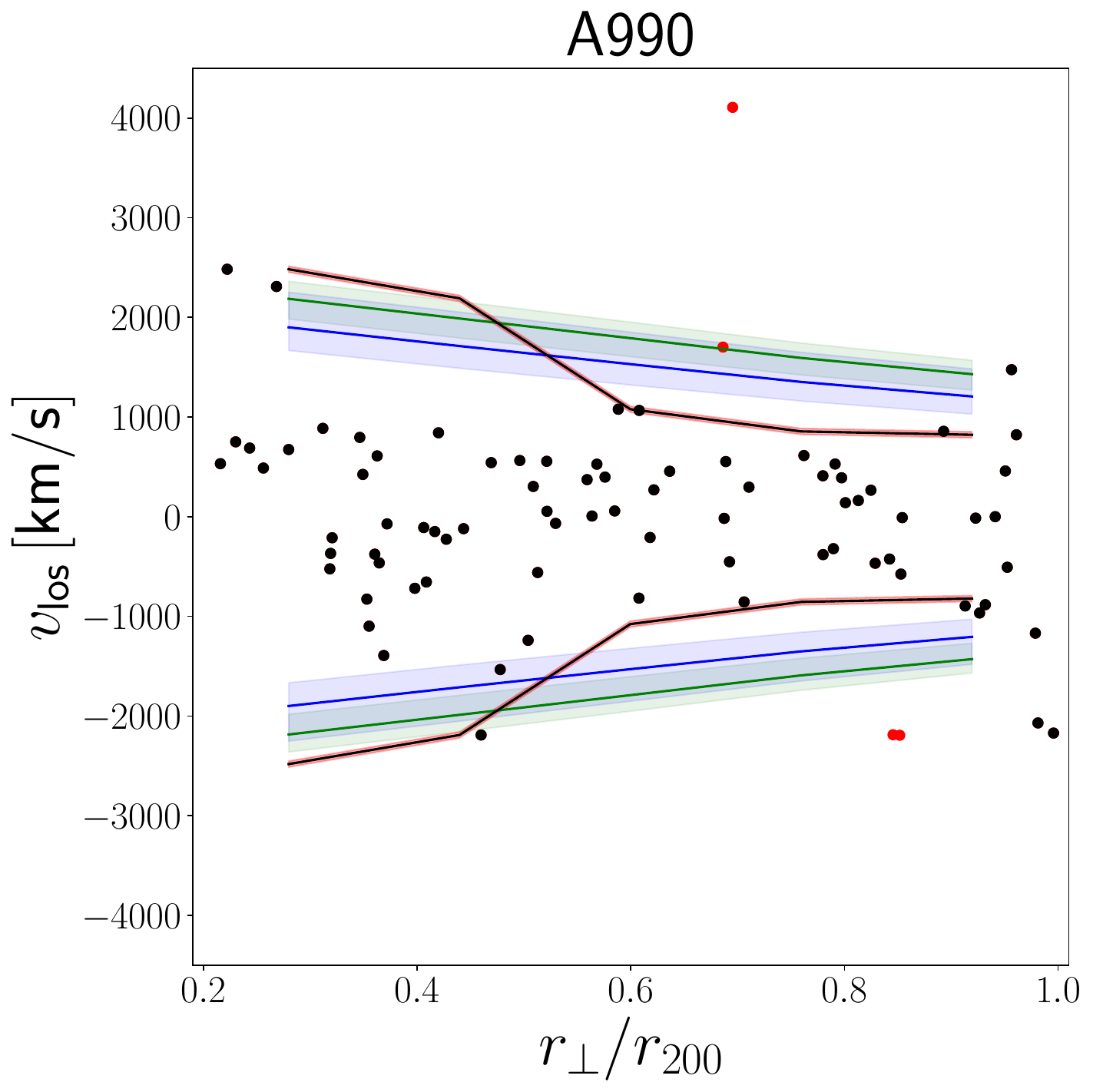}
\includegraphics[width=.19\textwidth,height=.12\textheight,keepaspectratio]{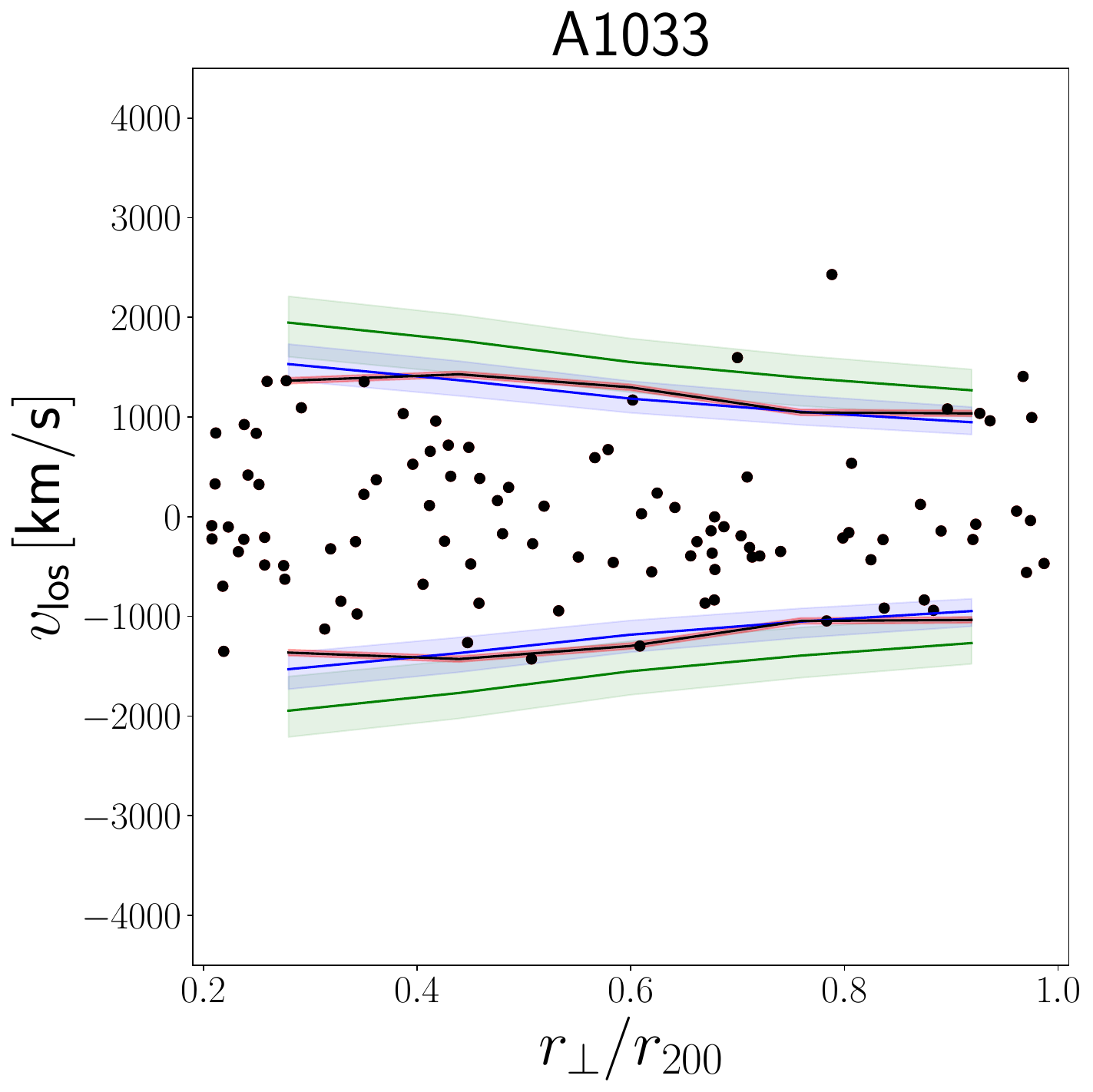}
\includegraphics[width=.19\textwidth,height=.12\textheight,keepaspectratio]{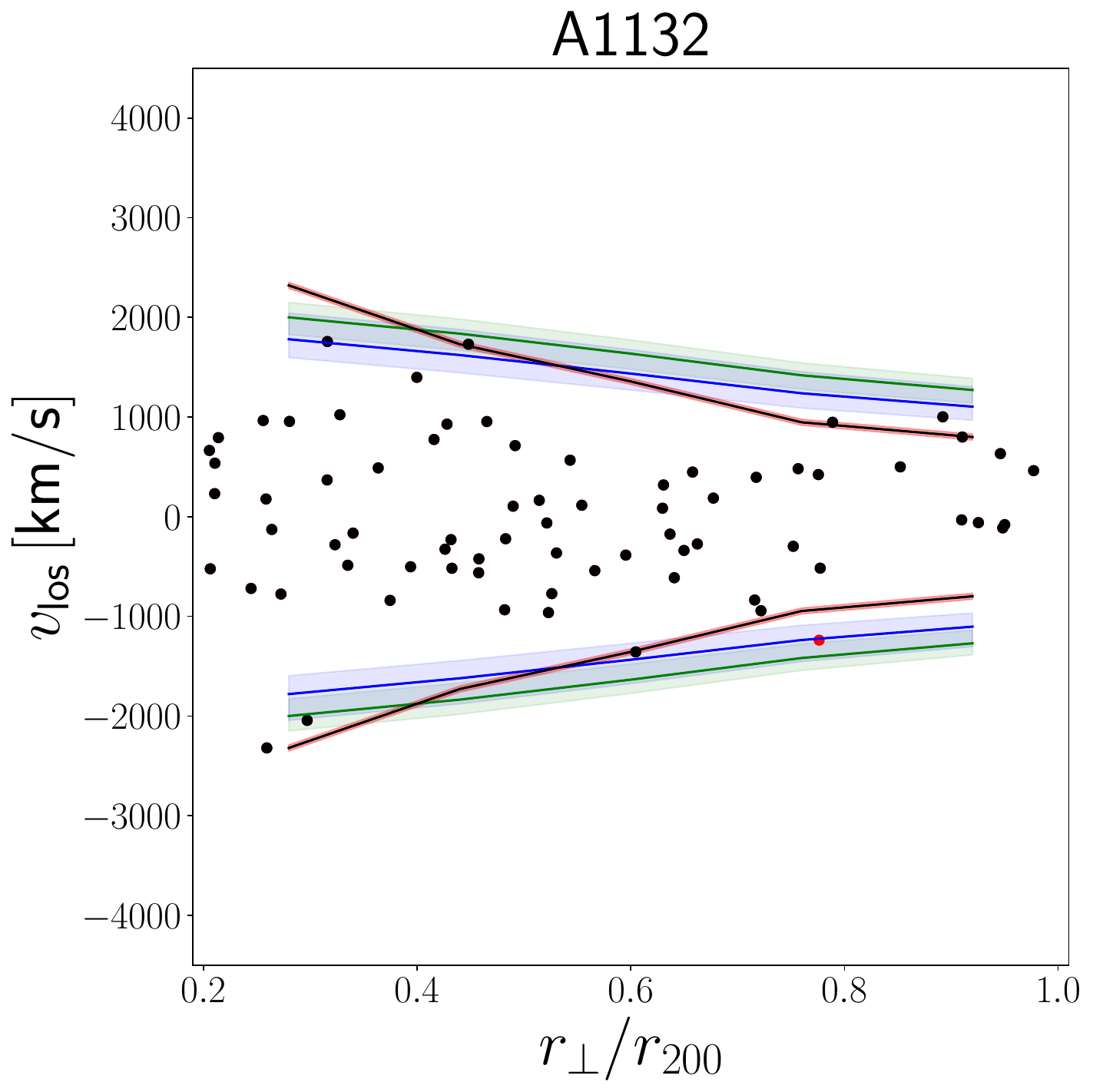}
\includegraphics[width=.19\textwidth,height=.12\textheight,keepaspectratio]{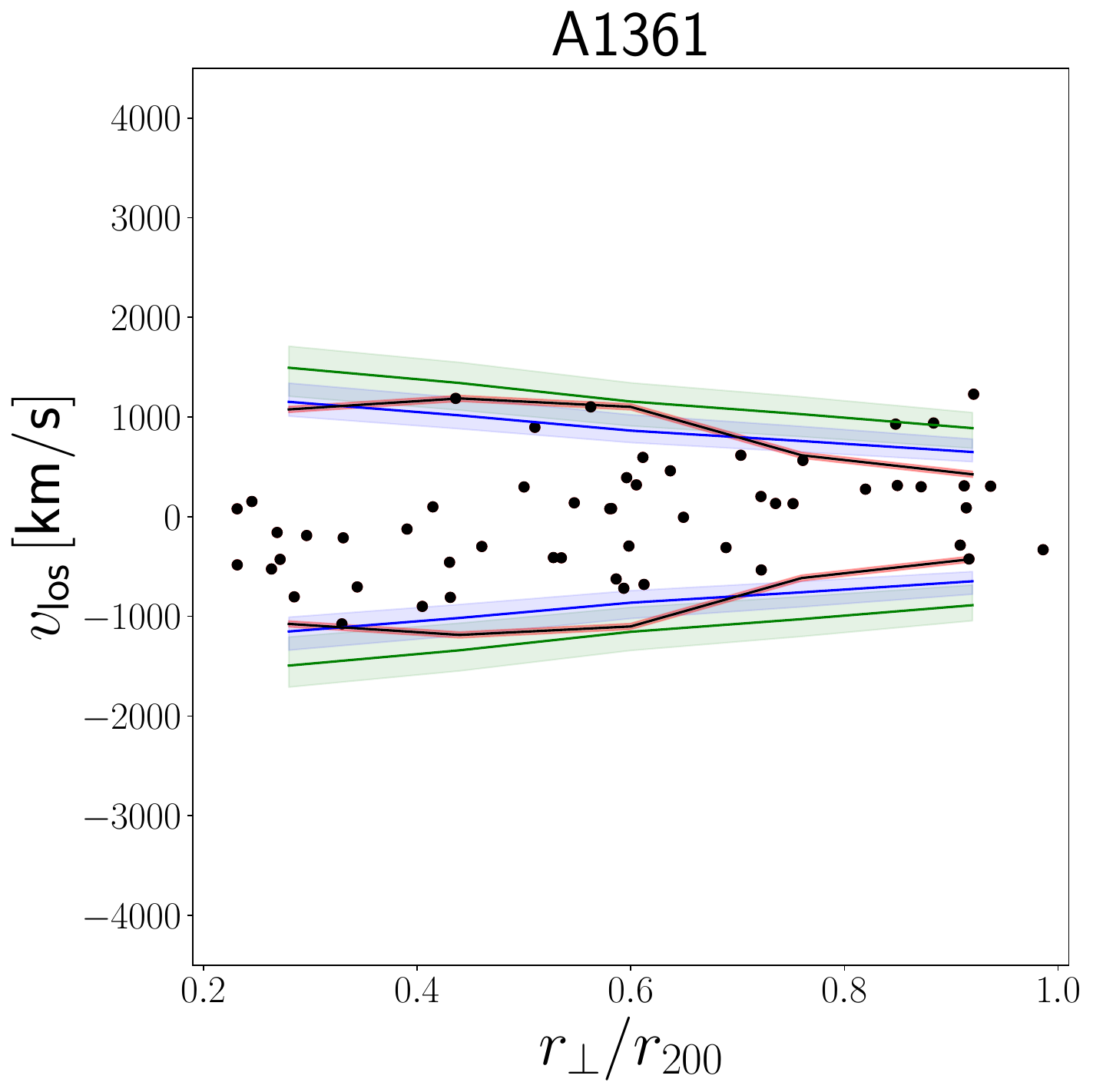}\\[-.7em]
\includegraphics[width=.19\textwidth,height=.12\textheight,keepaspectratio]{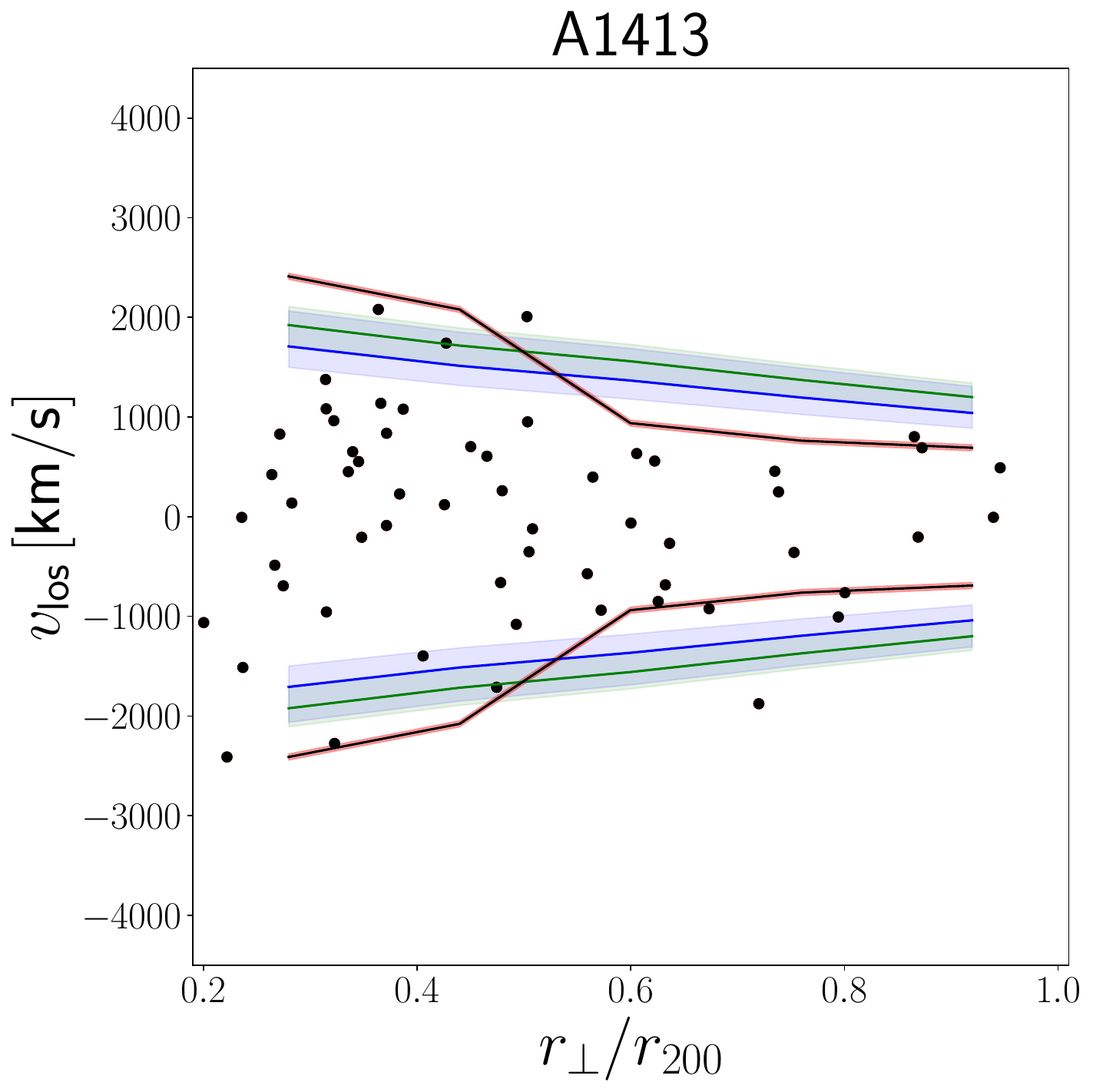}
\includegraphics[width=.19\textwidth,height=.12\textheight,keepaspectratio]{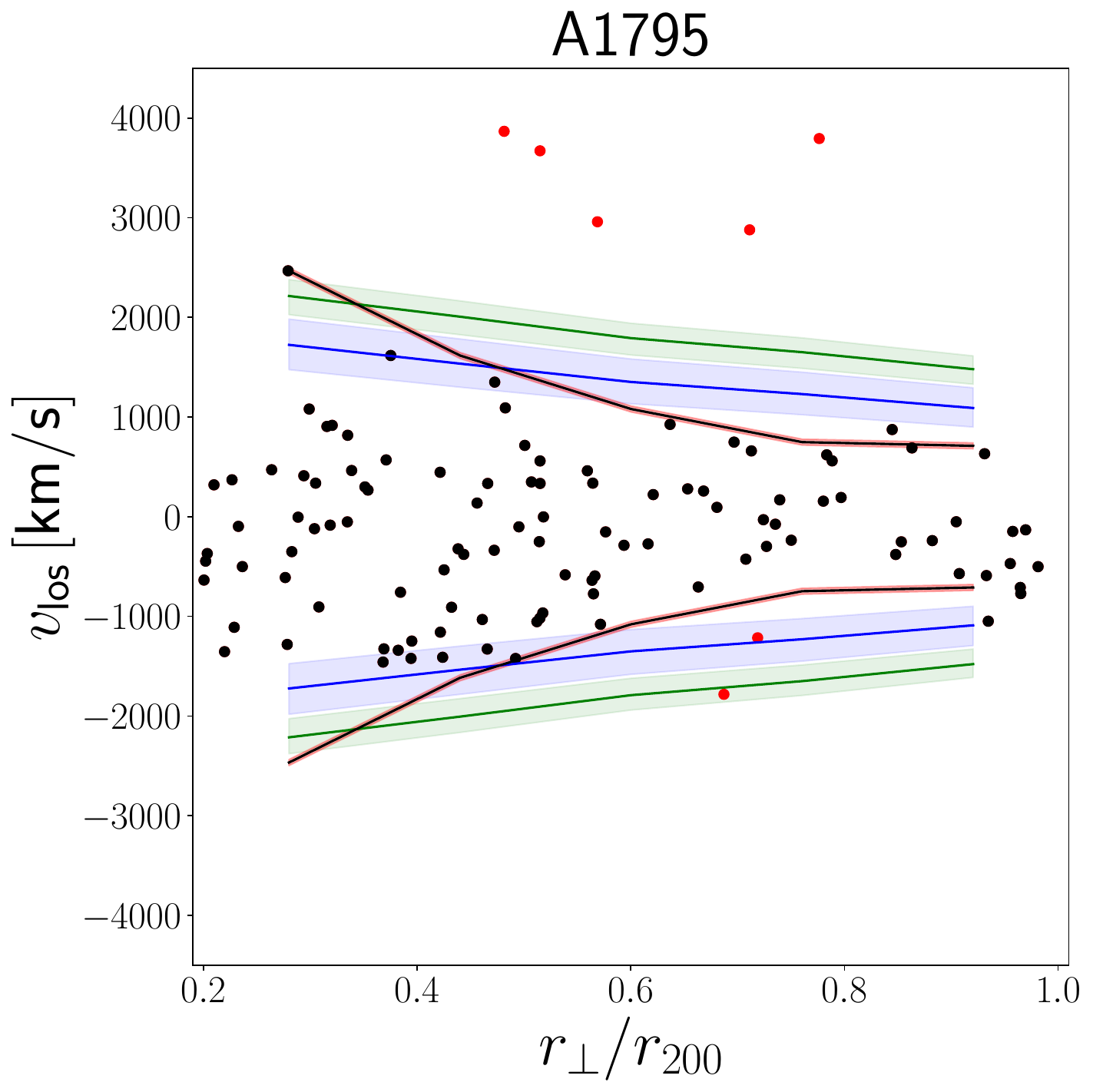}
\includegraphics[width=.19\textwidth,height=.12\textheight,keepaspectratio]{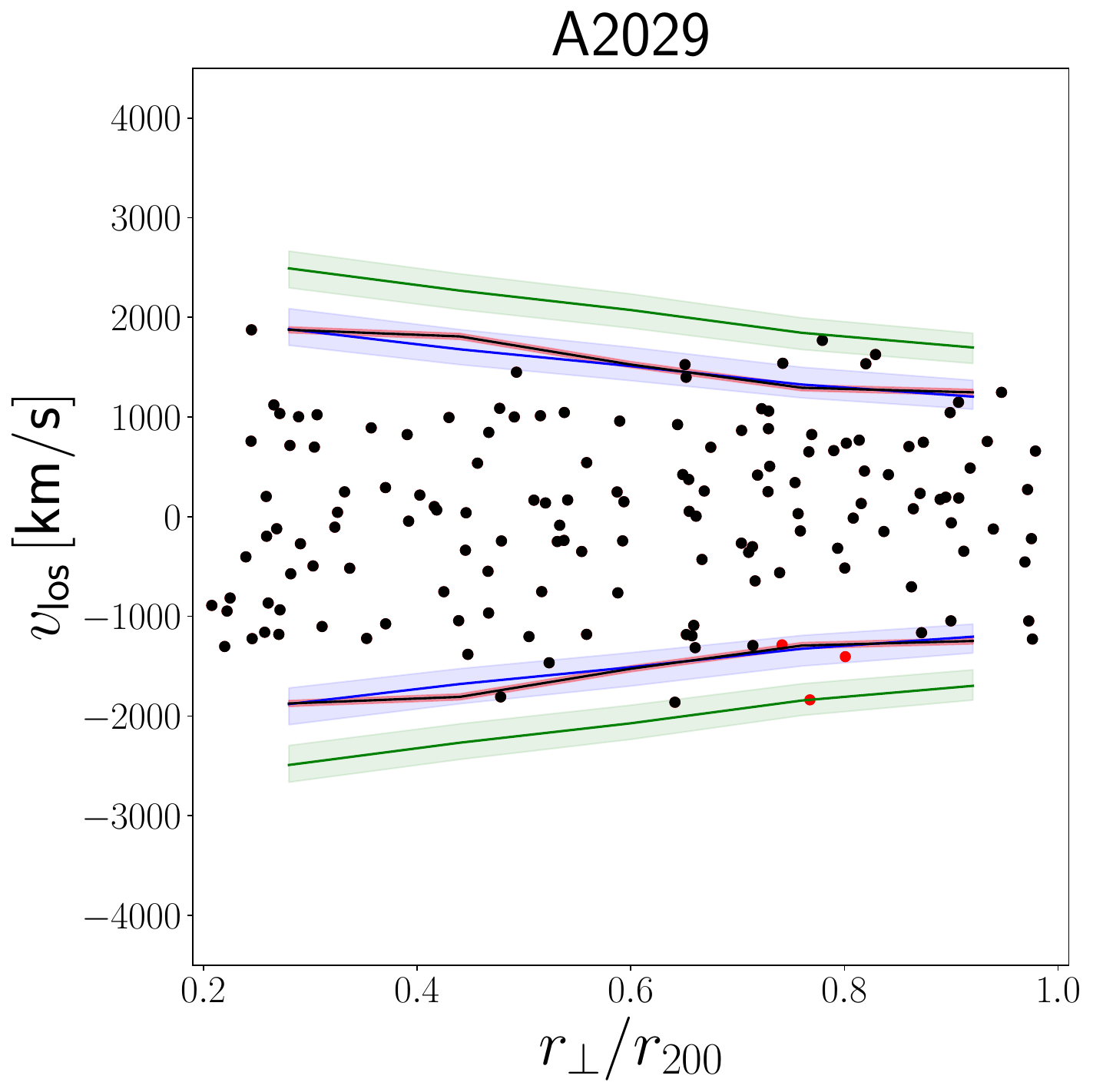}
\includegraphics[width=.19\textwidth,height=.12\textheight,keepaspectratio]{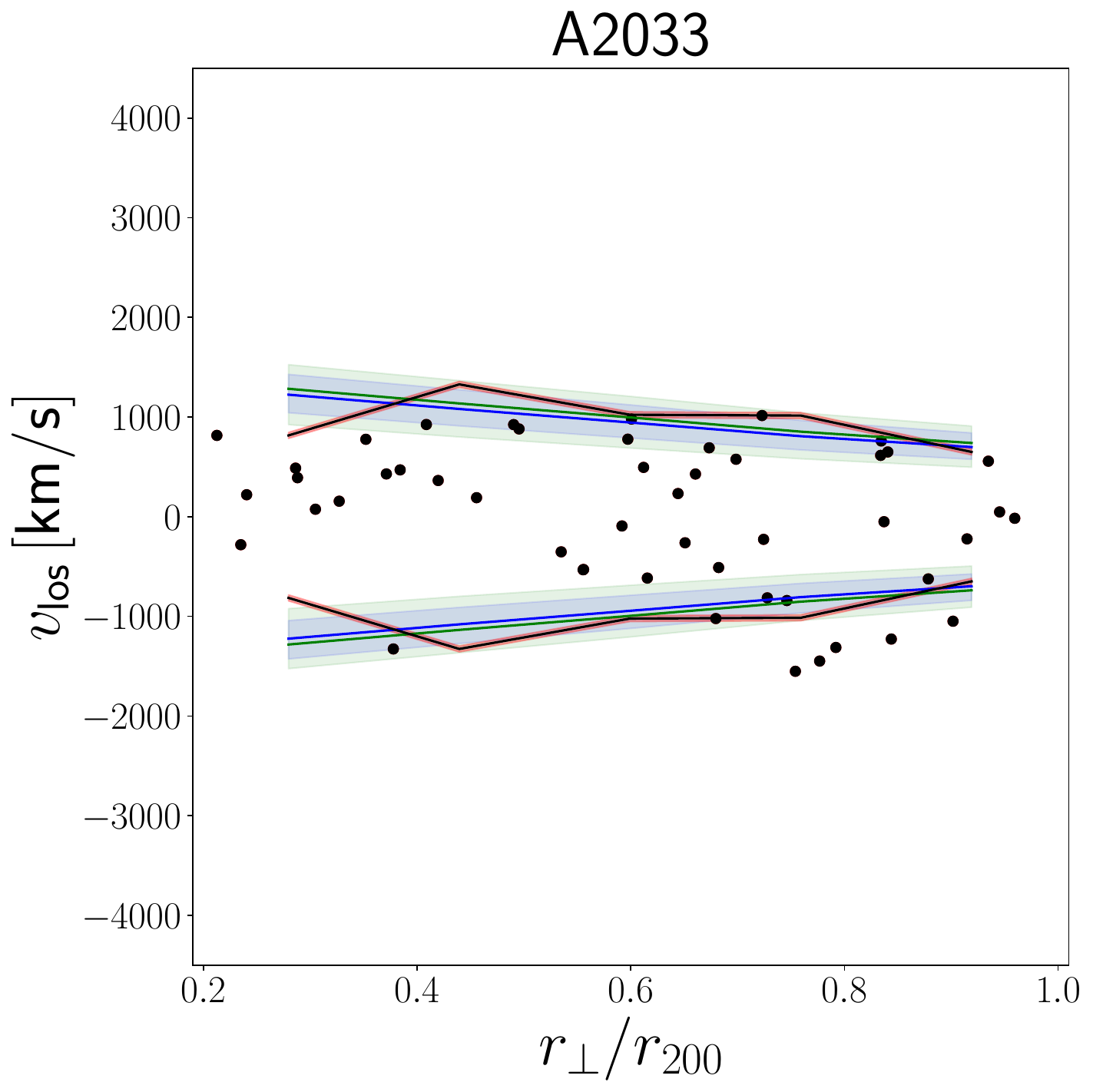}
\includegraphics[width=.19\textwidth,height=.12\textheight,keepaspectratio]{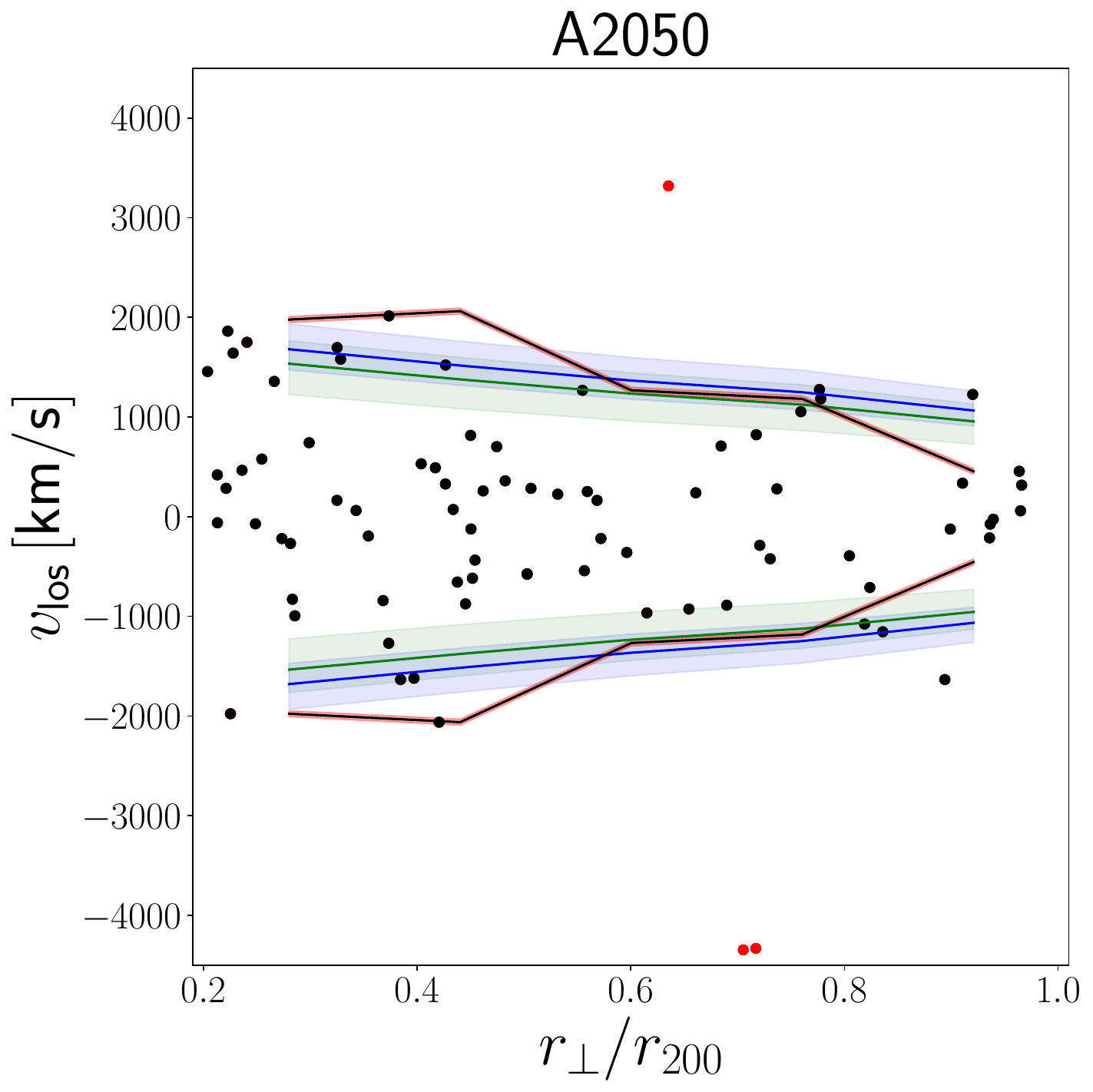}\\[-.7em]
\includegraphics[width=.19\textwidth,height=.12\textheight,keepaspectratio]{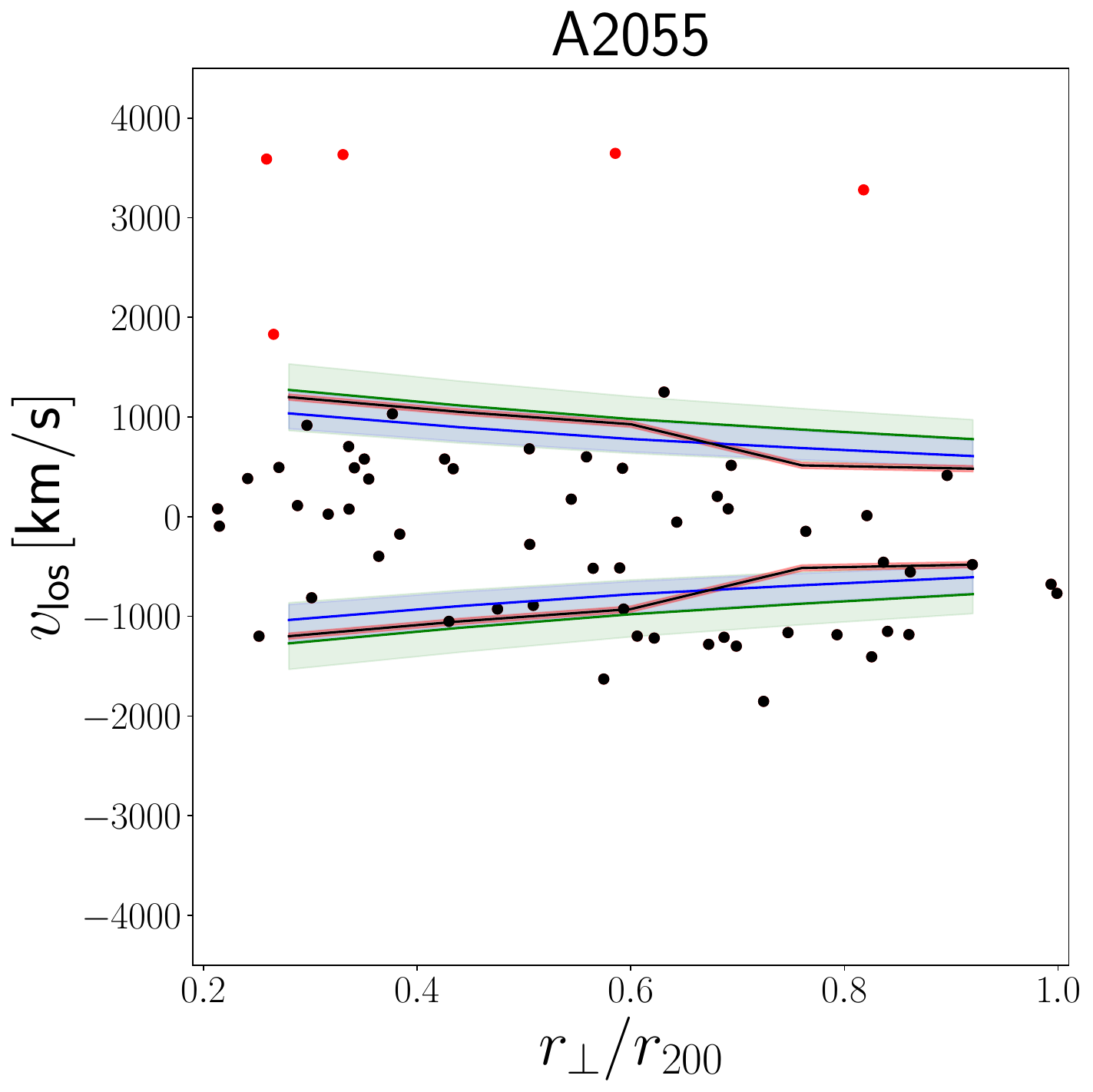}
\includegraphics[width=.19\textwidth,height=.12\textheight,keepaspectratio]{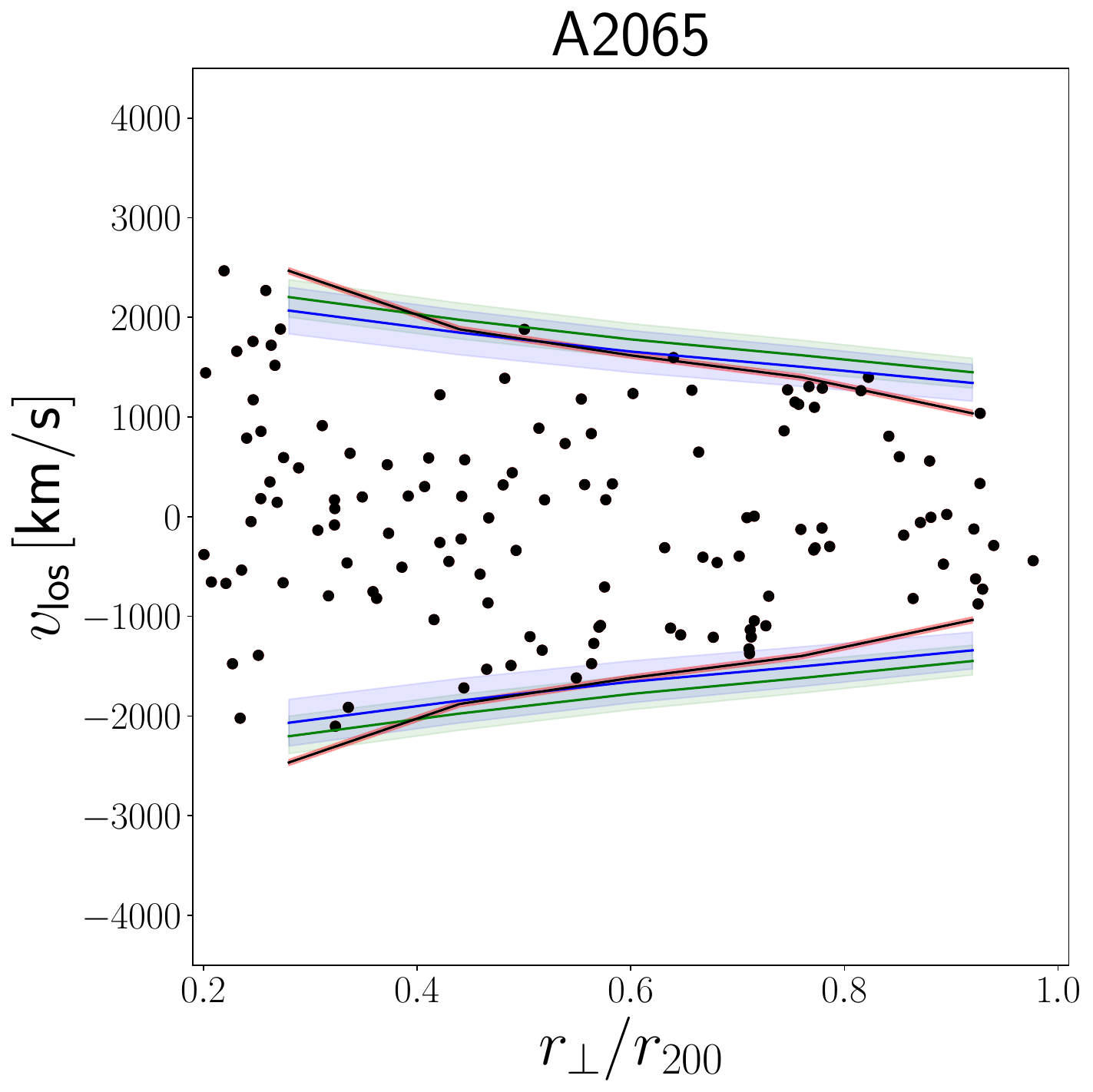}
\includegraphics[width=.19\textwidth,height=.12\textheight,keepaspectratio]{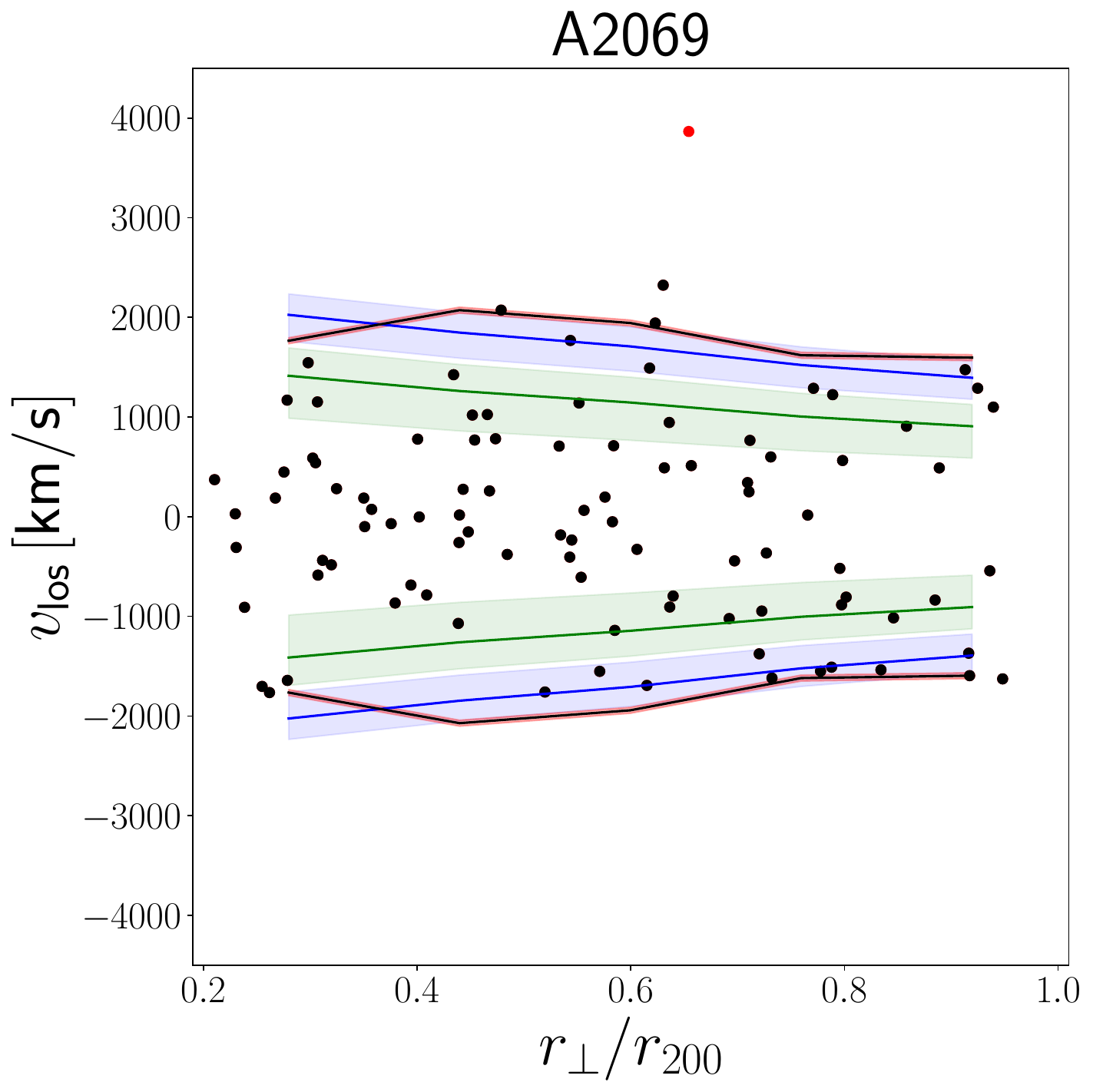}
\includegraphics[width=.19\textwidth,height=.12\textheight,keepaspectratio]{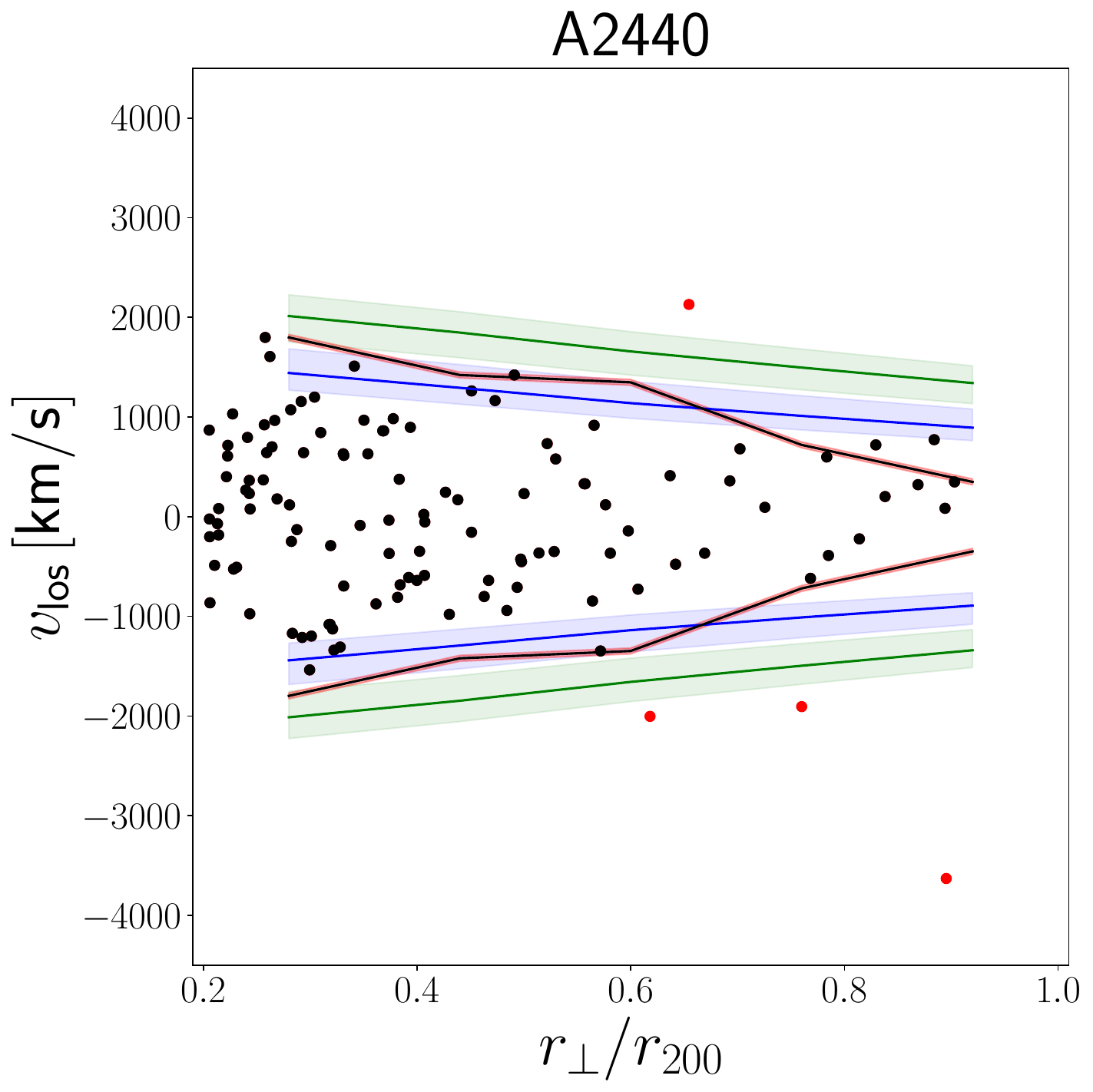}
\includegraphics[width=.19\textwidth,height=.12\textheight,keepaspectratio]{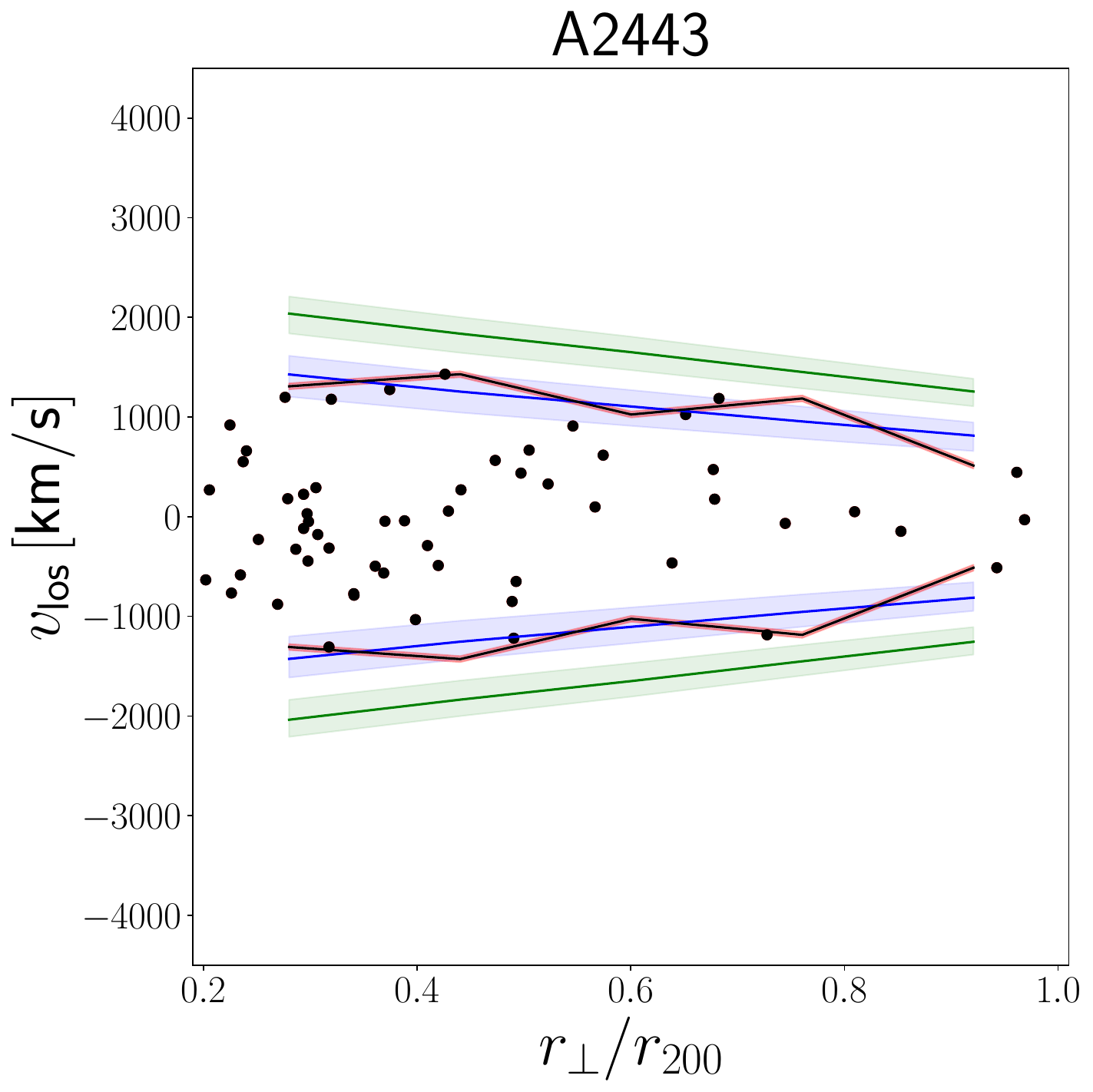}\\[-.7em]
\includegraphics[width=.19\textwidth,height=.12\textheight,keepaspectratio]{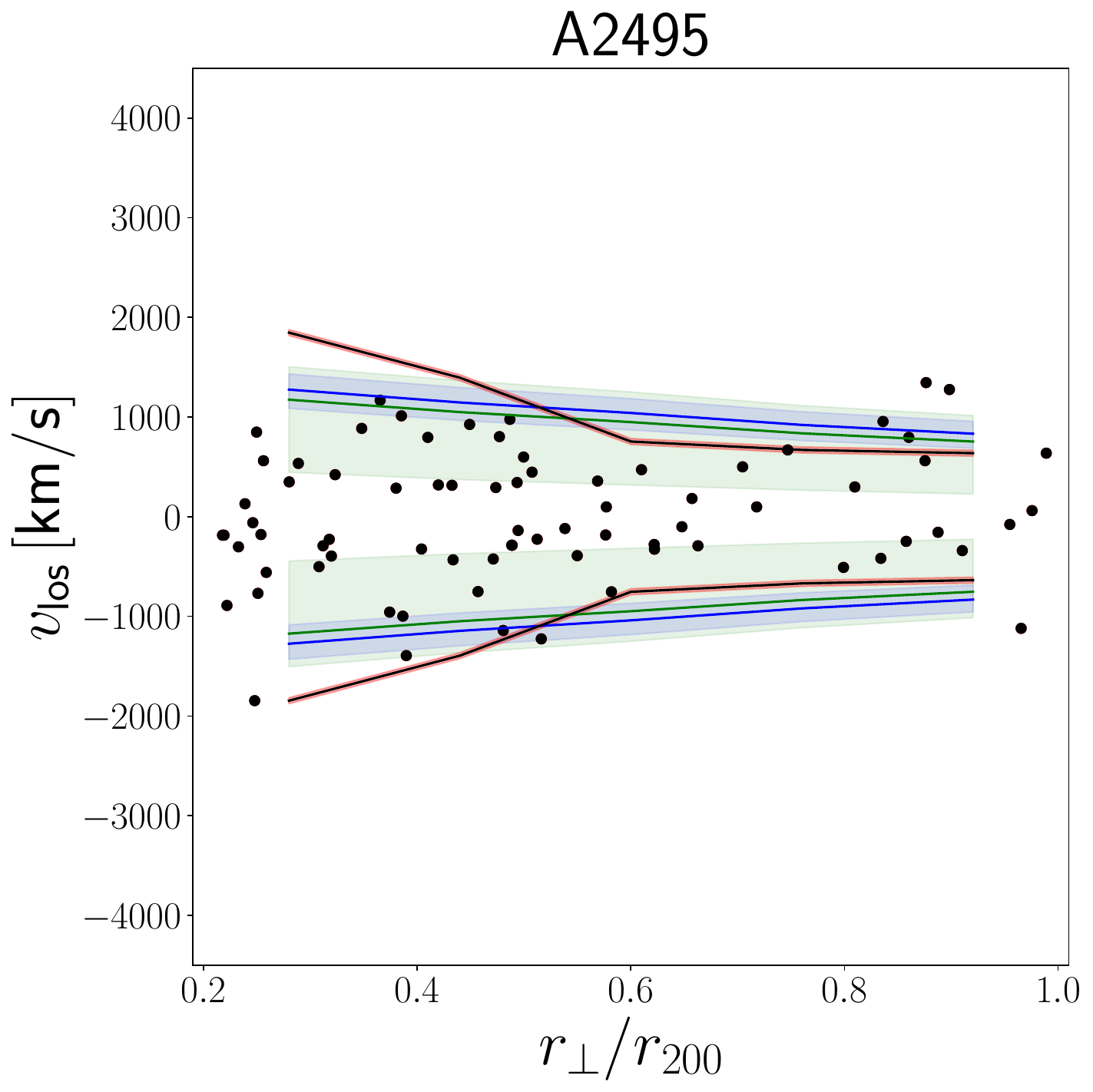}
\includegraphics[width=.19\textwidth,height=.12\textheight,keepaspectratio]{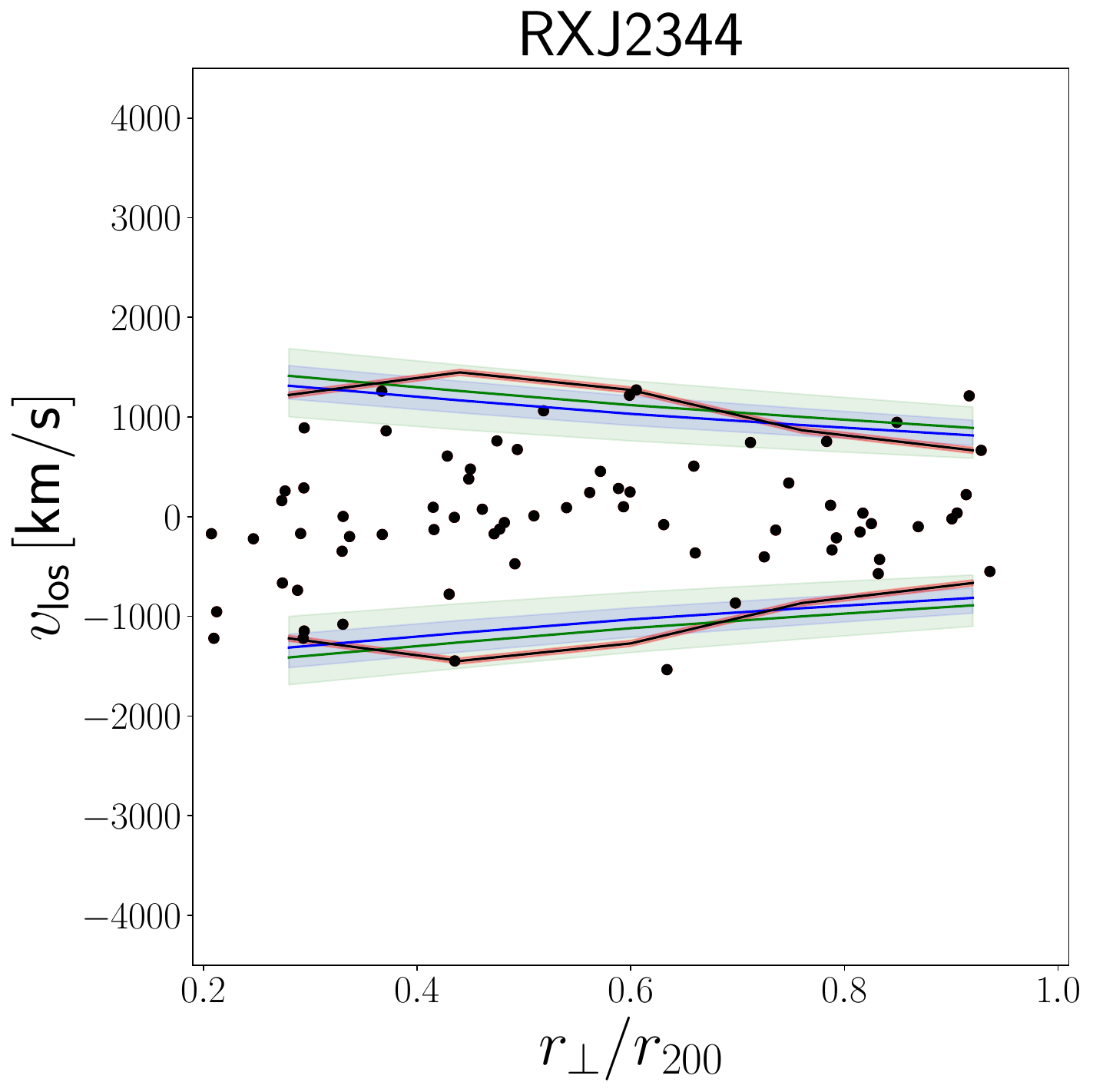}
\includegraphics[width=.19\textwidth,height=.12\textheight,keepaspectratio]{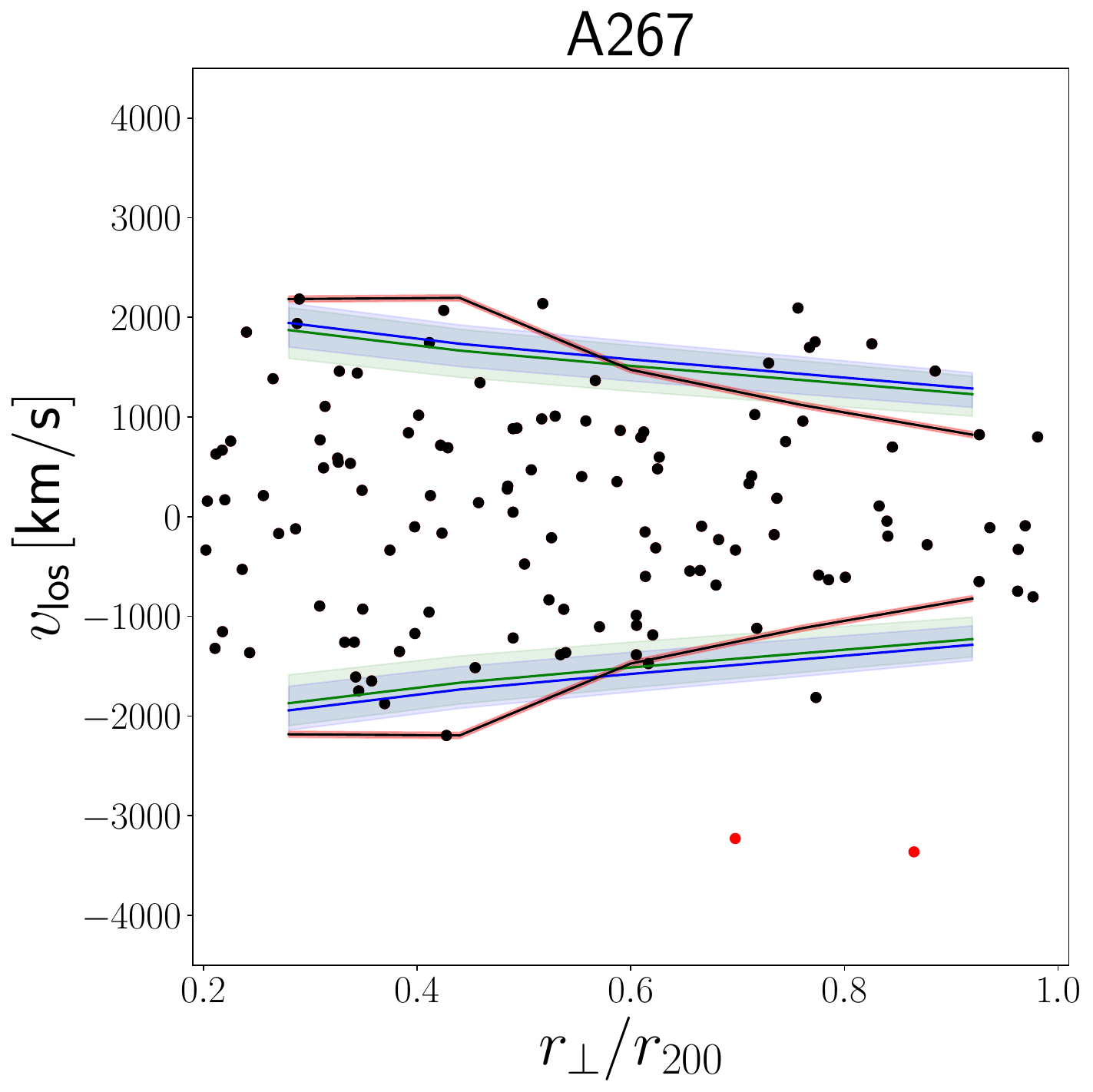}
\includegraphics[width=.19\textwidth,height=.12\textheight,keepaspectratio]{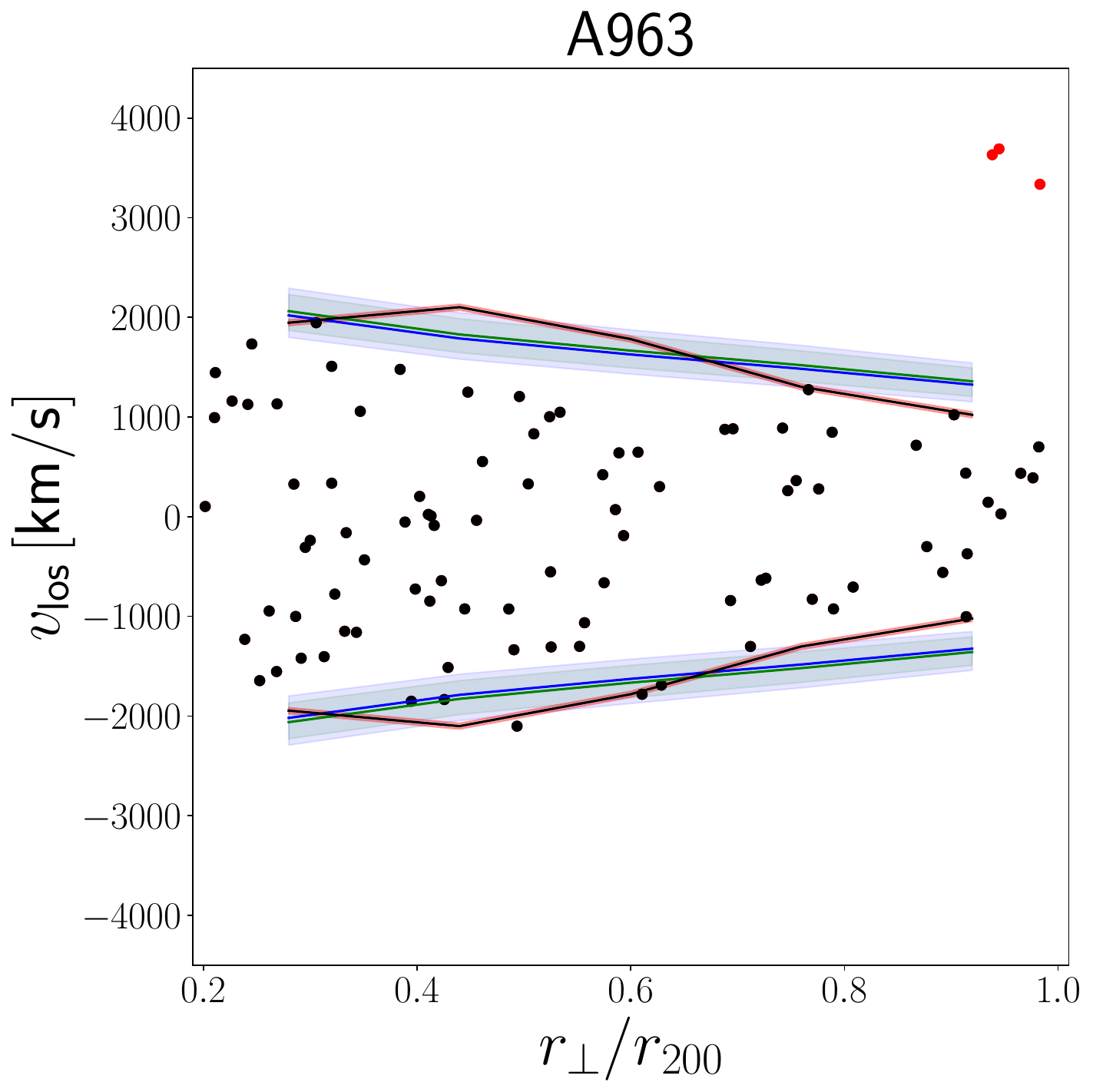}
\includegraphics[width=.19\textwidth,height=.12\textheight,keepaspectratio]{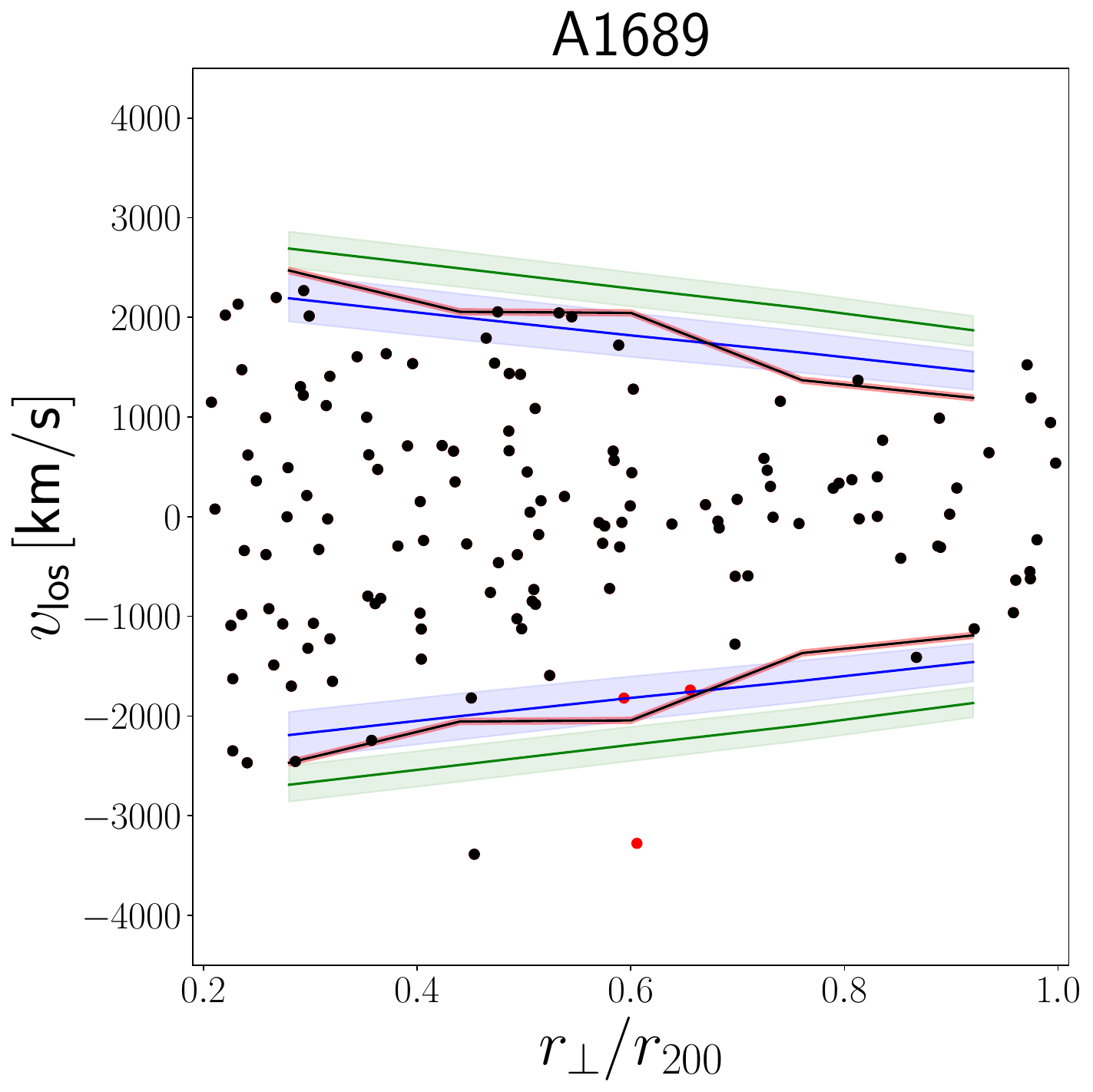}\\[-.7em]
\includegraphics[width=.19\textwidth,height=.12\textheight,keepaspectratio]{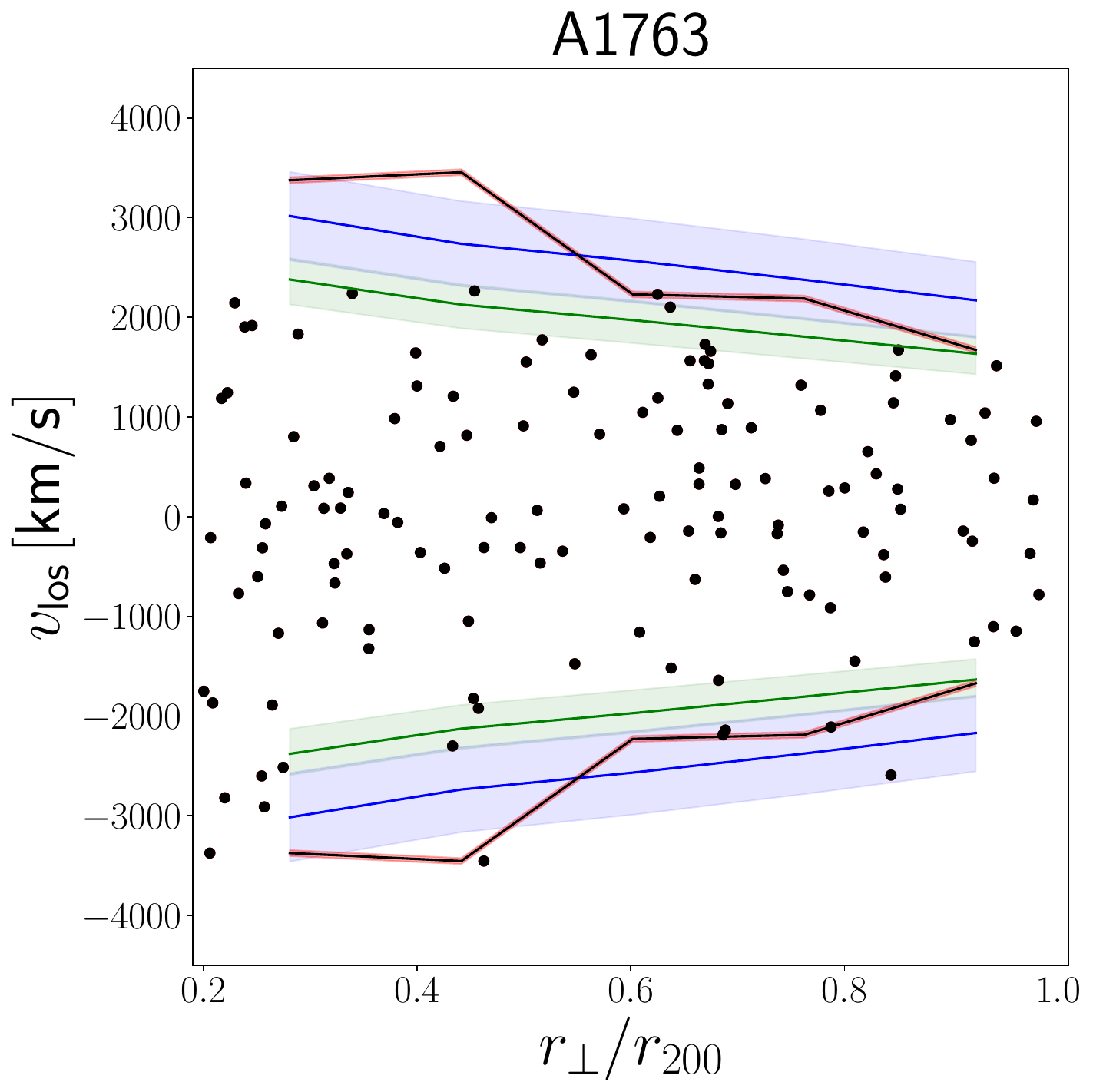}
\includegraphics[width=.19\textwidth,height=.12\textheight,keepaspectratio]{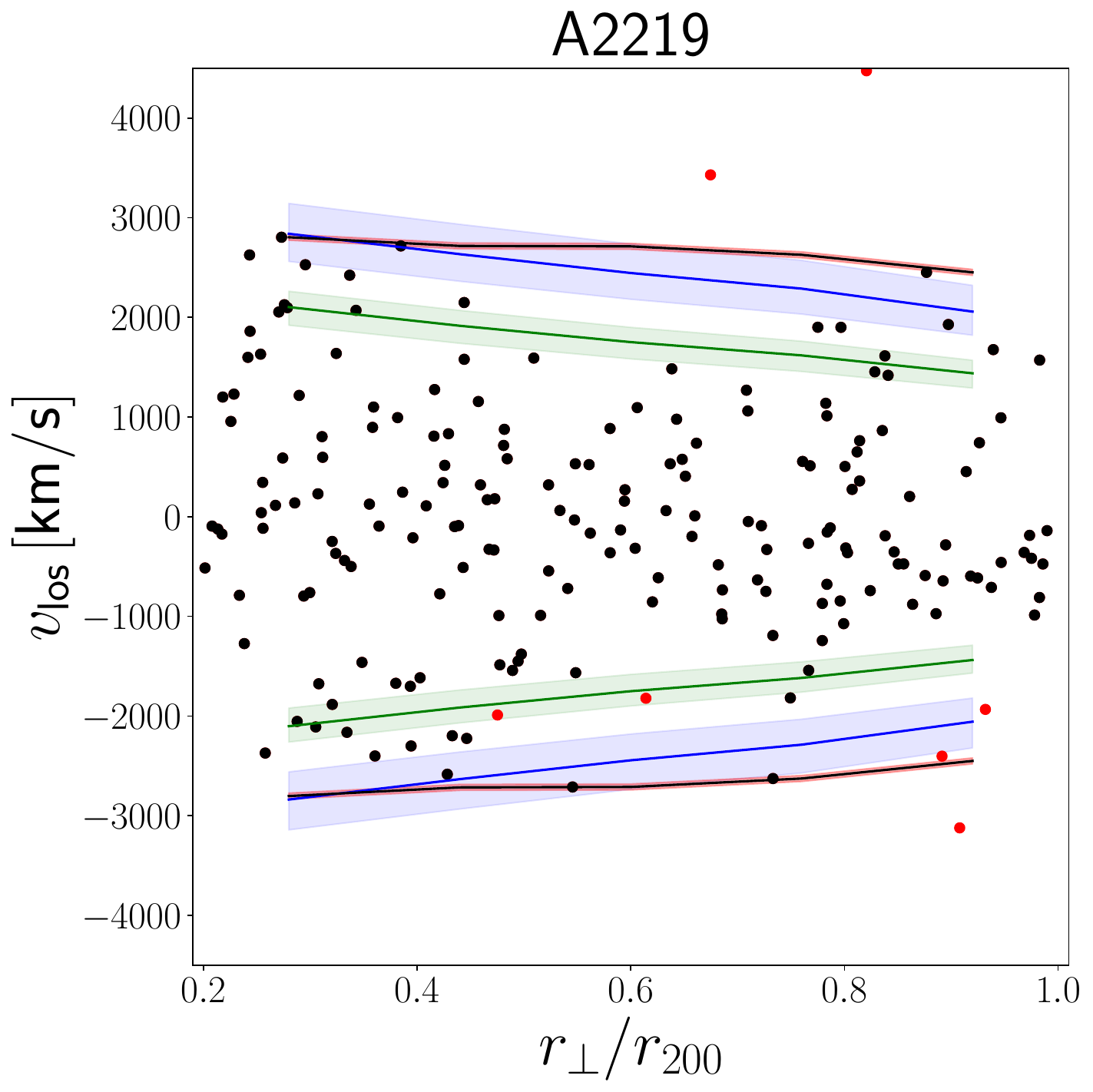}
\includegraphics[width=.19\textwidth,height=.12\textheight,keepaspectratio]{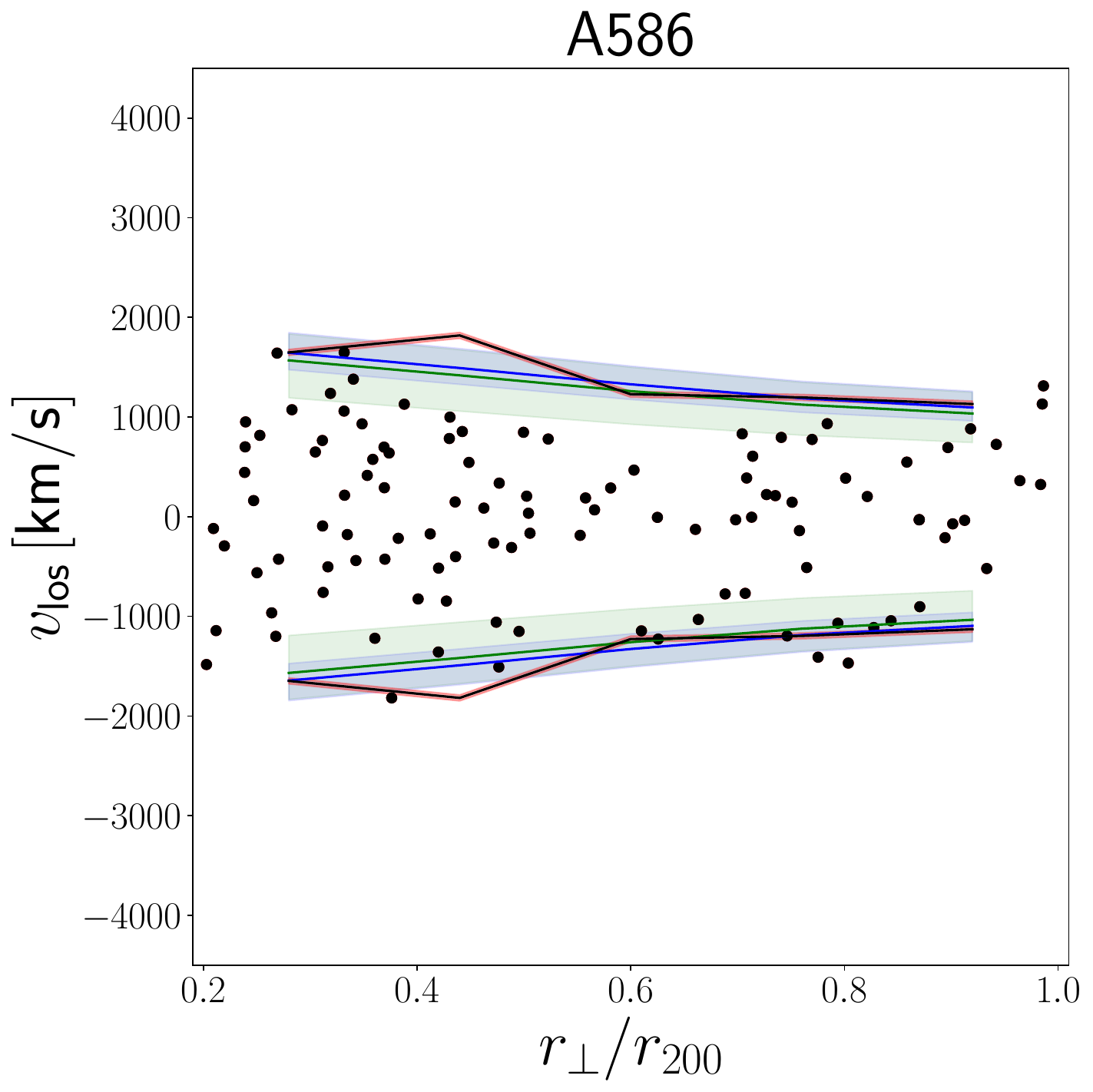}
\includegraphics[width=.19\textwidth,height=.12\textheight,keepaspectratio]{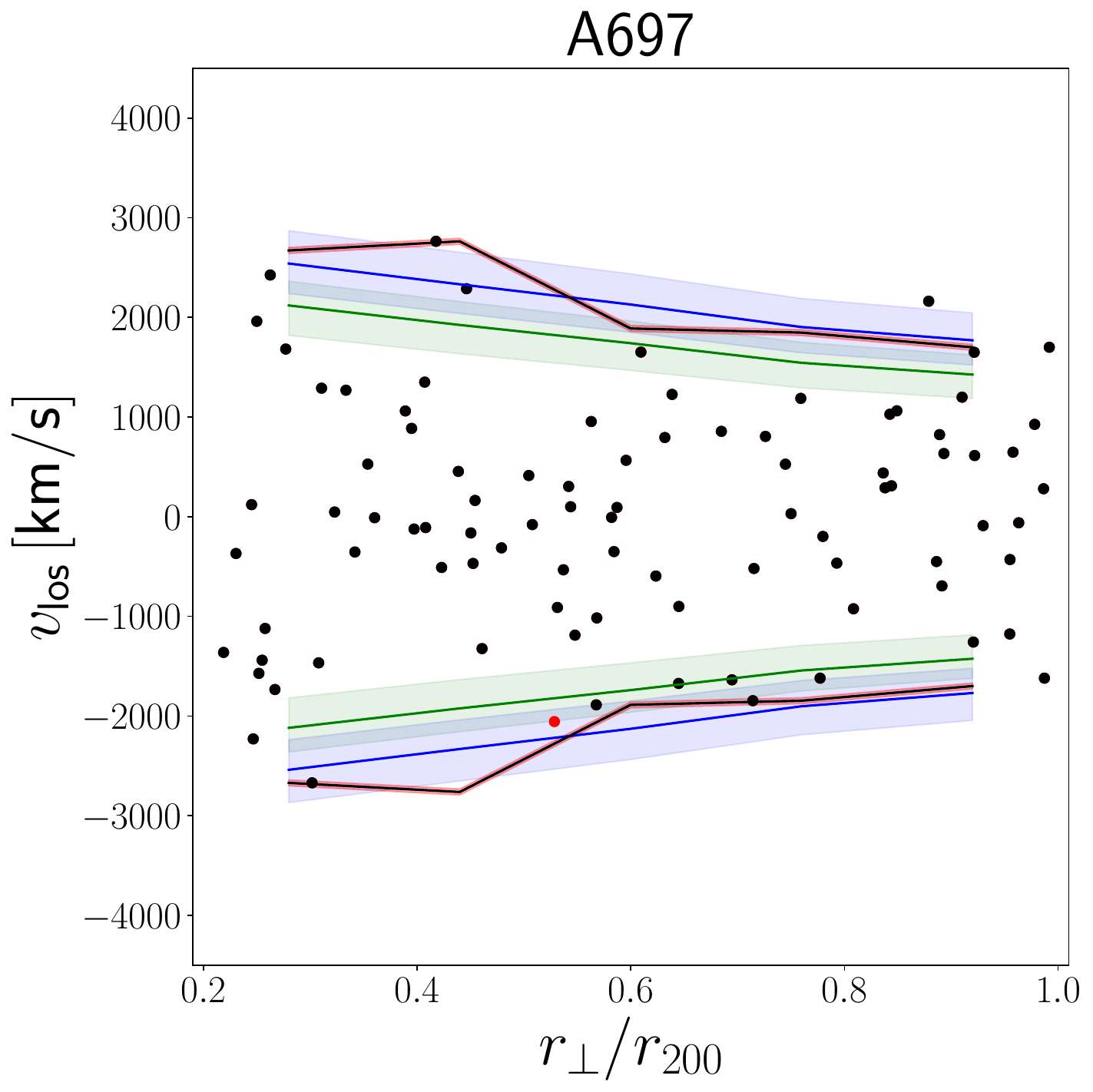}
\includegraphics[width=.19\textwidth,height=.12\textheight,keepaspectratio]{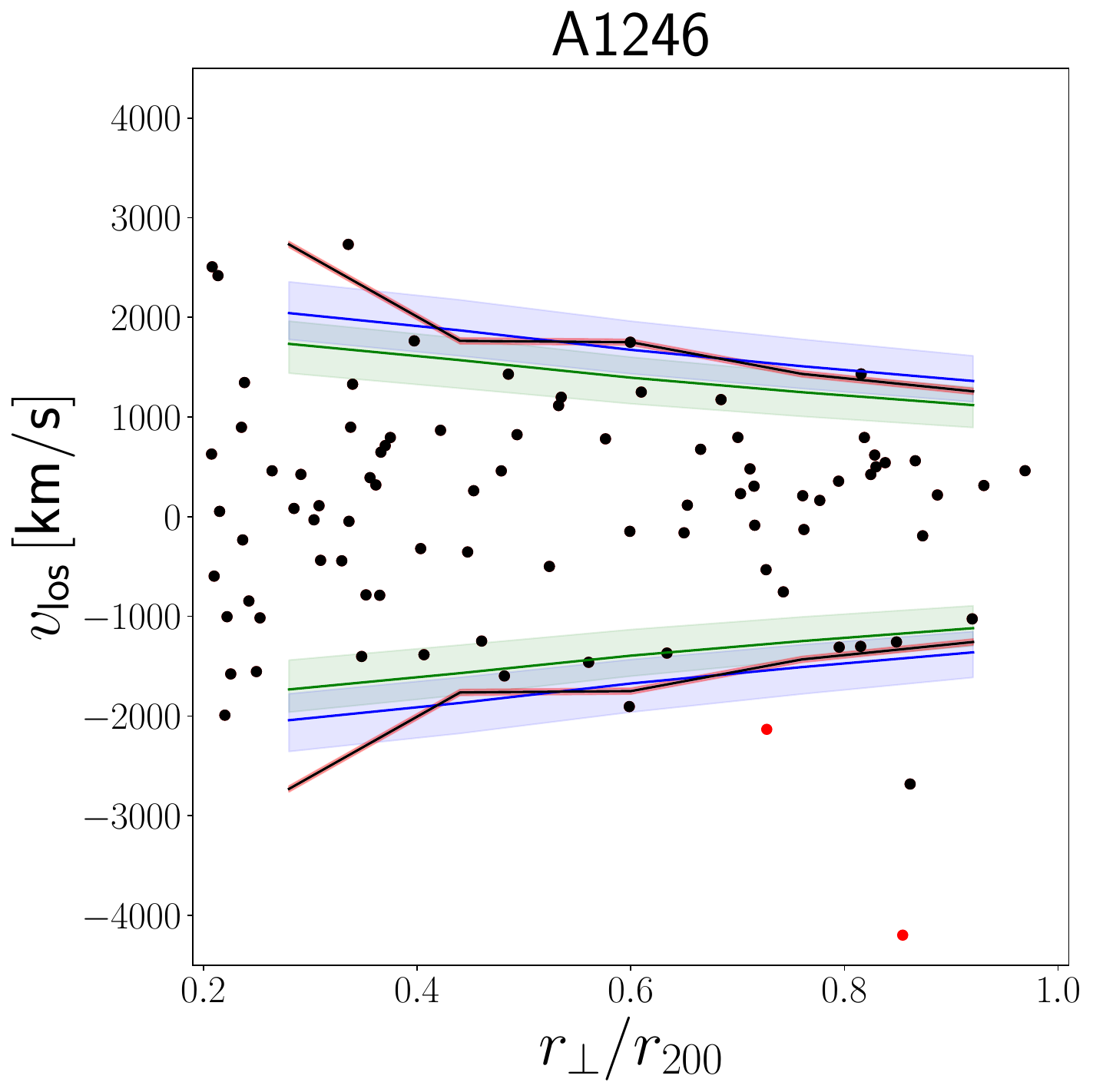}\\[-.7em]
\includegraphics[width=.19\textwidth,height=.12\textheight,keepaspectratio]{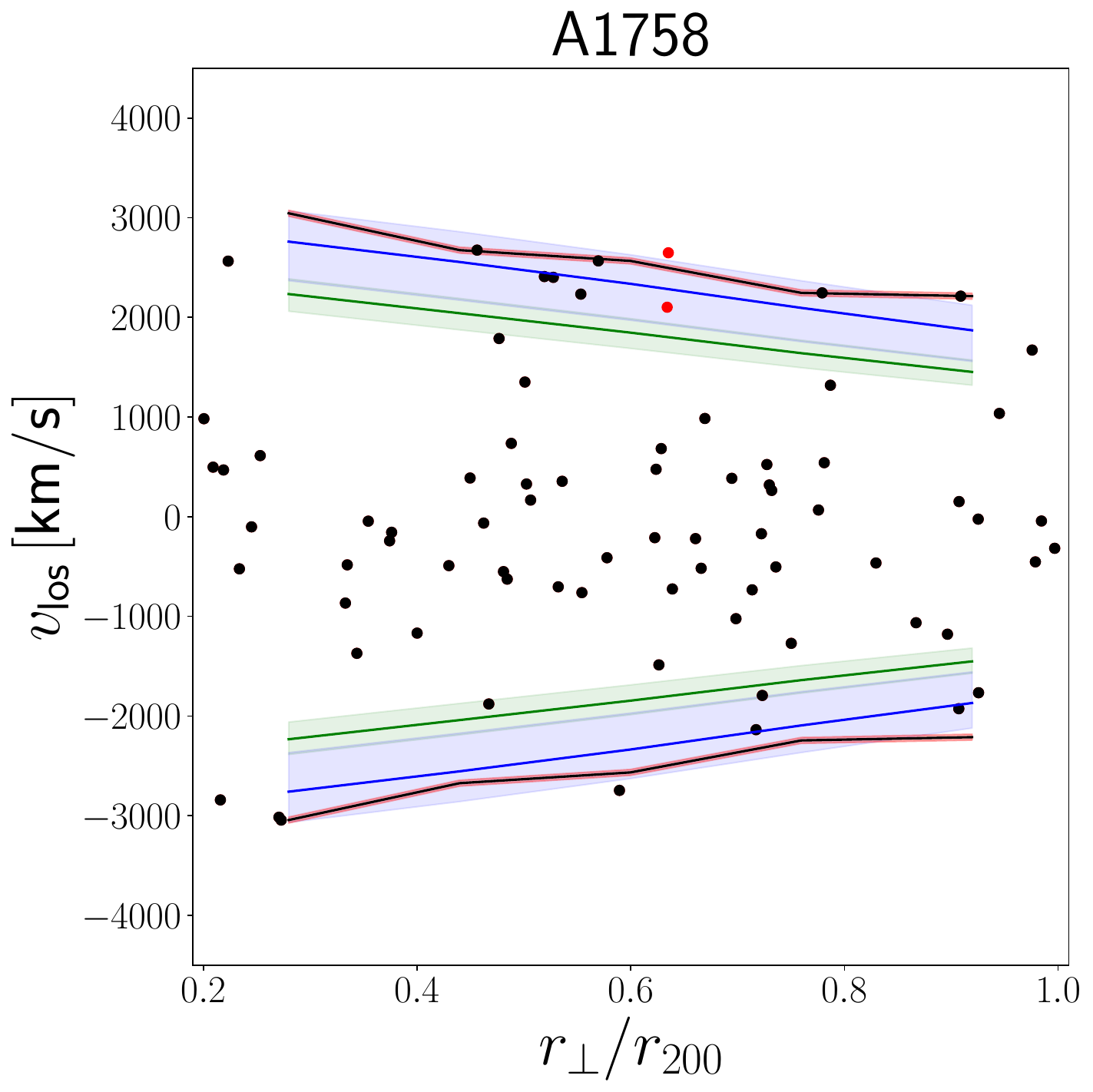}
\includegraphics[width=.19\textwidth,height=.12\textheight,keepaspectratio]{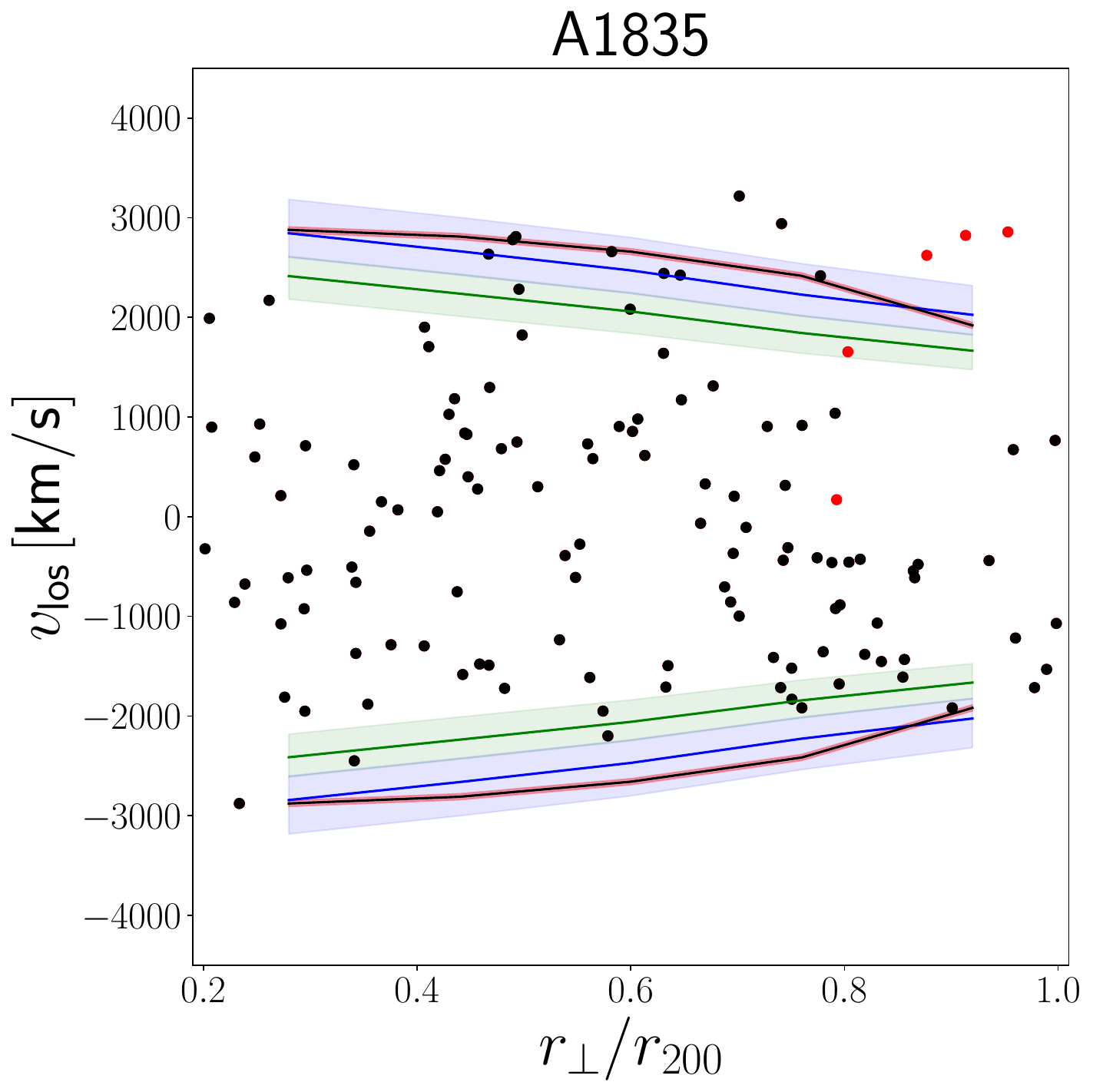}
\includegraphics[width=.19\textwidth,height=.12\textheight,keepaspectratio]{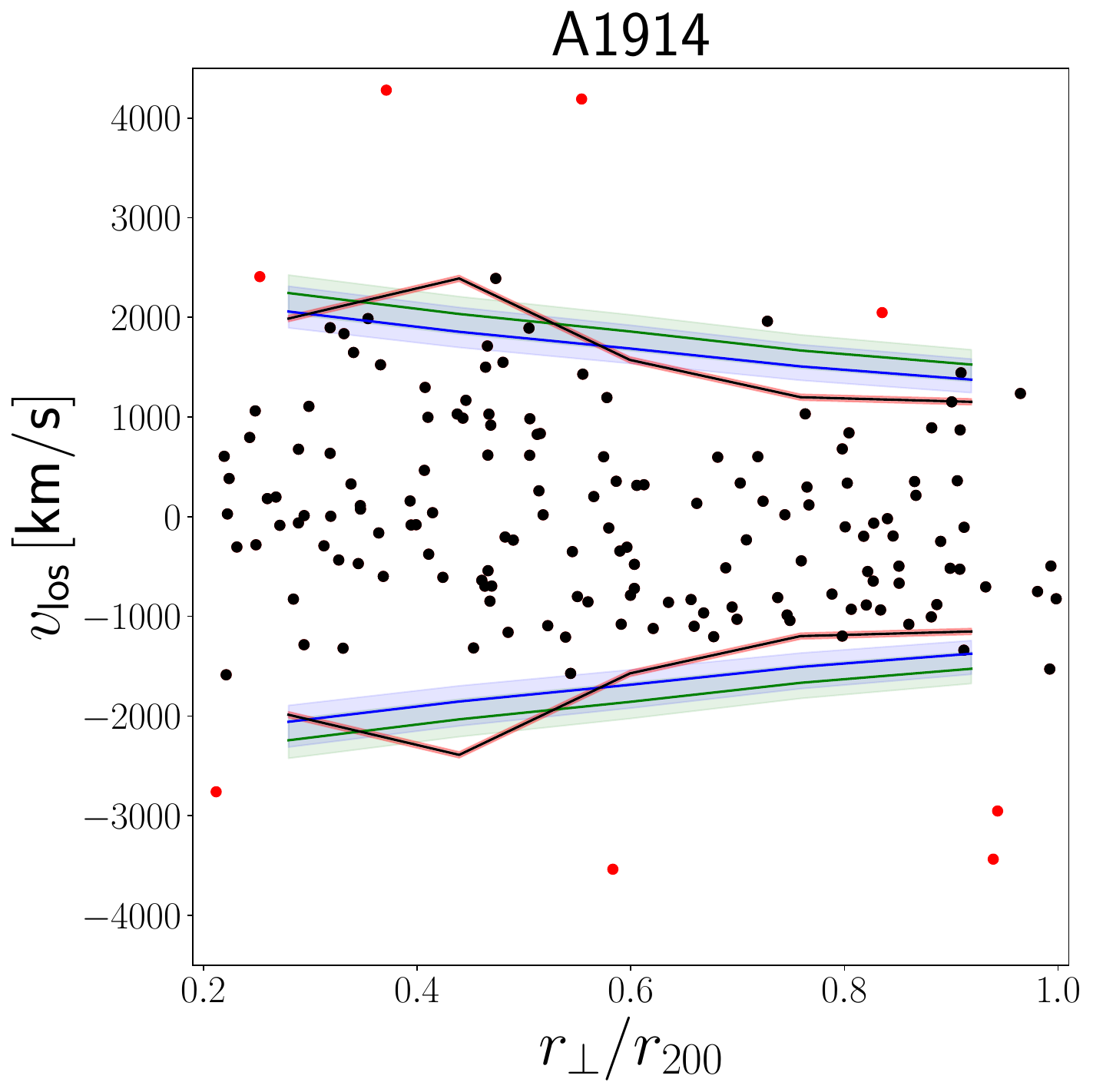}
\includegraphics[width=.19\textwidth,height=.12\textheight,keepaspectratio]{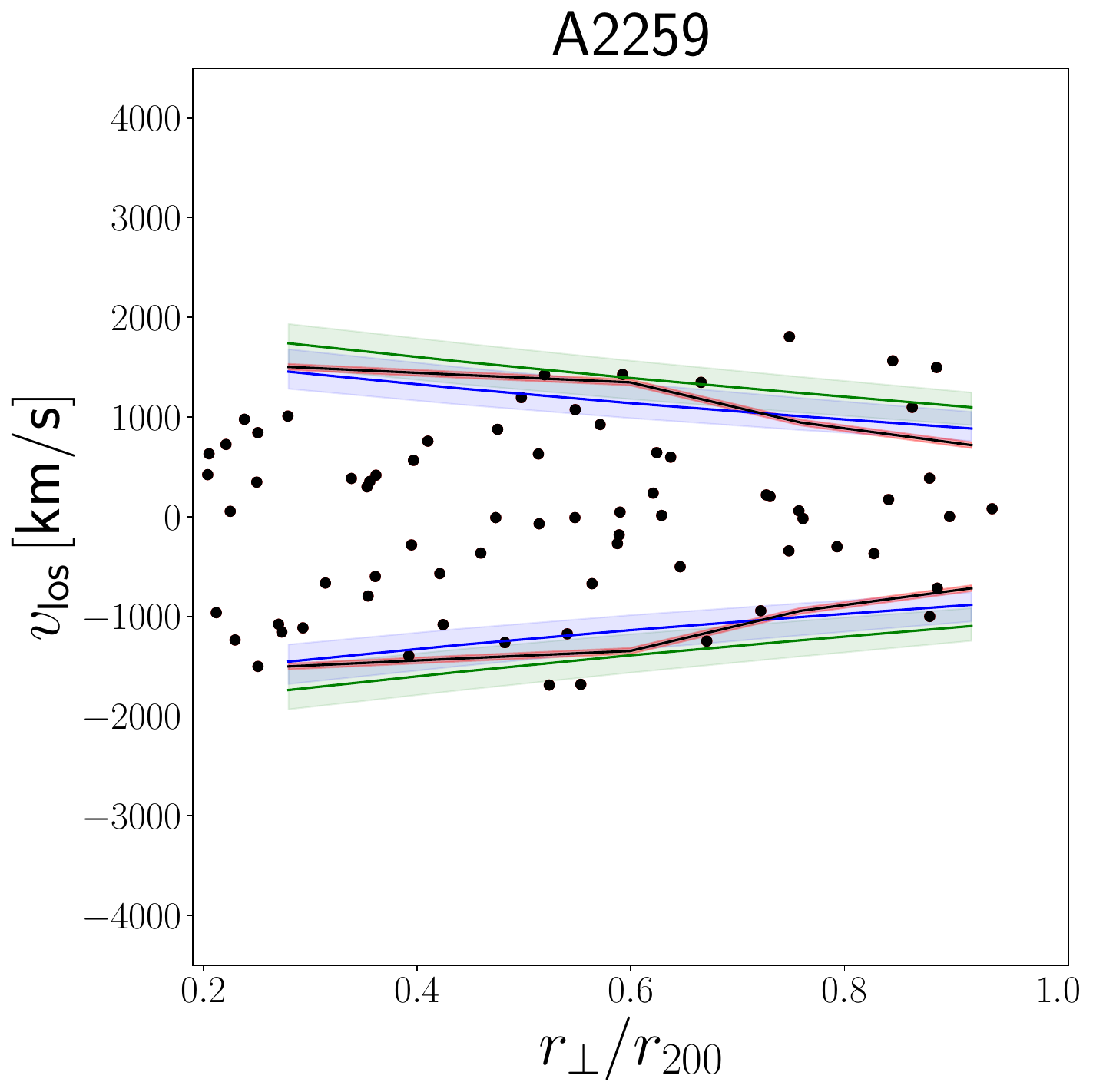}
\includegraphics[width=.19\textwidth,height=.12\textheight,keepaspectratio]{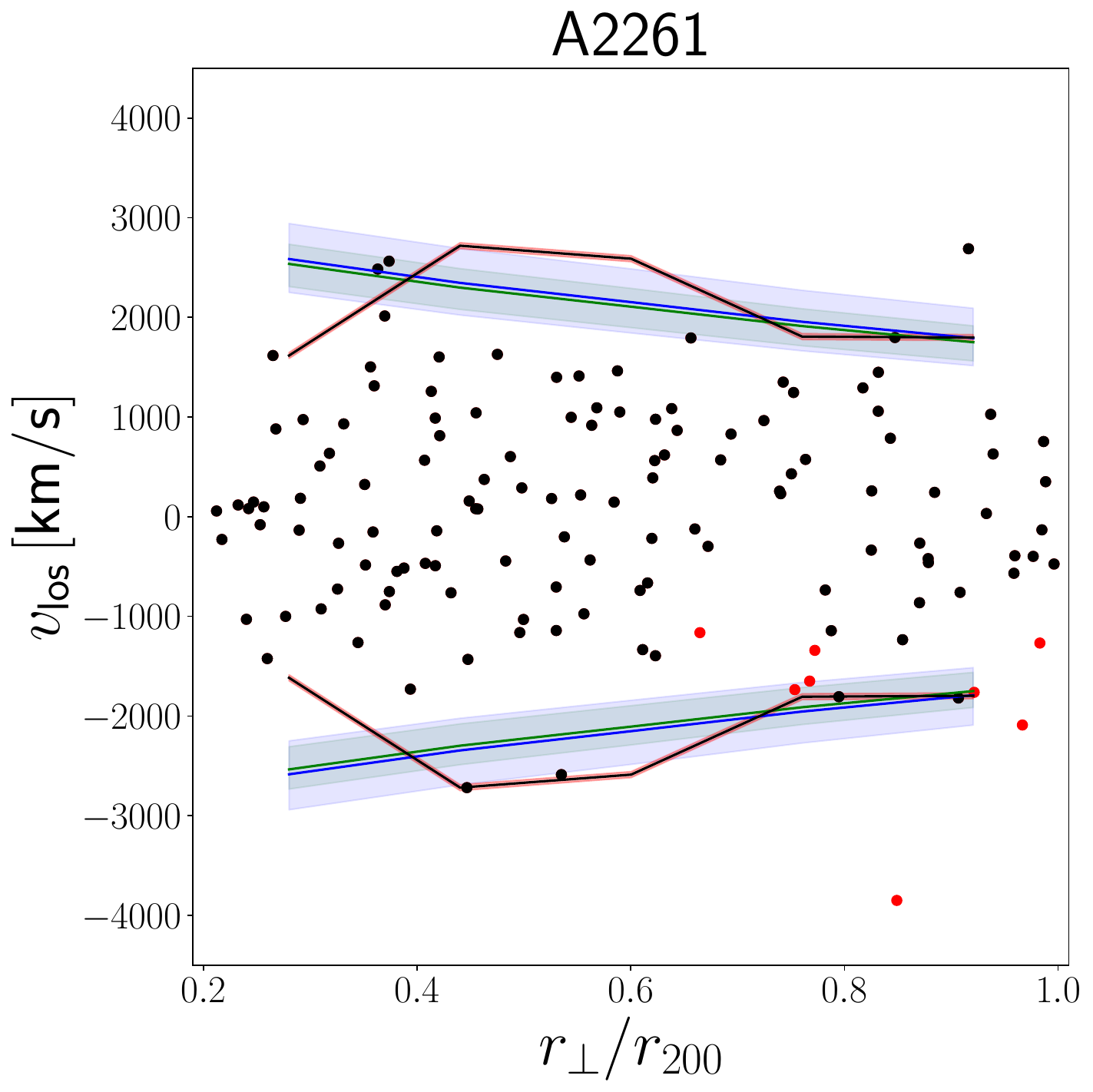}\\[-.7em]
\includegraphics[width=.19\textwidth,height=.12\textheight,keepaspectratio]{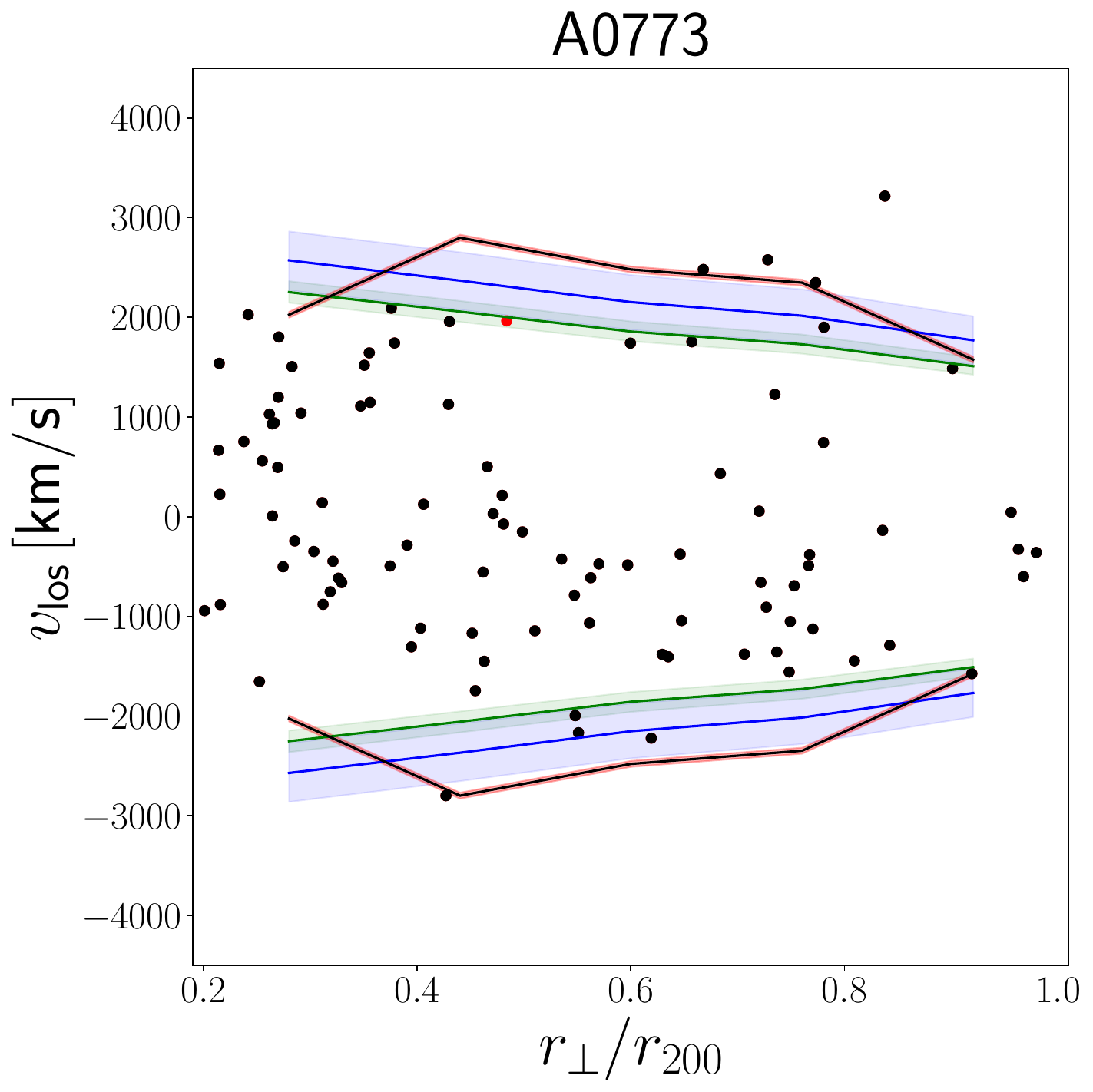}
\includegraphics[width=.19\textwidth,height=.12\textheight,keepaspectratio]{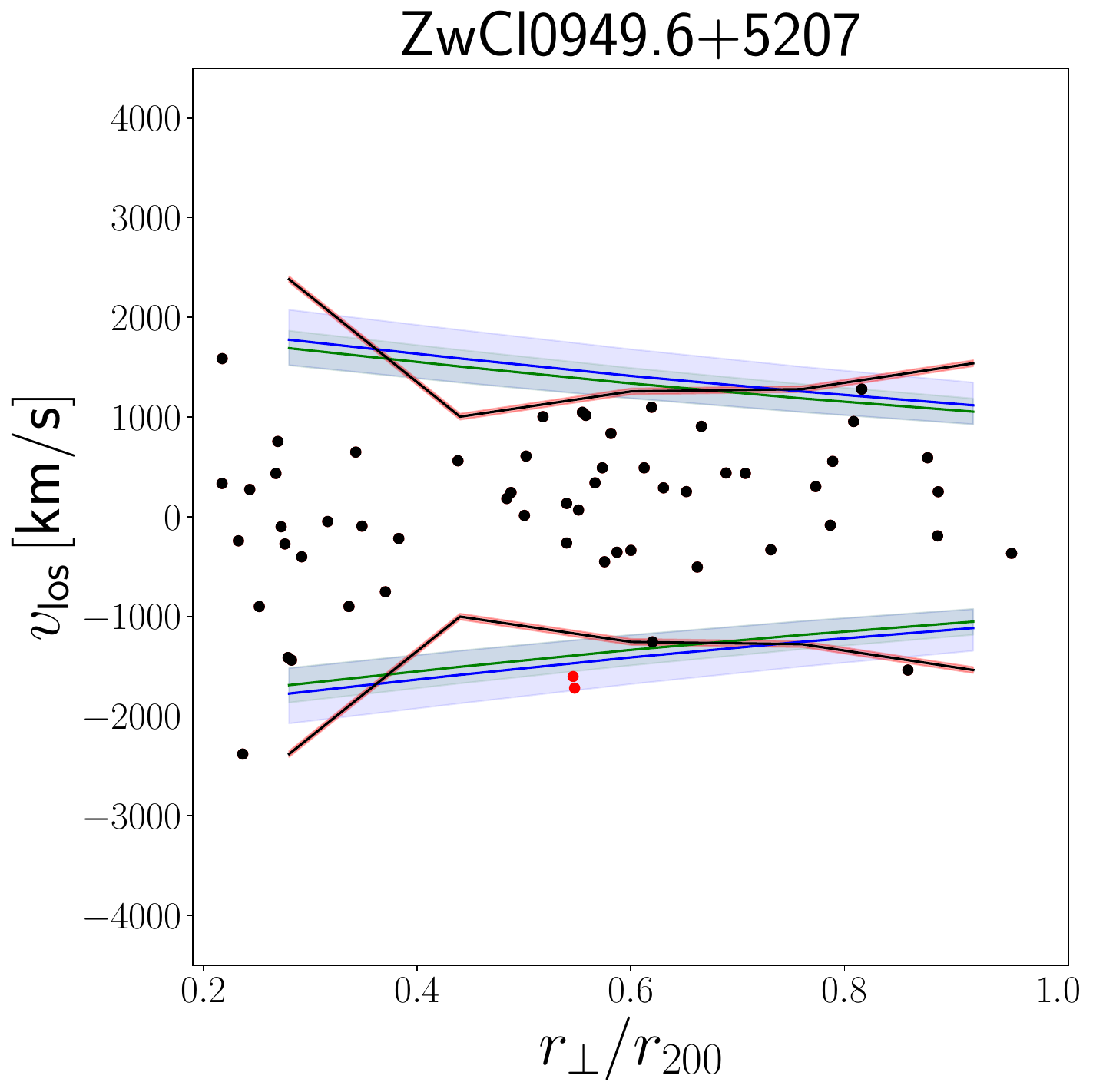}
\includegraphics[width=.19\textwidth,height=.12\textheight,keepaspectratio]{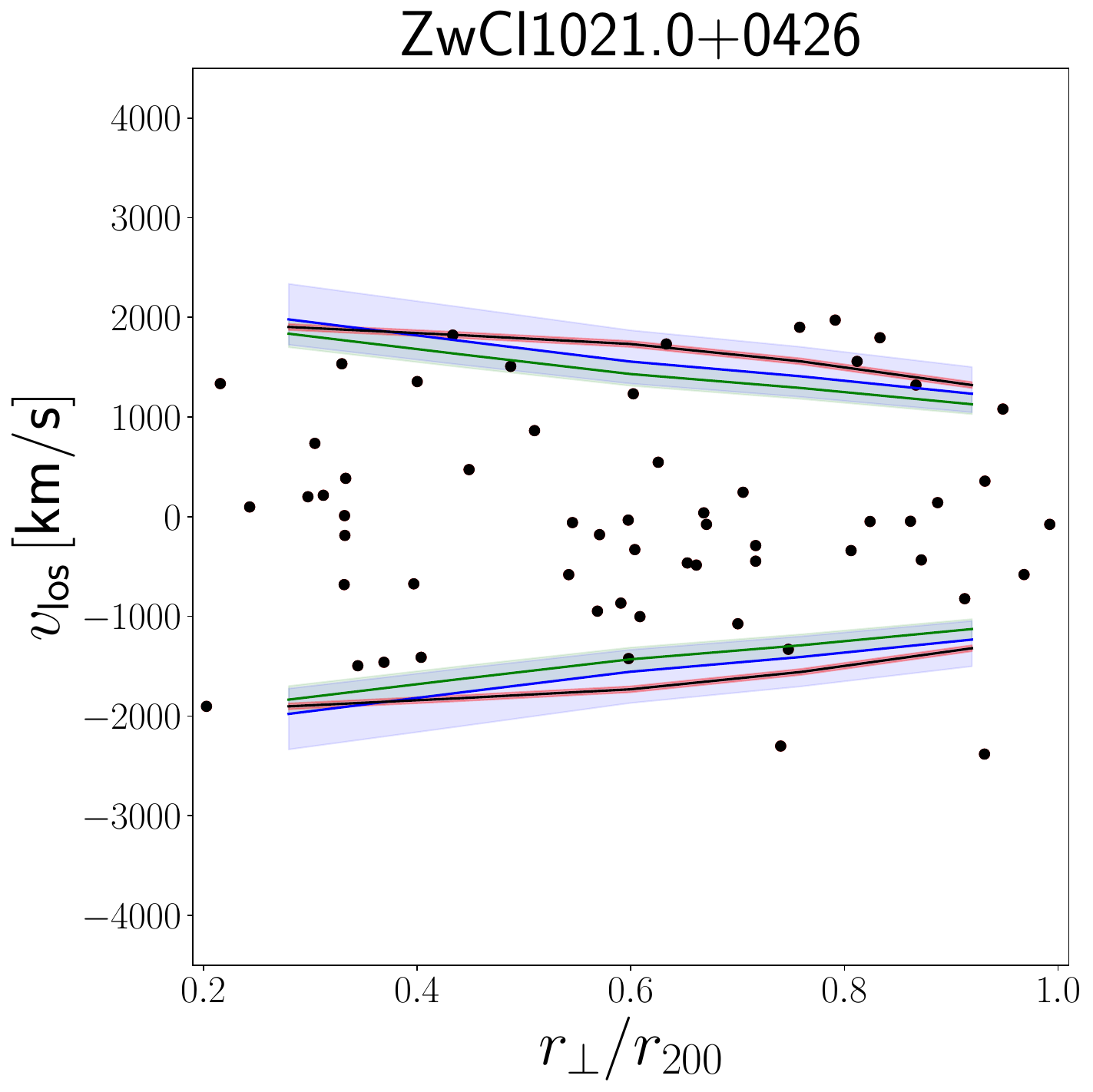}
\includegraphics[width=.19\textwidth,height=.12\textheight,keepaspectratio]{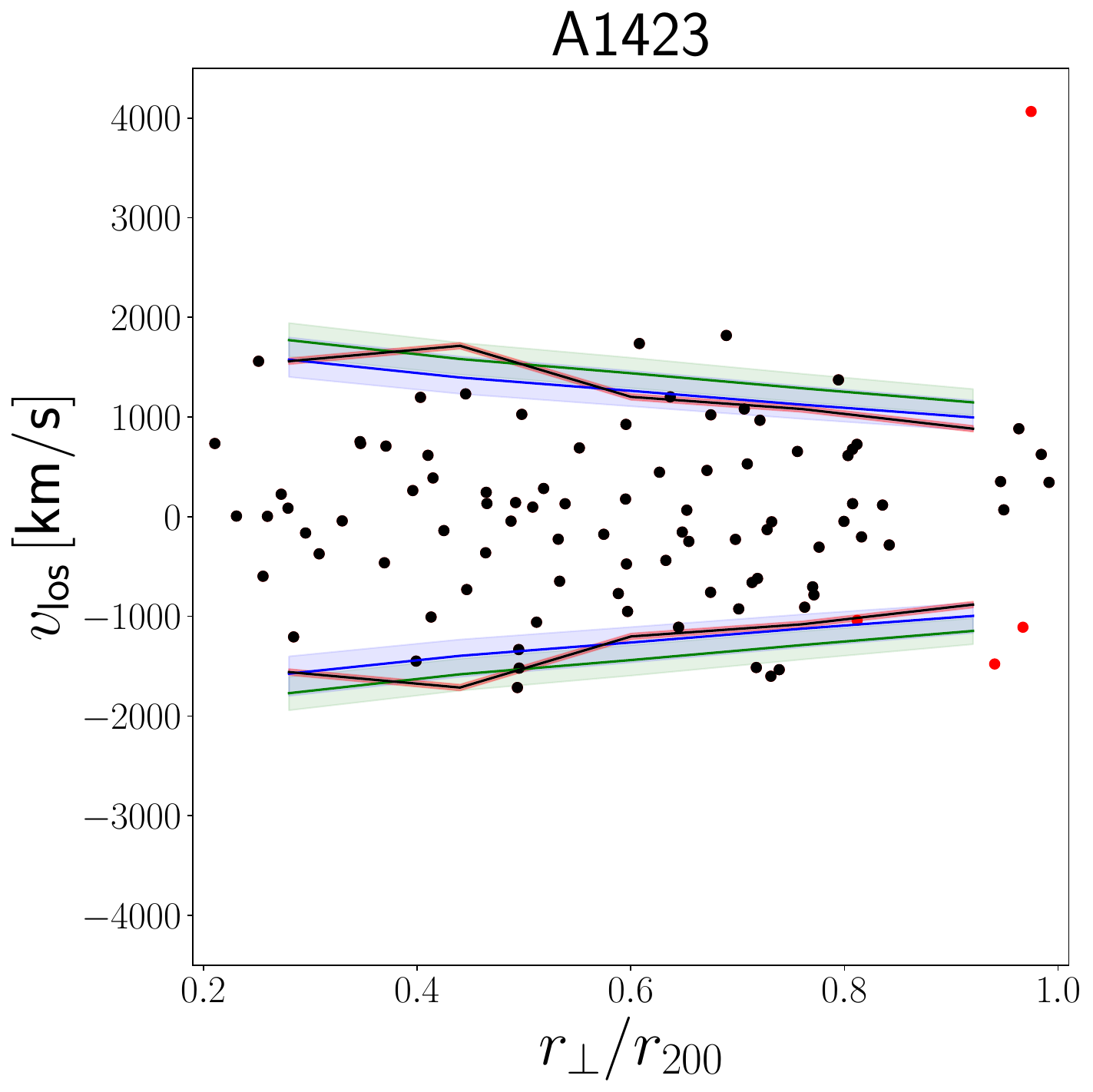}
\includegraphics[width=.19\textwidth,height=.12\textheight,keepaspectratio]{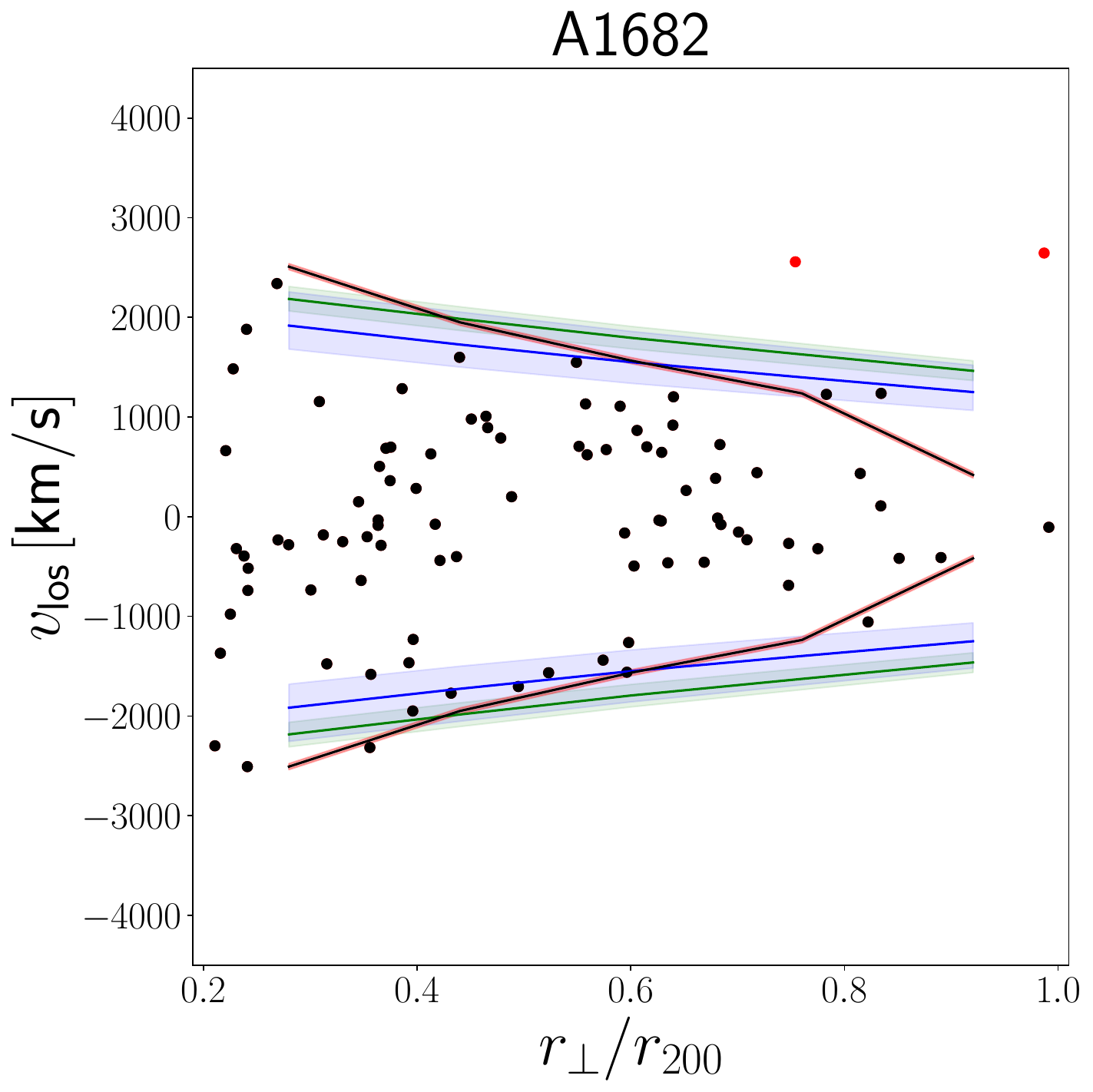}\\[-.7em]
\includegraphics[width=.19\textwidth,height=.12\textheight,keepaspectratio]{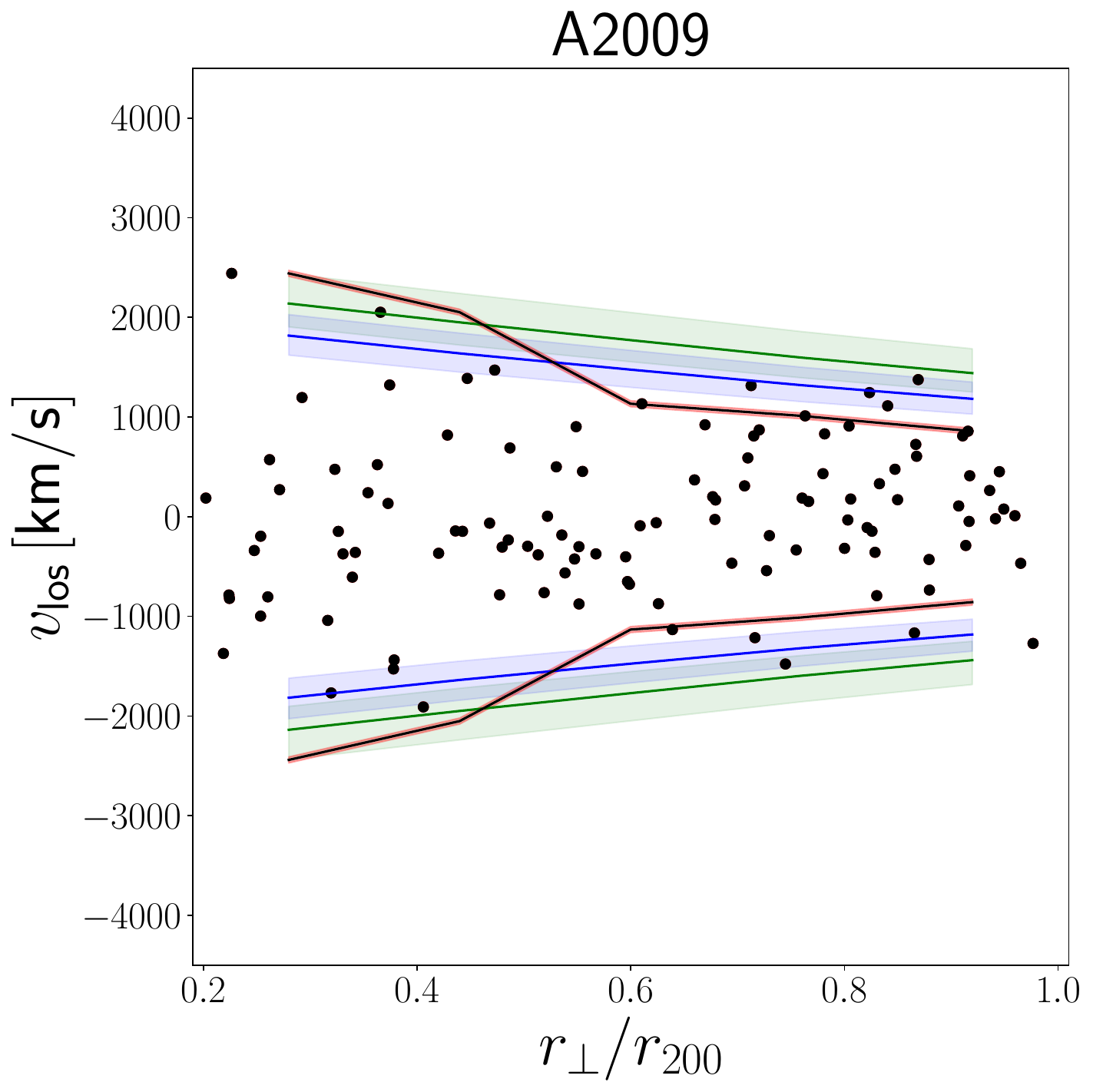}
\includegraphics[width=.19\textwidth,height=.12\textheight,keepaspectratio]{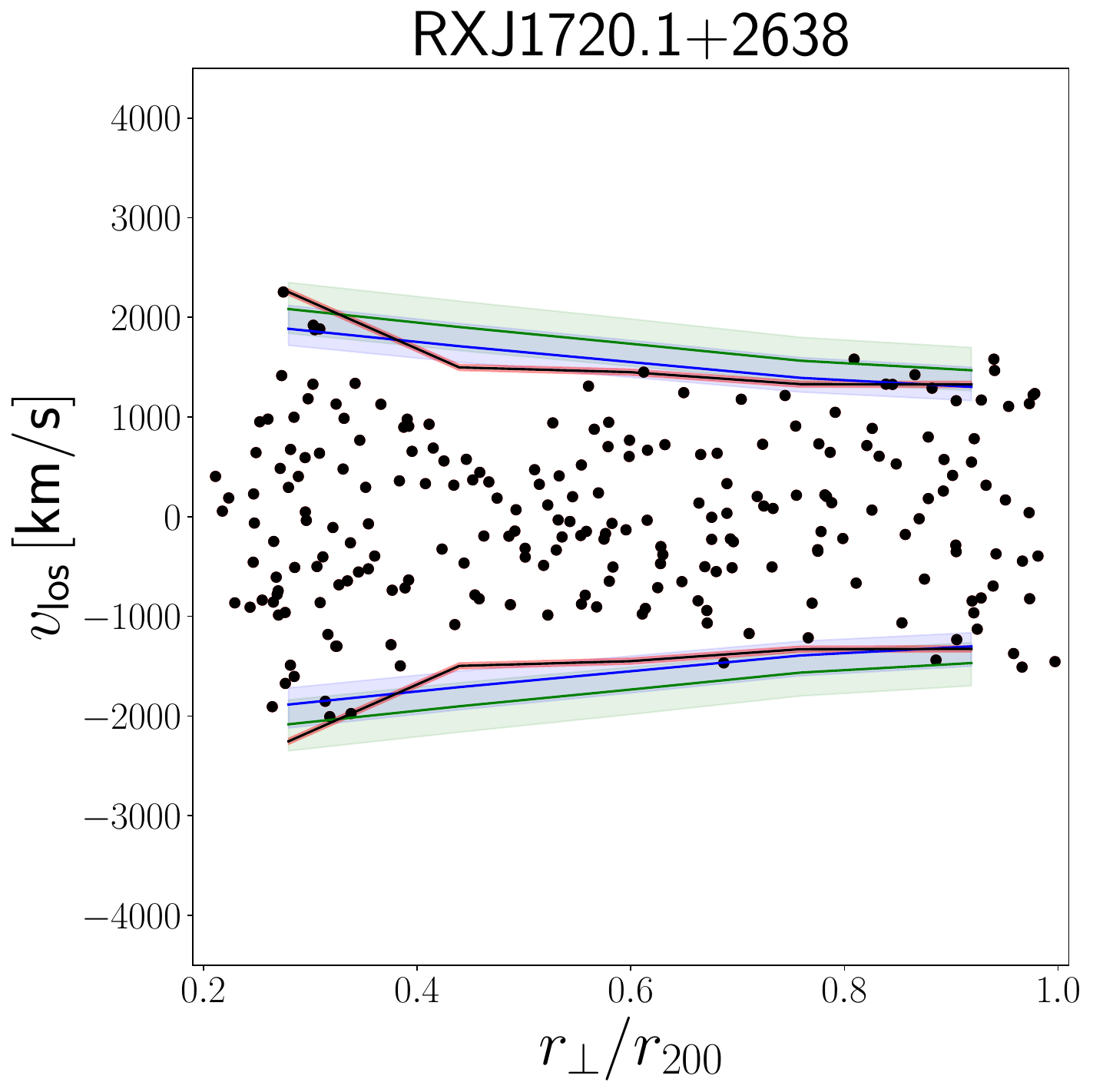}
\includegraphics[width=.19\textwidth,height=.12\textheight,keepaspectratio]{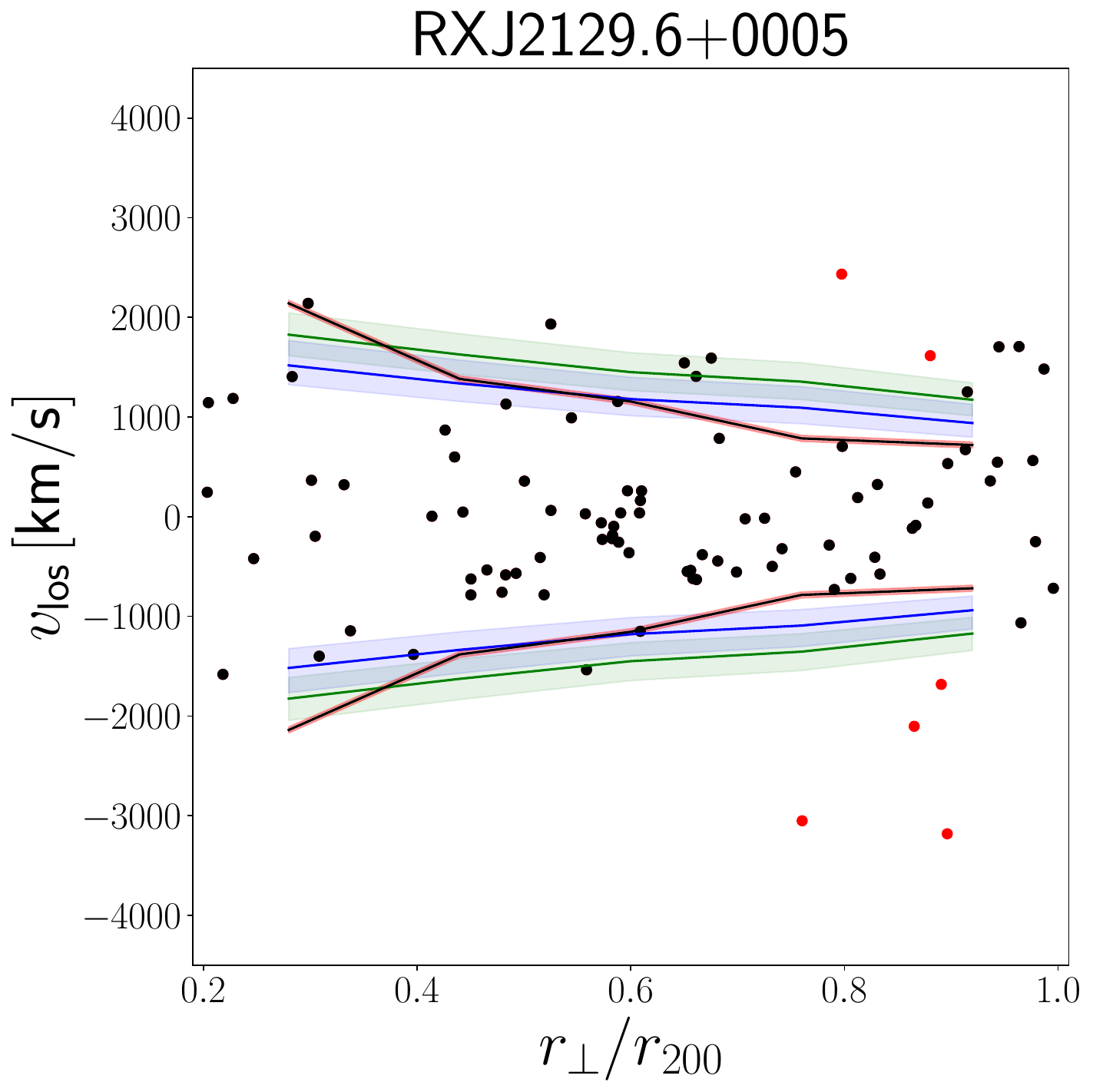}
\includegraphics[width=.19\textwidth,height=.12\textheight,keepaspectratio]{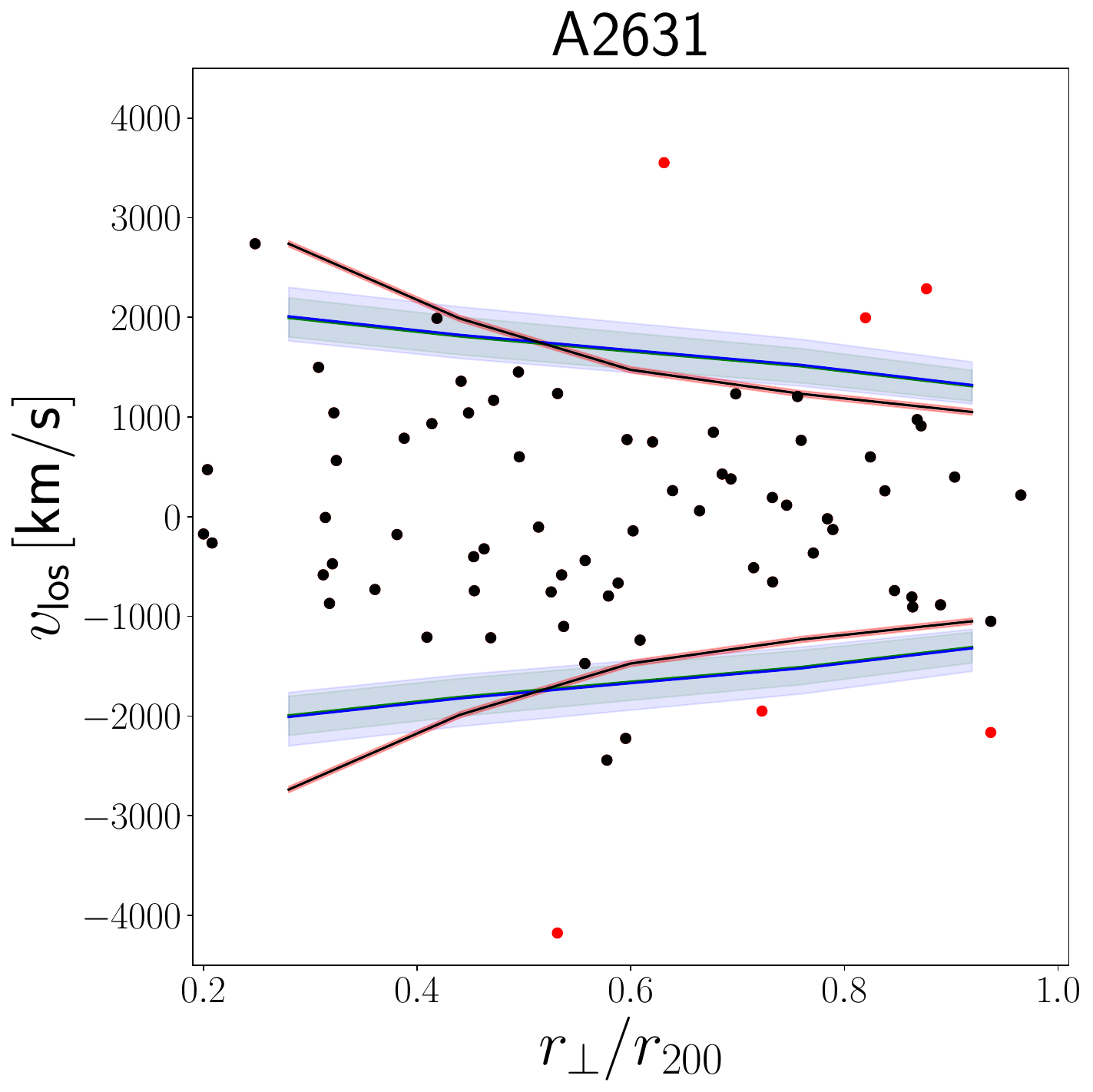}
\includegraphics[width=.19\textwidth,height=.12\textheight,keepaspectratio]{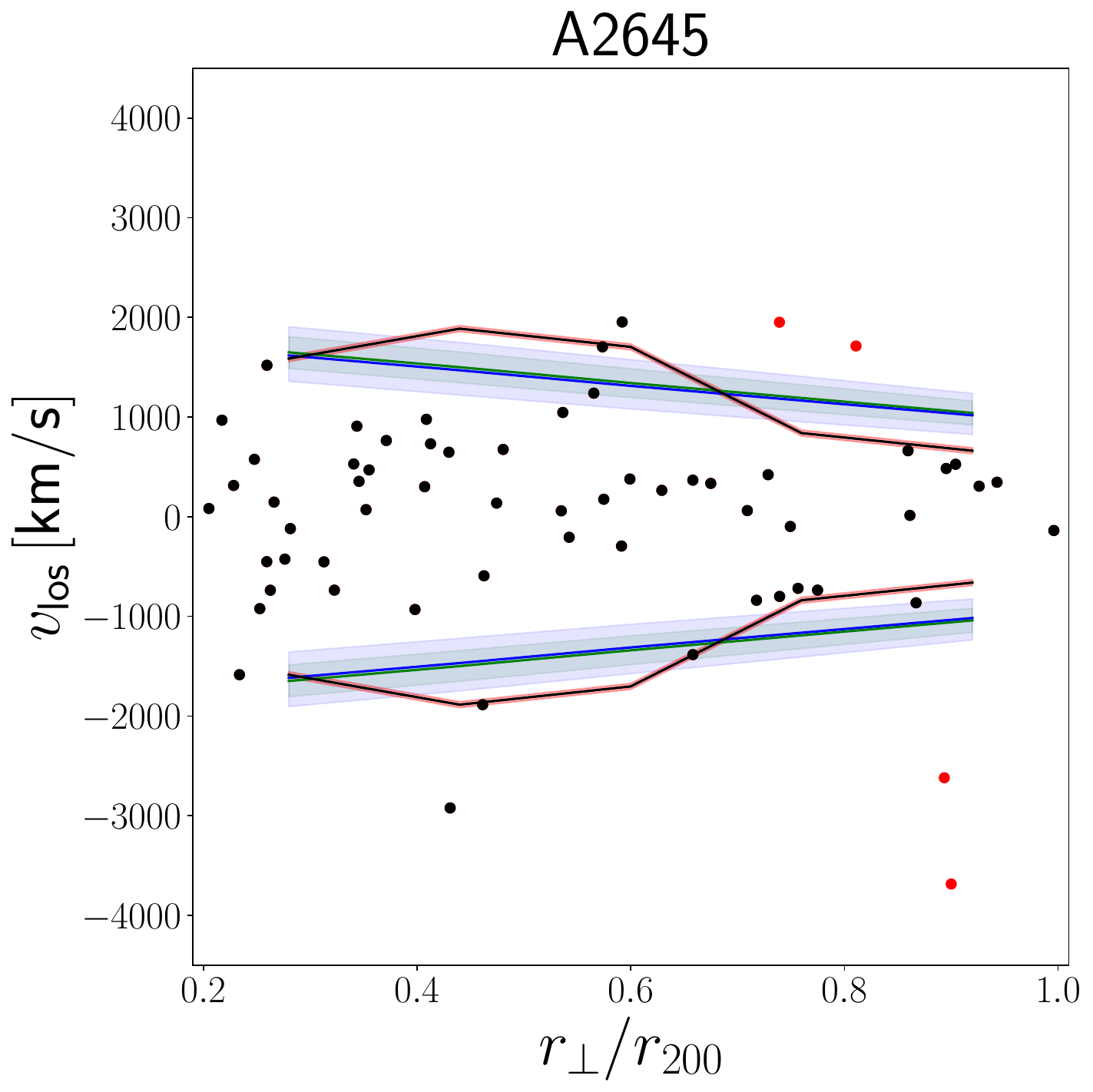}
\caption{\label{} Phase-space diagrams of 45 clusters in the sample (AS1063 not shown, see \citet{Rodriguez+2024}). \arnote{The black points indicate member galaxies, the red points indicate interlopers, the red lines indicate the identified phase-space edge, the blue lines indicate the dynamical fits, and the green lines indicate the suppressed lensing profiles.}}
\end{figure}

\end{document}